\documentclass[11pt]{amsart}

\usepackage{amsmath}
\usepackage{amssymb}
\usepackage{graphicx}

\usepackage{mathtools}
\mathtoolsset{showonlyrefs}
\usepackage{float}
\usepackage{subcaption}
\usepackage{bbold}
\usepackage{hyperref}
\usepackage{fullpage}
\usepackage{xcolor}
\hypersetup{
  colorlinks   = true, %Colours links instead of ugly boxes
  urlcolor     = purple, %Colour for external hyperlinks
  linkcolor    = violet, %Colour of internal links
  citecolor   = teal %Colour of citations
}
\usepackage[giveninits=true,uniquename=false,uniquelist=false,style=alphabetic,sortcites=false,natbib=true,backend=biber,datamodel=mrnumber, maxbibnames=99, block=space,maxnames=99,maxalphanames=99]{biblatex}
\DeclareFieldFormat{mrnumber}{%
  MR\addcolon\space
  \ifhyperref
{\href{http://www.ams.org/mathscinet-getitem?mr=#1}{\nolinkurl{#1}}}
    {\nolinkurl{#1}}}
\renewbibmacro*{doi+eprint+url}{%
  \iftoggle{bbx:doi}
    {\printfield{doi}}
    {}%
  \newunit\newblock
  \printfield{mrnumber}%
  \newunit\newblock
  \iftoggle{bbx:eprint}
    {\usebibmacro{eprint}}
    {}%
  \newunit\newblock
  \iftoggle{bbx:url}
    {\usebibmacro{url+urldate}}
    {}}

\newtheorem{theorem}{Theorem}[section]
\newtheorem{corollary}[theorem]{Corollary}
\newtheorem{rhp}[theorem]{Riemann-Hilbert problem}
\newtheorem{conjecture}[theorem]{Conjecture}

\newtheorem{proposition}[theorem]{Proposition}
\theoremstyle{definition}

\newtheorem{remark}[theorem]{Remark}

  \newcommand{\ee}{\mathrm{e}}
  \newcommand{\ii}{\mathrm{i}}
  \newcommand{\dd}{\mathrm{d}}
\numberwithin{equation}{section}
\title[$\beta =6$ Tracy-Widom distribution and Calogero-Painlev\'e system]{On $\beta =6$ Tracy-Widom distribution and the second Calogero-Painlev\'e system}

\author[A. R. Its]{Alexander Its}

\address[A. R. Its]{Department of Mathematical Sciences, Indiana University, Indianapolis, 402 N. Blackford St. Indianapolis, IN 46202, USA,
 \& St. Petersburg State University, 7/9 Universitetskaya nab. 199034 St. Petersburg, Russia.}
\email{{\tt aits@iu.edu}}

\author[A. O. Prokhorov]{Andrei Prokhorov}
\address[A. O. Prokhorov]{Department of Mathematical Sciences, University of 
Cincinnati, 2600 Clifton Ave,
Cincinnati, OH, 45221, USA \& St. Petersburg State University, 7/9 Universitetskaya nab. 199034 St. Petersburg, Russia}
\email{\tt prokhoai@ucmail.uc.edu}

\keywords{asymptotic analysis, Riemann-Hilbert problems, beta ensembles, Tracy-Widom distribution, Calogero-Painlev\'e systems.}

\subjclass[2020]{  33E17, 34E05,34M55, 34M55, 34M56,	37J35, 41A60, 60B20, 82B44}

\begin{document}

\begin{abstract}
The Calogero-Painlev\'e systems were introduced in 2001 by K. Takasaki as a natural generalization of the classical Painlev\'e equations to the case
of the several Painlev\'e ``particles''  coupled via the Calogero type interactions. In 2014, I. Rumanov discovered the remarkable fact that a particular case of the second Calogero-Painlev\'e II equation describes the Tracy-Widom distribution function for the general
beta-ensembles with even values of the parameter beta. Most recently, in 2017 work of M. Bertola, M. Cafasso, and V. Rubtsov, it was proven that all Calogero-Painlev\'e systems are Lax integrable, and hence their solutions admit a Riemann-Hilbert representation. This important observation has opened the door to rigorous, based on the Deift-Zhou nonlinear steepest descent method,
asymptotic analysis of the Calogero-Painlev\'e equations. This in turn yields the possibility of rigorous evaluation of the asymptotic behavior of the Tracy-Widom distributions for the values of beta beyond the classical $\beta =1, 2, 4.$
 In this work we shall start an asymptotic analysis of the Calogero-Painlev\'e system  with a special focus on the Calogero-Painlev\'e system corresponding to $\beta = 6$ Tracy-Widom distribution function. The principle technical challenge is the implementation of the nonlinear steepest descent approach beyond the
$2\times 2$ matrix dimension of the corresponding Riemann-Hilbert problem; in our case, it is $6\times 6$.
\end{abstract}

\maketitle
\begin{center}
	{\it Dedicated to Percy  Deift  on the occasion of his 80th birthday.}
\end{center}
\tableofcontents
 \section{Introduction and Main Result}
\subsection{Calogero-Painlev\'e system}
Denote $x_j(t)$, $j = 1, ..., n$ the positions of $n$ 1D particles.  The second Calogero-Painlev\'e model is the dynamical system given by the equations
\begin{equation}\label{CalPII}
x_k''=2x_k^3+tx_k + \theta +2g^2\sum_{j\neq k}\frac{1}{(x_k-x_j)^3}, \quad k=1,... n.
\end{equation}
These are Hamiltonian equations,
$$
x'_k = \frac{\partial H}{\partial y_k} = y_k,\quad x''_k = y'_k = -\frac{\partial H}{\partial x_k},
$$
 with the  Hamiltonian, 
\begin{equation}\label{HamCP}
H=\sum_{k=1}^n\left(\frac{y_k^2}{2}-\frac{x_k^4}{2}-\frac{x_k^2t}{2}-\theta x_k\right) + g^2\sum_{j< k}\frac{1}{(x_k-x_j)^2}. 
\end{equation}
System \eqref{CalPII}  can be thought of as a collection of the n independent Painlev\'e II particles interacting via the classical Calogero 
potential. This system was introduced by K. Takasaki in 2001 \cite{Tak}, together with five other similar systems associated with each of the remaining five Painlev\'e equations, as non-autonomous generalization of the Inozemtsev systems \cite{Ino}. The Calogero-Painlev\'e VI version
of the model had also showed up in a completely different context in the work of P. Etingof, W. L. Gan, and A. Oblomkov \cite{EGO} in connection
with their study of the generalized double affine algebras of higher rank. 
Our principal interest in system \eqref{CalPII} is its relation, discovered by I. Rumanov \cite{Rum1,Rum2,Rum3}, to the Tracy-Widom
distribution function of the general beta-ensemble of random matrices. 

\subsection{General beta-ensemble and Calogero-Painlev\'e systems}

Given $\beta >0$, Dyson's  $\beta$-ensemble is defined as a Coulomb gas of $N$ charged particles, that is as the space  
of $N$ one dimensional particles, $(\lambda_1,\ldots,\lambda_N)\in\mathbb{R}^N$ with the 
probability density given by the equation,  
\begin{equation}\label{densityeigenvalues}
	 p(\lambda_1,\ldots,\lambda_N)\dd\lambda_1\ldots \dd\lambda_N  = \frac{1}{Z_N} \prod_{1\leq j<k \leq N }|\lambda_j-\lambda_k|^\beta 
	 {\rm e}^{-\beta \sum_{j=1}^{N}V(\lambda_j)}\dd\lambda_1\ldots \dd\lambda_N ,
\end{equation}
\begin{equation}\label{partfunc}
Z_{N} = \int_{-\infty}^{\infty} ... \int_{-\infty}^{\infty}\prod_{1\leq j<k \leq N }|\lambda_j-\lambda_k|^\beta {\rm e}^{-\beta \sum_{j=1}^NV(\lambda_j)}\dd\lambda_1\ldots \dd\lambda_N. 
\end{equation} 
Here, $V(\lambda)$ has a meaning of external field which we will assume to be Gaussian, i.e. $V(\lambda) = \frac{\lambda^2}{2}$. The objects of interest are the gap probabilities in the large $N$ limit. We will be particularly concerned with the soft edge probability distribution
\begin{equation}
F_{\beta}(t) \equiv E_{\beta}^{\mbox{soft}}\Bigl(0; (t, \infty)\Bigr) = \lim_{N\rightarrow \infty}E_{\beta N}^{\mbox{soft}}\left(0;\Bigl(\sqrt{2N}
+\frac{t}{\sqrt{2}N^{1/6}}, \infty\Bigr)\right),
\end{equation}
where
\begin{equation}\label{mint}
E_{\beta N}^{\mbox{soft}}\Bigl(0; (t, \infty)\Bigr) = \int_{-\infty}^{t} ... \int_{-\infty}^{t} p(\lambda_1,\ldots,\lambda_N)\dd\lambda_1\ldots \dd\lambda_N.
\end{equation}
It was shown in \cite{Bourgade_erdos_yau} that this distribution is the same for a large class of potentials $V(\lambda)$. Cases $\beta = 1, 2, 4$ are known as Gaussian orthogonal (GOE), Gaussian unitary (GUE) and Gaussian symplectic
(GSE) ensembles. In fact, in these cases, the distribution \eqref{densityeigenvalues} describes the statistics of the eigenvalues of orthogonal, Hermitian, and symplectic random matrices, respectively, with i.i.d. matrix entries. The corresponding limiting edge distribution functions $F_{\beta}(s)$ are then becoming the classical {\it Tracy-Widom distributions} \cite{TW, Tracy_widom_orthogoanl}. They admit explicit representations either as the Airy kernel Fredholm determinants or in terms of the Hastings-McLeod solution of the second Painlev\'e
 equation. These representations, in turn, allow one to evaluate the asymptotic expansions of $F_{\beta}(t)$ as $t \to \pm\infty$,
  i.e. the so-called {\it tail asymptotics}.

 A principal issue is the asymptotic analysis of $F_{\beta}(s)$ beyond the classical values $\beta = 1, 2, 4$.
 The crux of the problem is that the orthogonal polynomial approach, which is the principal technique in the random matrix case,
 is not available for general $\beta$. However, several highly nontrivial conjectures concerning the general $\beta$ ensembles have been suggested. An excellent presentation of the state of the art in this area is given in the survey by P. Forrester \cite{For}.
 The current principal heuristic result concerning the asymptotic behavior of the generalized Tracy-Widom distribution
$F_{\beta}(t)$ was obtained in 2010 by G. Borot, B. Eynard, S. N. Majumdar and C. Nadal
and it reads as follows.
\begin{conjecture}[\cite{BEMN}]\label{bemncon1}
\begin{align}\label{con1}
F_{\beta}(t) =& \exp\left(-\beta\frac{|t|^3}{24}
+ \frac{\sqrt{2}}{3}\Bigl(\beta/2-1\Bigr)|t|^{3/2}\right.\\
&\left.+ \frac{1}{8}\Bigl(\beta/2 + 2/\beta-3\Bigr)\log|t| 
+ \chi  + O\left(\frac{1}{|t|^{3/2}}\right)\right), \quad t \to -\infty.
\end{align}
The constant term $\chi$ is also explicitly predicted. Indeed, it is claimed  that
\begin{align}\label{conconst}
\chi =& \frac{\beta}{2}\left(\frac{1}{12}-\zeta'(-1) \right) + \frac{\gamma}{6\beta}
 -\frac{\log2\pi}{4}-\frac{\log (\beta/2)}{2}\\ &
  +\left(\frac{17}{8}-\frac{25}{24}\Bigl(\beta/2 + 2/\beta \Bigr)\right)\log 2
 +\int_{0}^{\infty}\frac{1}{\ee^{\beta t/2}-1} \left(\frac{t}{\ee^t-1} -1 +\frac{t}{2} -\frac{t^2}{12}\right)\frac{\dd t}{t^2},
 \end{align}
 where $\zeta(z)$ is Riemann's zeta-function and $\gamma$ denotes Euler's constant.
 \end{conjecture}
 \begin{conjecture}[\cite{BN}]
     \begin{align}\label{expconst}
F_\beta(t)= 1-\frac{\Gamma\left(\frac{\beta}{2}\right)\exp\left(-\frac{2\beta }{3}t^\frac{3}{2}\right)}{(4\beta)^\frac{\beta}{2}2\pi t^{\frac{3\beta}{4}}}\left(1+O\left(\dfrac{1}{t^{\frac{3}{2}}}\right)\right),\quad t\to \infty.
\end{align}
 \end{conjecture}
 \noindent
Formulae \eqref{con1} -- \eqref{expconst} have been derived in \cite{BEMN,BN} within the framework of the
so-called loop-equation technique by performing the relevant double scaling limit directly in the formal large $N$
expansion of the multiple integral in \eqref{mint}.
\begin{remark}
For even $\beta$ the integral above was evaluated in \cite{BEMN} as $$
\int_{0}^{\infty}\frac{1}{\ee^{\beta t/2}-1} \left(\frac{t}{\ee^t-1} -1 +\frac{t}{2} -\frac{t^2}{12}\right)\frac{\dd t}{t^2}=-\sum_{m=1}^{\beta/2-1}\frac{2m}{\beta}\log\Gamma\left(\frac{2m}{\beta}\right)+\frac{2}{\beta}\zeta'(-1)-\frac{\log(\beta/2)}{6\beta}$$
$$
+\frac{\beta}{8}\log(2\pi)-\frac{\gamma}{6\beta}-\frac{\beta}{2}\left(\frac{1}{12}-\zeta'(-1) \right). 
$$
\end{remark}
\begin{remark} Rigorous derivation of \eqref{con1} -- \eqref{conconst},  for the basic classical  case $\beta =2$ is given
by  P. Deift et al in \cite{DIK} and in \cite{BBDiF}. In \cite{BBDiF} the cases $\beta =1, 4$ are also done. Both papers use
the Riemann-Hilbert method which is available to the classical cases of $\beta = 1, 2, 4$. The formula \eqref{expconst} for that cases was known already in \cite{TW,Tracy_widom_orthogoanl}.
\end{remark}
\begin{remark} The leading asymptotic term in \eqref{con1} for arbitrary $\beta$  has been rigorously obtained 
in \cite{RRV} with the help of the analysis of the certain stochastic Schr\"odinger operator.
\begin{align}\label{con1rough}
F_{\beta}(t) =& \exp\left(-\beta\frac{|t|^3}{24}(1+o(1))\right), \quad t \to -\infty.
\end{align}
We shall
mention this paper again in the next paragraph. 
Similarly the right tail estimate for arbitrary $\beta$ was obtained in \cite{dumaz_virag} 
    \begin{align}
F_{\beta}(t)  =1-t^{-\frac{3\beta}{4} + o[(\log t)^{-1/2}]}\ee^{-\frac{2\beta}{3} t^{3/2}}, \quad t\rightarrow \infty.
\end{align}
\end{remark}
\begin{remark}
    Similarly to continuous ensemble \eqref{densityeigenvalues}, the discrete $\beta$ ensemble was introduced in \cite{Borodin_Gorin_Guionnet}. The largest particle behavior is described by the same Tracy-Widom $\beta$ distribution. The corresponding universality was shown in \cite{GH}.
\end{remark}

It is remarkable, that paper \cite{BEMN},  while giving a such detailed formulae for the asymptotics of the
distribution function $F_{\beta}(t)$ does not actually  produce any description of the object itself (for the finite values of $t$). 
The latter has been done  by A. Bloemendal and B. Virag \cite{BV}. Inspired by the pioneering work of E. Dumitriu and A. Edelman \cite{DE} and by subsequent works \cite{VV} and \cite{RRV},  Bloemendal and Virag \cite{BV}
have connected the analysis of the generalized Tracy-Widom distribution $F_{\beta}(t)$ with the study of stochastic Schr\"odinger operators.  In particular, 
it has been proven in \cite{BV} that the  Tracy-Widom distribution function  $F_{\beta}(t)$, for any $\beta$, 
can be expressed in terms of the solution of a certain linear PDE. In more detail, the result of \cite{BV} can be formulated
as follows.

Consider the  partial differential equation,
\begin{equation}\label{bv1}
\frac{\partial F}{\partial t} +\frac{2}{\beta}\frac{\partial^2F}{\partial x^2} + (t-x^2)\frac{\partial F}{\partial x} = 0,\quad (x, t)\in {\mathbb{R}}^2 
\end{equation}
supplemented by the  boundary conditions,
\begin{equation}\label{bv2}
F(x, t)  \to 1, \quad \mbox{as}\quad x, t \to \infty,\quad \mbox{together}
\end{equation}
and 
\begin{equation}\label{bv3}
F(x, t)  \to 0, \quad \mbox{as}\quad x \to -\infty, \quad \mbox{for fixed $t$.} 
\end{equation}
\begin{theorem}[\cite{BV}]\label{th1}
The boundary value problem \eqref{bv1}, \eqref{bv2}, \eqref{bv3} has a unique bounded smooth solution. Moreover,
equation,
\begin{equation}\label{bv4}
F_{\beta} (t) = \lim_{x \to \infty} F(x, t),
\end{equation}
determines the Tracy-Widom distribution function for the general value of the parameter $\beta >0$.
\end{theorem}

The boundary value problem \eqref{bv1} -- \eqref{bv3} is an efficient tool for numerical evaluation of Tracy-Widom distribution, see \cite{trogdon_zhang}. 
The next step in the studying of $F_{\beta} (t)$  has been done in the works  of I. Rumanov, \cite{Rum1,Rum2,Rum3},  who,
in the case of even values of the parameter $\beta$,  has 
reduced the analysis of the  Bloemendal-Virag  equation   to the analysis  of an auxiliary system of nonlinear
ODEs. For the first nontrivial case, $\beta =6$, Rumanov has obtained 
the following representation for $ F_{6}(t)$,
$$
\log F_{6}(3^{-2/3}t) = \frac{1}{3}\int_{t}^{\infty}\Bigl(u^4(s) + su^2(s) -(u'(s))^2\Bigr)\dd s 
$$
\begin{equation}\label{betarum}
+\frac{2}{3}\int_{t}^{\infty}\alpha(s)\dd s-\frac{1}{3}\int_{t}^{\infty}\frac{u'}{u}(s)\Bigl(1 + \gamma(s)\Bigr)\dd s
+ \log\left(\frac{1-\gamma(t)}{2}\right).
\end{equation}
Here, the function $u(t)$ is the Hastings-McLeod second Painlev\'e transcendent, i.e. the solution of the
second Painlev\'e equation,
\begin{equation}\label{p20}
u'' = tu + 2u^3, 
\end{equation}
uniquely determined by the condition,
\begin{equation}\label{p21}
u\sim \frac{1}{2\sqrt{\pi}t^\frac{1}{4}}\ee^{-\frac{2}{3} t^{\frac{3}{2}}}, \quad t \rightarrow + \infty,
\end{equation}
while the  pair of the functions $(\gamma(t), \alpha(t))$ is defined as the solution of the system of the first order ODEs,
\begin{equation}\label{gammaeq}
\gamma' = \frac{2}{3}\alpha \gamma + \frac{u'}{u}\frac{(1+\gamma)(2-\gamma)}{3},
\quad \alpha' = \alpha\left(\frac{2}{3}\alpha + \frac{u'}{u}\frac{2-\gamma}{3}\right)-\frac{t}{6}(1+\gamma)-
\frac{u^2}{3}(3+\gamma).
\end{equation}
with the asymptotic conditions,
\begin{equation}\label{gammalphaplus}
\gamma \sim -1 + \frac{1}{8\pi t^{\frac{3}{2}}}\ee^{-\frac{4}{3}t^{\frac{3}{2}}}, \quad \alpha \sim \frac{3}{16\pi t}\ee^{-\frac{4}{3}t^{\frac{3}{2}}}, 
\quad t \rightarrow \infty.
\end{equation}

It should be mentioned that Rumanov's derivation of the above results is based on certain heuristic assumptions
which have  been highlighted and some of them proven
in \cite{GIKM}. However, the proof of a  key assumption  that  the asymptotic conditions \eqref{gammalphaplus} determine 
a unique smooth solution of system \eqref{gammaeq}, {\it has not been yet found}{\footnote{ Although, there are very strong numerical
evidence produced by YuQi Li from East China Normal University that this statement is true (see \cite{L}).}}.

In more details, the relation of Rumanov's equations \eqref{gammaeq}, \eqref{gammalphaplus} and the Bloemendal -Vir\'ag boundary value problem \eqref{bv1} -- \eqref{bv3} can be described as follows (here we follow the interpretation of Rumanov's results given in \cite{GIKM}).

Let $u(t)$ $\gamma(t)$, $\alpha(t)$ be the objects defined in \eqref{p20} -- \eqref{gammalphaplus}. Also let $\Psi_0(x,t)$ be the fundamental solution of the first equation of Painlev\'e~II Lax pair
\begin{gather}\label{laxint1}
\dfrac{d\Psi_0}{dx}={L}\Psi_0,\qquad
{L}=\frac{x^2}{2}\sigma_3+x\begin{pmatrix}0&u\\ u&0\end{pmatrix}+\begin{pmatrix}-\dfrac{t}{2} -u^2&-u'\\u'&\dfrac{t}{2} + u^2\end{pmatrix}
\end{gather}
with the asymptotic condition
\begin{align}
    \Psi_0(x,t)\simeq \ee^{\left(\frac{x^3}{6}-\frac{xt}{2}\right)\sigma_3},\quad x\to\infty.
\end{align}
Define, in addition, the function $\kappa(t)$ by the formulae
\begin{gather}\label{kappaint}
\frac{\kappa'}{\kappa} = -\frac{1}{3}\left(u^4+tu^2-\left(u'\right)^2\right)-\frac{2}{3}\alpha -\frac{u'}{u}\left(\frac{1-2\gamma}{6}\right),
\end{gather}
and
\begin{align}
    \kappa\sim {}{\sqrt{2}\pi^{\frac{1}{4}}t^\frac{1}{8}}\ee^{\frac{1}{3}t^{\frac{3}{2}} }, \quad t \rightarrow + \infty.
\end{align}
Then (see \cite{GIKM})  the solution of boundary value problem \eqref{bv1}, \eqref{bv2}, \eqref{bv3} for $\beta=6$ is given by
\begin{align}\label{bv_solution}
F\big(3^{-1/3}x,3^{-2/3}t\big) = \kappa \ee^{\frac{x^3}{6} -\frac{xt}{2}}\left[ u^{-\frac{1}{2}}\left(\frac{1+\gamma}{2}x -\alpha\right)\Psi_{0,12}(x,t)+u^{\frac{1}{2}}\Psi_{0,22}(x,t)\right].
\end{align}

We are now going to explain the appearance in the picture  of the Calogero-Painlev\'e system. To this end,   let us pass from the triple of the functions $(u(t), \gamma(t), \alpha(t))$ to the triple $(Q_1(t), Q_2(t), Q_3(t))$,
according to the equations,
$$
e_1 = -\frac{4\alpha\gamma}{1-\gamma^2} -\frac{u'}{u}\frac{1+\gamma}{1-\gamma}, \quad
e_2 = \frac{4}{1-\gamma^2}\left[-\alpha^2 +u^2  -(1+\gamma)\left(\frac{t}{2} +u^2 +\alpha\frac{u'}{u}\right)\right],
$$
$$
e_3 =- \frac{4}{1-\gamma^2}\left[2\alpha\left(\frac{t}{2} +u^2\right) +\alpha^2 \frac{u'}{u}
+ u'u +\frac{1+\gamma}{2} \right],
$$
where $e_k$ denote the  first three symmetric functions of $Q_k$, i.e., 
$$
e_1 = Q_1 + Q_2 + Q_3,   \quad e_2 = Q_1Q_2 + Q_1Q_3 + Q_2Q_3, \quad   e_3 = Q_1Q_2Q_3.
$$
The opposite relations are given by
$$
\gamma=\mathop{\sum}_{k=1}^3\dfrac{-3Q_k'+2\sum_{j\neq k}(Q_k-Q_j)^{-1}}{\prod_{j\neq k}(Q_k-Q_j)},\quad $$
$$\alpha=\mathop{\sum}_{k=1}^3\dfrac{(-3Q_k'+2\sum_{j\neq k}(Q_k-Q_j)^{-1})\sum_{j\neq k}Q_j}{2\prod_{j\neq k}(Q_k-Q_j)}-\frac{u'}{2u}({1+\gamma}).
$$
Functions $Q_k(t)$ satisfy the system of equations
\begin{equation}\label{CalPIIQ}
Q_k''=\frac{2}{9}Q_k^3-\frac{2}{9}tQ_k + \frac{1}{9} -\frac{8}{9}\sum_{j\neq k}\frac{1}{(Q_k-Q_j)^3}, \quad k=1,2,3.
\end{equation}
To get it in the form \eqref{CalPII} we put 
$$
x_k(t) :=-6^{-1/3} Q_k\left(-\left(\frac{9}{2}\right)^{1/3} t\right), \quad k = 1, 2, 3.
$$
Rumanov in \cite{Rum3}  has shown that, in terms of $x_k(t)$,  the system of the three equations \eqref{p20}, \eqref{gammaeq}  becomes a particular case of the {\it Calogero-Painlev\'e system}
of the three 1D interacting particles, with $\theta = -\frac{1}{6}$ and $g^2 = -\frac{1}{9}$, i.e.,
\begin{equation}\label{cc6}
x''_k =2x_k^3+tx_k  -\frac{1}{6} -\frac{2}{9}\sum_{j\neq k}\frac{1}{(x_k-x_j)^3}, \quad k=1, 2, 3.
\end{equation}

The Tracy-Widom distribution $F_6(t)$ can be expressed directly in terms of the
solution $x_k(t)$ of the  Calogero-Painlev\'e  system \eqref{cc6}; indeed, one has that{\footnote{In fact, without specification of the 
particular solution to be taken, Rumanov in \cite{Rum3} is presenting similar formula for any even $\beta$; the formula involves the
Calogero-Painlev\'e system for $n$ particles with $\theta = -\frac {n-2}{2n}$ and $g^2 = -\frac{1}{n^2}$ if $\beta =2n$.}
\begin{equation}\label{betacc1}
\log F_{6}(-2^{-1/3}t) = \int_{-\infty}^t \left(H-\frac{3t^2}{8} +\frac{1}{2}\sum_{k=1}^3 x_k \right)\dd t,
\end{equation}
where,
$$
H=\sum_{k=1}^3\left(\frac{y_k^2}{2}-\frac{x_k^4}{2}-\frac{x_k^2t}{2} +\frac{1}{6} x_k\right)-\frac{1}{9}\sum_{j< k}\frac{1}{(x_k-x_j)^2},
$$
is the Hamiltonian of  the system \eqref{cc6} ( cf. \eqref{HamCP}). 

\subsection{The setting of the asymptotic problem  for the Calogero-Painlev\'e system }
The Borot-Nadal conjecture \eqref{expconst} in the case of $\beta =6$ reads
    \begin{align}\label{expconst6}
F_6(t)\sim 1-\frac{\ee^{-{4 t^\frac{3}{2}}{}}}{(24)^3\pi t^{\frac{9}{2}}}\left(1+O\left(\dfrac{1}{t^{\frac{3}{2}}}\right)\right),\quad t\to \infty.
\end{align}  
It is important to notice that extended version of asymptotics \eqref{p21} and \eqref{gammalphaplus}
\begin{align}
  &  u=\frac{1}{2\sqrt{\pi}t^\frac{1}{4}}\ee^{-\frac{2}{3} t^{\frac{3}{2}}}\left(1-\frac{5}{48t^{\frac{3}{2}}}+\frac{385}{4608t^3}+O\left(\frac{1}{t^{\frac{9}{2}}}\right)\right),\quad t\to\infty,\\&
  \gamma = -1 + \frac{1}{8\pi t^{\frac{3}{2}}}\ee^{-\frac{4}{3}t^{\frac{3}{2}}}\left(1-\frac{59}{24t^{\frac{3}{2}}}+\frac{12085}{1152t^3}+O\left(\frac{1}{t^{\frac{9}{2}}}\right)\right),\quad t\to\infty,\\&
  \alpha =\frac{3}{16\pi t}\ee^{-\frac{4}{3}t^{\frac{3}{2}}}\left(1-\frac{29}{24t^{\frac{3}{2}}}+\frac{4477}{1152t^3}+O\left(\frac{1}{t^{\frac{9}{2}}}\right)\right),\quad t\to\infty
\end{align}
yield the asymptotics \eqref{expconst6} for the right tail of the Tracy-Widom distribution
for $\beta=6$.
The asymptotic conditions \eqref{p21}, \eqref{gammalphaplus} at $t =\infty$ are transformed into the
following asymptotic condition for the functions $x_k(t)$ at $t = -\infty$
\begin{equation}\label{x1ascond}
\begin{gathered} 
\sum_{k=1}^{3}\frac{ x_k' + \frac{1}{3} \sum_{j\neq k}^3\frac{1}{x_k-x_j}}{\prod_{j\neq k}^{3}(x_k-x_j)}\sim1-\frac{\mathrm{e}^{-2\sqrt{2}(-t)^{\frac{3}{2}}}}{12\pi\sqrt{2}(-t)^\frac{3}{2}},\quad t\to-\infty,\\
\sum_{k=1}^{3}\frac{\left( x_k' + \frac{1}{3} \sum_{j\neq k}^3\frac{1}{x_k-x_j}\right)\sum_{j\neq k}x_j}{\prod_{j\neq k}^{3}(x_k-x_j)}\sim-\frac{\mathrm{e}^{-2\sqrt{2}(-t)^{\frac{3}{2}}}}{12\pi t},\quad t\to-\infty.
\end{gathered}
\end{equation}
Moreover we get that up to permutation
\begin{equation}\label{x1as}
\begin{gathered} 
x_1 =  -\sqrt{\frac{|t|}{2}} + \frac{13}{36t} + O(|t|^{-5/2}), \quad 
x_{2,3} = -\sqrt{\frac{|t|}{2}} \mathbin{\pm}\frac{\ii}{2^{\frac{3}{4}}|t|^{\frac{1}{4}}} + \frac{4}{9t} + O(|t|^{-7/4}).
\end{gathered}
\end{equation}
The  Borot-Eynard-Majumdar-Nadal conjecture \eqref{con1}  in the case of $\beta =6$ reads

\begin{equation}\label{con16}
\log F_6(t) = -\frac{1}{4}|t|^3 +\frac{2\sqrt{2}}{3}|t|^{\frac{3}{2}} +\frac{1}{24}\log|t| + \chi + O(t^{-3/2}), \quad t \to -\infty.
\end{equation}
{\it The ultimate goal} of our study is to prove this formula using the representation 
\eqref{betarum} for the function $F_6(t)$. To this end, one needs to find the asymptotics at $t = -\infty$ of the
solution  $(\gamma(t), \alpha(t))$) of the system \eqref{gammaeq}  which behaves at  $t = \infty$ 
as it is indicated in \eqref{gammalphaplus}.  
Taking into account the known behavior of the Hastings-McLeod Painlev\'e function $u(t)$ as $t \to -\infty$,   i.e.,
\begin{equation}\label{p22}
u = \sqrt{\frac{|t|}{2}}\left( 1 +\frac{1}{8} t^{-3} + O(t^{-6})\right), \quad t \to -\infty,
\end{equation}
one can check that the system \eqref{gammaeq}  admits the formal solution with the asymptotics,
\begin{equation}\label{gammaalphaminus1}
\gamma = \frac{1}{\sqrt{2}}|t|^{-\frac{3}{2}} -\frac{21}{8}t^{-3} +\frac{1707}{64\sqrt{2}}|t|^{-\frac{9}{2}}
+ O(t^{-6}), 
\quad
\alpha  = \frac{1}{\sqrt{2}}|t|^{\frac{1}{2}} +\frac{1}{8}t^{-1} -\frac{37}{64\sqrt{2}}|t|^{-\frac{5}{2}}
+ O(t^{-4}), 
\end{equation} 
as $t \to -\infty$. 
It is significant, that the asymptotics \eqref{p22}, \eqref{gammaalphaminus1}  generate via \eqref{betarum}   the asymptotic formula \eqref{con16}
(without  though the constant term $\chi$). Hence, a key  
question to address is to show that the solution  $(\gamma(t), \alpha(t))$) of the system \eqref{gammaeq} 
fixed by the behavior  \eqref{gammalphaplus} at  $t = \infty$  has indeed the asymptotics \eqref{gammaalphaminus1},
at $t = -\infty$. In other words,  one needs to find a {\it  connection formulae }
for the solution  $(\gamma(t),\alpha(t))$ of  system  \eqref{gammaeq}. 

Notice, that the $t =-\infty$  asymptotic formulae \eqref{p22}--\eqref{gammaalphaminus1}, in terms of the Calogero-Painlev\'e coordinates $x_j$, become
the following  asymptotic relations at $t = \infty$ up to permutation,
\begin{equation}\label{x2as}
x_1 = \frac{1}{6t} + \frac{8}{27t^{\frac{5}{2}}} + O(t^{-4}),
\quad x_{2,3} =\mathbin{\pm}\frac{\ii\sqrt{2}}{2t^{\frac{1}{4}}} +\frac{1}{6t} + O(t^{-\frac{7}{4}}),\quad t \to \infty.
\end{equation}
Therefore, we arrive at the following {\it connection  problem} for the Calogero-Painlev\'e system \eqref{cc6}:

{\it Show that there is a unique solution,   $(x_1(t), x_2(t), x_3(t))$, of \eqref{cc6} which is smooth for all real $t$ and which behaves
at $t = -\infty$ and $t = \infty$ as it is indicated in equations \eqref{x1as} and \eqref{x2as}, respectively, up to permutation.}

In this paper we address this problem using the   Riemann-Hilbert representation of the solutions of the  Calogero-Painlev\'e system
which was found in the recent  work \cite{BCR} and which we will now describe in some details.  

\subsection{The Riemann-Hilbert representation of the Calogero-Painlev\'e particles.}\label{rhp1}

In \cite{BCR}, it is shown (in fact, for the general case of $n$ particles) that  the system \eqref{cc6} is Lax-pair integrable,
and the following Riemann-Hilbert representation of its general solution takes place. 
\begin{rhp}\label{mainrhp} 
Denote $\Gamma_j$ , $j = 1, 2, 3, 4, 5, 6$ the rays $\arg z = \frac{\pi}{6}(2j-1)$, $j =1, 2, 3$  and
$\arg z = -2\pi + \frac{\pi}{6}(2j-1)$, $j =4, 5, 6$  oriented toward infinity, and also denote 
$\,{\mathbb{R}}_- $ the half  real line,  $z<0$ oriented to the left (see Figure \ref{gammamainrhp}).  Let now set the  Riemann-Hilbert  problem
consisting in finding the $6\times6$ matrix valued function $\Psi(z;t)$ such that
\begin{itemize}
\item $\Psi(z)$ is holomorphic on ${\mathbb{C}}\setminus \{{\mathbb{R}}_-\cup \Bigl(\cup_{j=1}^6 \Gamma_j\Bigr)\}$, and the 
boundary values, $\Psi_{\pm}(z)$,  are finite and satisfy the following jump conditions,
\begin{equation}\label{jumbs}
\Psi_+(z) = \Psi_-(z) S_j, \quad z \in \Gamma_j,  \quad j = 1, ..., 6, \quad \Psi_+(z) = \Psi_-(z)\ee^{-\frac{2\pi \ii}{3}}I_6
\quad z \in {\mathbb{R} }_-.
\end{equation}
\item The behavior of \,$\Psi(z)$ as $ z \to  \infty, 0$ is described by the equations,
\begin{equation}\label{Psiinfty}
\Psi(z) = \left( I_6 + \frac{m_1}{z} +  O\left(\frac{1}{z^2}\right)\right) z^{I_2\otimes \Lambda_1}
\ee^{\frac{\ii}{2}\left(\frac{z^3}{3} +  tz\right)\sigma_3\otimes I_3},\quad z \to \infty,
\end{equation}
and
\begin{equation}\label{Psizero}
\Psi(z) =G_0\Bigl( I_6 + O(z)\Bigr)z^ {\frac{1}{6}\sigma_3\otimes I_3}E, \quad z \to 0,\quad |\arg z| < \frac{\pi}{6}.
\end{equation}
\end{itemize}
Here, 
\begin{equation}\label{eq:Lambda_1}
\sigma_3 = \begin{pmatrix}1&0\cr0&-1\end{pmatrix}, \quad \Lambda_1 =
\begin{pmatrix}\frac{1}{3}&0&0\cr
 0&\frac{1}{3}&0\cr
 0&0&-\frac{2}{3}\end{pmatrix},
\end{equation}
and $S_j$ are the block-triangular matrices,
\begin{equation}\label{stokes}
\begin{gathered}
  S_1=\left(\begin{array}{cc}
                            I_3&A\\
                            0&I_3
                            \end{array} \right),\quad  S_2=\left(\begin{array}{cc}
                            I_3&0\\
                            B&I_3
                            \end{array} \right),\quad   S_3=\left(\begin{array}{cc}
                            I_3&C\\
                            0&I_3
                            \end{array} \right),
\\
  S_4=\left(\begin{array}{cc}
                            I_3&0\\
                          \ee^{\ii\pi\Lambda_1}A\ee^{-\ii\pi\Lambda_1} &I_3
                            \end{array} \right),\quad  S_5=\left(\begin{array}{cc}
                            I_3&  \ee^{\ii\pi\Lambda_1}B\ee^{-\ii\pi\Lambda_1}\\
                           0&I_3
                            \end{array} \right),\quad   S_6=\left(\begin{array}{cc}
                            I_3&0\\
                              \ee^{\ii\pi\Lambda_1}C\ee^{-\ii\pi\Lambda_1}&I_3
                            \end{array} \right).
                            \end{gathered}
\end{equation}
We denote by $I_n$ the identity matrix in $\mathbb{R}^n$.

\end{rhp}
\begin{figure}[h]
	\centering
\includegraphics[scale=1]{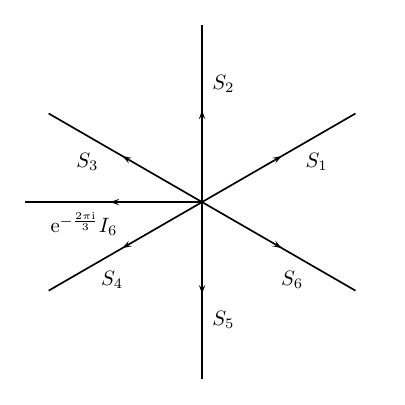}
	\caption{Contour for the Riemann-Hilbert problem \ref{mainrhp}}
	\label{gammamainrhp}
\end{figure} %
\noindent Here and further in the paper we use Kronecker product notation
\begin{align}
    A\otimes B=\begin{pmatrix}
        a_{11}B&\cdots&a_{1n}B\\
         \vdots & \ddots &           \vdots \\
         a_{n1}B&\cdots&a_{nn}B\\
    \end{pmatrix}.
\end{align}
The $3\times3$ matrices $A$, $B$, and $C$ are satisfying  certain algebraic relations (see equations \eqref{cyclic2} in  Section \ref{RH}) 
which are the consequence of the  cyclic equation,

\begin{equation} \label{cyclic}
E^{-1}\ee^{-\frac{\pi \ii}{3}\sigma_3\otimes I_3}E = \ee^{-\frac{2\pi \ii}{3}}S_1S_2S_3
S_{4}S_{5}S_6,
\end{equation} 
which also determines the matrix factor $E$ in \eqref{Psizero}.  

Having solution $\Psi(z)$, the solutions $x_k(t)$ of the Calogero-Painlev\'e system are  determined as the eigenvalues of the $3\times3$ 
matrix 
\begin{equation}\label{qquant}
 q_0(t) =-\ii \hat{m}_{1, 12},
\end{equation}
where $\hat{m}_{1, 12}$ denote the $3\times3$ up-right block of the coefficient  $6\times 6$ matrix $m_1$ from the expansion
\eqref{Psiinfty}.

The first principal question now is which Riemann-Hilbert data, i.e. the matrices $A$, $B$ and $C$, yield
the solution of the Calogero-Painlev\'e system \eqref{cc6} that corresponds to  the Tracy-Widom function $F_6(t)$.
{\it The main result }of this paper is the following partial answer to this question.

\begin{theorem}\label{ccCon2}  Consider the Riemann-Hilbert problem \ref{mainrhp} with matrices $A,B,C,E$ satisfying cyclic condition \eqref{cyclic}.
\begin{enumerate}
\item The solution $\Psi(z)\equiv \Psi(z, t)$ and the corresponding $3\times3$ matrix $q_0(t)$ 
from \eqref{qquant} exist as meromorphic functions of $t$.

\item Assume that matrices $A, B, C$ are parametrized by $b_{11}, b_{22}, b_{21}, b_{31}$ and $a_{33}\in\{\ii,\ee^{-\frac{5\pi\ii}{6}}\}$ using formulas \eqref{ansatzminf}, \eqref{eq:monodromy_minus_infty} with $b_{12}=b_{23}=1$.
Then the eigenvalues $\Bigl(x_1(t), x_2(t), x_3(t)\Bigr)$
of the  matrix $q_0(t)$ form a family 
of solutions of the Calogero-Painlev\'e system \eqref{cc6} whose behavior as 
$t \to -\infty$ is given, up to permutation, by the asymptotic formulae \eqref{x1as} for $a_{33}=\ii$, and by
\begin{equation}\label{x3as}
\begin{aligned}
x_1&=\sqrt{\frac{|t|}{2}}-\frac{19}{36t}+O(|t|^{-5/2}),\\
x_{2,3}&=\sqrt{\frac{|t|}{2}}\mathbin{\pm}
\frac{\ii}{2^{3/4}|t|^{1/4}}-\frac{11}{18t}+O(|t|^{-7/4}).
\end{aligned}
\end{equation}
for $a_{33}=\ee^{-\frac{5\pi\ii}{6}}$.
In \eqref{x1as}, \eqref{x2as}, and \eqref{x3as}, the signs label two distinct particles.
For general complex monodromy parameters, conjugate displayed asymptotic
series do not by themselves assert the exact identity $x_2(t)=\overline{x_3(t)}$; exponentially small
terms are not specified by these expansions.

\item Assume that matrices $A, B, C$ are parametrized by $a_{13}, a_{23}, a_{33}, b_{13}$ using formulas \eqref{ansatzpinf}, \eqref{eq:monodromy_plus_infty} with $b_{12}=b_{23}=1$. Then the eigenvalues $\Bigl(x_1(t), x_2(t), x_3(t)\Bigr)$
of the matrix $q_0(t)$ form the solutions of the Calogero-Painlev\'e system \eqref{cc6} whose behavior as
$t \to \infty$ is given by the asymptotic formulae \eqref{x2as} up to permutation.
\end{enumerate} 
\end{theorem}
The proof of this theorem is based on the Deift-Zhou nonlinear steepest descent method. As it has already been indicated in the Abstract,
the principle technical challenge is the  high matrix size of the associated  Riemann-Hilbert problem, i.e., Riemann-Hilbert problem \ref{mainrhp}. Indeed,
the higher than 2 matrix dimension  leads to a considerable complications in two basic steps of the Deift-Zhou method, i.e. the contour deformation and the analysis
of the local parametrices. We want to highlight specifically the difficulties related to the local parametrices.  Indeed, in $2\times 2$ case and  in generic
situation the model Riemann-Hilbert problems corresponding to local parametrices always admit explicit solutions in classical special functions.
Beyond the $2\times 2$  matrix setting this is  generally not the case any more. However, the specifics of the Riemann-Hilbert problem \ref{mainrhp} has
allowed us to solve all the relevant local model Riemann-Hilbert problems explicitly, in terms of the  parabolic cylinder, Bessel, Airy, and confluent hypergeometric functions.

By considering the choice \eqref{S1S2}  of the RH data corresponding to   intersection of parametrization \eqref{ansatzpinf}, \eqref{eq:monodromy_plus_infty} and \eqref{ansatzminf}, \eqref{eq:monodromy_minus_infty}  we arrive at the
following corollary which is our key result.

\begin{corollary} The choice \eqref{S1S2} of the RH data with $a_{33}=\ii$ generates the  solution of the Calogero-Painlev\'e  system \eqref{cc6} with  both, the asymptotics \eqref{x1as} as $t\to -\infty$ and 
the asymptotics \eqref{x2as}  as $t\to \infty$.
\end{corollary}
For $a_{33}=\ee^{-\frac{5\pi\ii}{6}}$ the corresponding solution also satisfies \eqref{x2as} as $t\to\infty$, but it has the asymptotics \eqref{x3as} as $t\to-\infty$, so it does not generate $F_6(t)$.
\begin{remark}
In our original version we claimed that $a_{33}=\ee^{-\frac{5\pi\ii}{6}}$ also yields \eqref{x1as} as $t\to-\infty$. After checking our calculations with the help of Claude, a mistake was revealed, and the correct statement concerning the asymptotics as $t\to-\infty$ for $a_{33}=\ee^{-\frac{5\pi\ii}{6}}$ is presented in the current version of the paper.
\end{remark}
{\bf For the data \eqref{S1S2} with $a_{33}=\ii$, the still open questions are:} 
\begin{itemize}

\item To show that the choice \eqref{S1S2} corresponding to intersection of parametrizations \eqref{ansatzpinf}, \eqref{eq:monodromy_plus_infty} and \eqref{ansatzminf}, \eqref{eq:monodromy_minus_infty} guarantees
that the Riemann-Hilbert problem is solvable for all real $t$ and that the corresponding solution $(x_1(t), x_2(t), x_3(t))$
of the Calogero-Painlev\'e system exists and is smooth for all real $t$.
\item  To show that the choice  \eqref{S1S2} yields the solution $(x_1(t), x_2(t), x_3(t))$ which indeed generates the Tracy-Widom distribution $F_6(t)$ and
hence to prove the BEMN conjecture \eqref{con1} for $\beta =6$ (without the constant term $\chi$). The first issue here is to show that 
the above determined  solution $(x_1(t), x_2(t), x_3(t))$  is the unique solution of \eqref{cc6} that satisfies
simultaneously  the asymptotic  conditions \eqref{x1as} and \eqref{x2as}. It should be noticed that the asymptotics \eqref{x1as} by itself do not define the solution $(x_1(t), x_2(t), x_3(t))$
uniquely--there is an “unseen” exponentially small term in it, and hence it can not be easily transformed
back to the asymptotics \eqref{gammalphaplus} for the functions $\gamma(t)$ and $\alpha(t)$. Presumably, the second asymptotics, i.e.
formula \eqref{x2as}, fixes the solution uniquely. 
The second issue is to show  that the  Bloemendal-Virag 
function $F(x,t)$  defined by the solution $(x_1(t), x_2(t), x_3(t))$ via \eqref{bv_solution} (and automatically satisfying \eqref{bv1}) is bounded, smooth and satisfies 
the boundary conditions \eqref{bv2}, \eqref{bv3}. 
\item To evaluate rigorously $\chi$.
\item To extend the result to arbitrary even $\beta$.
\end{itemize}

Another open problem is to evaluate the asymptotic behavior of the solutions of the Calogero-Painlev\'e system corresponding to other monodromy
data beyond the special cases (\ref{ansatzpinf}) and (\ref{ansatzminf}). Apart from the intrinsic importance for the general theory of the 
Calogero-Painlev\'e systems this evaluation  would help to resolve the above mentioned uniqueness question
of the solution $(x_1(t), x_2(t), x_3(t))$ satisfying simultaneously  the asymptotic  conditions \eqref{x1as} and \eqref{x2as}. 
Indeed, in the pure Painlev\'e II, i.e. $\beta =2$, case, such complete asymptotic description had been long ago obtained by A. Kapaev in \cite{Kapaev}
(see also \cite{FIKN}), and it shows that it is only for the $\beta=2$  analog of the special choice  (\ref{ansatzpinf}) and (\ref{ansatzminf}) of the
corresponding monodromy data,  the $\beta=2$ analog  of our special solution has simultaneously the analogs of the asymptotics
 \eqref{x1as} and \eqref{x2as} {\footnote{This special solution for Painlev\'e II equation is presented in detail in Appendix \ref{app:beta_two}}}.
 We expect the same to happen in the $\beta=6$ case which we are studying. However, the implementation
 of the Deift-Zhou method  in the case of general Calogero-Painlev\'e monodromy data is theoretically possible but technically much more challenging.   This is yet 
 another manifestation of the above mentioned principal difficulties with extension of the Deift-Zhou nonlinear steepest descent method
 to matrix dimensions beyond $2\times2$. 
 
There is one more open question which we want to emphasize. As we already mentioned before, asymptotics \eqref{x2as} of the solution $(x_1 (t) , x_2(t), x_3(t))$ of the Calogero-Painlev\'e system  yield (without the constant $\chi$)  the BEMN asymptotics \eqref{con16} for the Tracy-Widom function $F_6(t)$ as $t \to -\infty$.
At the same time, asymptotics \eqref{x1as} does not reproduce BN asymptotics \eqref{expconst6}
of $F_6(t)$ as $t \to \infty$ . Indeed,  direct  substitution of \eqref{x1as} into \eqref{betacc1} only produces the estimate,
\begin{align}\label{expconstweak}
F_6(t) = 1 + O\left(\frac{1}{t^p}\right), \quad t \to \infty,
\end{align}
with some $p>0$ depending on the number of terms kept in \eqref{x1as} (since the expansion proceeds in powers of $(-t)^{-3/4}$, $p$ is a multiple of $\frac34$). Moreover, if we calculate more terms in the asymptotic formulae \eqref{x1as} , the value 
of the exponent $p$ will increase. In fact, one can conjecture that substitution of our Calogero-Painlev\'e  solution 
into \eqref{betacc1}  would lead to the exponential type estimate,
\begin{equation}\label{F6}
F_6(t) = 1 + O\left(\frac{1}{t^{\infty}}\right), \quad t \to \infty.
\end{equation}
The question is how to make it as precise as in \eqref{expconst6}.} In Appendix \ref{app:beta_two} we show how the similar problem can be resolved in the case of $\beta=2$.
Also, in this Appendix, following the $\beta=2$ experience we shall outline a strategy how we think the  right tail problem can be resolved.
We intend to address all these questions, including the uniqueness question discussed above, in the forthcoming publication. We also want to  point out that the reconstruction of the asymptotics \eqref{expconst} from the  asymptotics \eqref{x2as} of the  solution $(x_1(t), x_2(t), x_3(t))$
is equivalent to reconstruction of the asymptotics \eqref{gammalphaplus} of the functions $\gamma(t)$ and $\alpha(t)$ and, in fact, is closely
related to the resolution of the issues indicated  in the second open question above.
\subsection{Plan of the paper} In the next Section \ref{Lax},  we shall present, following \cite{BCR}, the isomonodromy
 Lax pair for the Calogero-Painlev\'e  system \eqref{cc6}. Then, in the beginning of Section \ref{RH}, following again
 \cite{BCR}, we will transform the Lax formalism for the Calogero-Painlev\'e system  to its Riemann-Hilbert formalism. The main body of the paper consists of  Sections \ref{RH1} and \ref{RH2} where we perform an asymptotic analysis of the Bertola-Cafasso-Rubtsov Riemann-Hilbert problem as $t \to \infty$ and $t \to -\infty$, respectively. Section~\ref{RH3} treats the second root $a_{33}=\ee^{-\frac{5\pi\ii}{6}}$ as $t\to-\infty$. The specific structure of the jump matrices  announced in Theorem \ref{ccCon2} will be fixed in a process of application of  the Deift-Zhou
 nonlinear steepest descent method which will eventually lead us to the proof of the Theorem. In fact, and we will say more about that  later on, the Riemann-Hilbert 
 we are dealing with is a non-abelian version of the Flaschka-Newell Lax pair for the second Painlev\'e equation. The explicit formulas for parametrices are given in the Appendices \ref{parab}-\ref{confluent}. We provide comparison with $\beta=2$ case in Appendix \ref{app:beta_two}.

 \subsection*{Acknowledgements}  AI is partially supported by NSF grant DMS:1955265, by RSF grant No. 22-11-00070. AP was supported by NSF MSPRF grant DMS-2103354 and RSF grant 22-11-00070.

\section{Lax pair of the Calogero-Painlev\'e System}\label{Lax}
\subsection{Matrix Painlev\'e equation}
The following Lax pair considered in \cite{BCR} is the quantization of Flaschka-Newell pair for the second Painlev\'e equation
\begin{equation}\label{lpq}
\dfrac{dW}{dz}=U(z)W(z),\quad \dfrac{dW}{dt}=V(z)W(z),
\end{equation}

$$
U(z)=U_2z^2+U_1z+U_0+\dfrac{U_{-1}}{z},\quad V(z)=V_1z+V_0, 
$$

$$
U_2=\left(\begin{array}{cc}
\dfrac{\ii I_3}{2}&0\\
0&-\dfrac{\ii I_3}{2}
\end{array}\right),\quad U_1=\left(\begin{array}{cc}
0&q\\
q&0
\end{array}\right),\quad U_0=\left(\begin{array}{cc}
\ii q^2+\dfrac{\ii tI_3}{2}&-\ii p\\
\ii p&-\ii q^2-\dfrac{\ii tI_3}{2}
\end{array}\right),$$
$$U_{-1}=\left(\begin{array}{cc}
0&-\theta I_3\\
-\theta I_3&0
\end{array}\right),
$$

$$
V_1=U_2,\quad V_0=U_1.
$$
Here $q$ and $p$ are $3\times 3$ matrices, compared to standard Flaschka-Newell Lax pair. The compatibility condition takes form of matrix Painlev\'e II equation
\begin{equation}\label{PIIv} 
\dfrac{dq}{dt}=p,\quad \dfrac{dp}{dt}=2q^3+tq+\theta I_3.
\end{equation} 
The symmetry of the coefficient matrix $$
-U(-z)=({\sigma}_1\otimes I_3 )U(z)({\sigma}_1\otimes I_3).
$$
implies the symmetry of solution
\begin{equation}   \label{sym}         
W(-z)=(\sigma_1\otimes I_3)W(z)(\sigma_1\otimes I_3).
\end{equation}
We have the following asymptotic at infinity
\begin{equation} \label{atinfty} 
W(z)=\left(I_6+\dfrac{n_1}{z}+O\left(\dfrac{1}{z^2}\right)\right)\ee^{(\ln z+\ii\pi \varepsilon)(I_2 \otimes [q,p])}\ee^{\frac{\ii}{2}\left(\frac{z^3}{3}+tz\right)(\sigma_3\otimes I_3)},
\end{equation} 
where
$$
\varepsilon=\begin{cases}
	0&\mathrm{Im}(z)\geq 0\\
	1&\mathrm{Im}(z)<0\\
\end{cases},\quad 
\hat{n}_{1,12}=\ii q.
$$
The asymptotic at the origin has form 
\begin{equation} \label{atzero} 
W(z)=G_0\left(I_6+O\left(z\right)\right)\ee^{(\ln z+\ii\pi \varepsilon){(-\theta \sigma_3\otimes I_3)}}(K\otimes I_3)^{-1}, \quad z\to 0,
\end{equation} 
with
\begin{equation} \label{K}
G_0=(K\otimes I_3) \begin{pmatrix}
 r_1&0\\
 0&r_2\\
\end{pmatrix},\quad
K=\dfrac{I_2-\ii\sigma_2}{\sqrt{2}}=\dfrac{1}{\sqrt{2}}\left(\begin{array}{cc}1&-1\\1&1
\end{array}\right),\quad K\sigma_3=\sigma_1 K,\quad \sigma_3K=-K\sigma_1.
\end{equation}
We can see that the symmetry \eqref{sym} is preserved in equations \eqref{atzero}, \eqref{atinfty}. 
We would like to notice that the choice of $g=\pm\frac{\ii}{3}$ makes the operator $\mathrm{Id}+\mathrm{ad}_{[q,p]}$ not invertible and to find $m_1$ we would need to use standard procedure presented in \cite{FIKN} and not in \cite{BCR}. The $3\times 3$ matrices $r_1$ and $r_2$ satisfy the equations
$$
\dfrac{dr_1}{dt}=qr_1,\quad \dfrac{dr_2}{dt}=-qr_2.
$$

\subsection{Calogero-Painlev\'e system of equations}
The commutator $[p,q]$ is preserved by dynamics \eqref{PIIv}, so it is part of monodromy data. We will choose it from Calogero-Moser space
\begin{equation}\label{def:commutator}
[p,q]=\ii g(I_3-v^Tv)=-\ii g\left(\begin{array}{ccc}
0&1&1\\
1&0&1\\
1&1&0
\end{array}\right), \quad v=(1,1,\ldots,1).
\end{equation}
Below we list the results from \cite{BCR}. It is possible to conjugate matrices $q$ and $p$ in such a way that
$$
q=\mathcal{O}X\mathcal{O}^{-1},\quad X=\left(\begin{array}{ccc}
x_1&0&0\\
0&x_2&0\\
0&0&x_3
\end{array}\right),
$$

$$
p=\mathcal{O}Y\mathcal{O}^{-1}, \quad Y=\left(\begin{array}{ccc}
y_1&\dfrac{\ii g}{(x_1-x_2)}&\dfrac{\ii g}{(x_1-x_3)}\\
\dfrac{\ii g}{(x_2-x_1)}&y_2&\dfrac{\ii g}{(x_2-x_3)}\\
\dfrac{\ii g}{(x_3-x_1)}&\dfrac{\ii g}{(x_3-x_2)}&y_3
\end{array}\right).
$$
The eigenvectors of $q$ forming matrix $\mathcal{O}$ satisfy the equation $
\dfrac{d\mathcal{O}}{dt}=\mathcal{O}F
$
with
$$
F=\left(\begin{array}{ccc}
f_1&-\dfrac{\ii g}{(x_1-x_2)^2}&-\dfrac{\ii g}{(x_1-x_3)^2}\\
-\dfrac{\ii g}{(x_2-x_1)^2}&f_2&-\dfrac{\ii g}{(x_2-x_3)^2}\\
-\dfrac{\ii g}{(x_3-x_1)^2}&-\dfrac{\ii g}{(x_3-x_2)^2}&f_3
\end{array}\right),
$$
$$
f_1=-\dfrac{2\ii g}{3}\dfrac{1}{(x_2-x_3)^2}+\dfrac{\ii g}{3}\dfrac{1}{(x_1-x_3)^2}+\dfrac{\ii g}{3}\dfrac{1}{(x_2-x_1)^2},
$$
$$
f_2=-\dfrac{2\ii g}{3}\dfrac{1}{(x_1-x_3)^2}+\dfrac{\ii g}{3}\dfrac{1}{(x_2-x_3)^2}+\dfrac{\ii g}{3}\dfrac{1}{(x_2-x_1)^2},
$$
$$
f_3=-\dfrac{2\ii g}{3}\dfrac{1}{(x_2-x_1)^2}+\dfrac{\ii g}{3}\dfrac{1}{(x_1-x_3)^2}+\dfrac{\ii g}{3}\dfrac{1}{(x_2-x_3)^2}.
$$
Taking it into account, one can rewrite the compatibility conditions \eqref{PIIv} as
$$
\dfrac{dX}{dt}=Y+[X,F],\quad \dfrac{dY}{dt}=2X^3+tX+\theta I_3+[Y,F].
$$
It implies that eigenvalues of matrix $q$ satisfy Calogero Painlev\'e system of equations \eqref{CalPII}.

\section{Riemann-Hilbert problem}\label{RH}
The case of $\beta=6$ Tracy-Widom law corresponds to $g^2=-\frac{1}{9},\,\theta=-\frac
{1}{6}$. We choose $g=\frac{\ii}{3},\, \theta=-\frac
{1}{6}$ from now on. Consider the following Riemann-Hilbert problem.

\begin{figure}[h]

	\centering
	
	\includegraphics[scale=1]{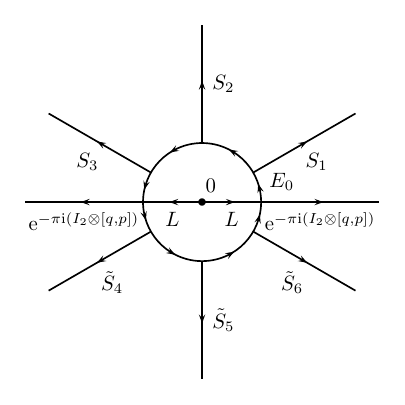}
	\caption{Contour for the Riemann-Hilbert problem \ref{rhplax}}
	\label{psi0}
\end{figure}

\begin{rhp}\label{rhplax}
Consider the contour $\Gamma$ shown on Figure \ref{psi0}. The $6\times 6$ matrix valued function $W(z)$ satisfies the following conditions 
\begin{itemize}
    \item $W(z)$ is holomorphic on $\mathbb{C}\setminus \Gamma$.
    \item $W(z)$ has finite boundary values on the contour $\Gamma$ and satisfies the jump condition  indicated on Figure \ref{psi0} with
    $$
L=(K\otimes I_3)\ee^{- \frac{\pi \ii(\sigma_3\otimes I_3)}{6}}(K\otimes I_3)^{-1},
$$
$$
\widetilde{S}_1=\left(\begin{array}{cc}
I_3&\widetilde{A}\\
0&I_3
\end{array} \right),\quad  \widetilde{S}_2=\left(\begin{array}{cc}
I_3&0\\
\widetilde{B}&I_3
\end{array} \right),\quad   \widetilde{S}_3=\left(\begin{array}{cc}
I_3&\widetilde{C}\\
0&I_3
\end{array} \right),
$$
$$
\widetilde{S}_4=\left(\begin{array}{cc}
I_3&0\\
\widetilde{A} &I_3
\end{array} \right),\quad  \widetilde{S}_5=\left(\begin{array}{cc}
I_3&\widetilde{B}\\
0&I_3
\end{array} \right),\quad   \widetilde{S}_6=\left(\begin{array}{cc}
I_3&0\\
\widetilde{C}&I_3
\end{array} \right).
$$
Matrix $E_0$ is chosen based on cyclic condition
$$
\widetilde{S}_1\widetilde{S}_2\widetilde{S}_3\ee^{-\ii\pi(I_2\otimes [q,p])}\widetilde{S}_4\widetilde{S}_5\widetilde{S}_6\ee^{-\ii\pi(I_2\otimes [q,p])}=E_0(K\otimes I_3)\ee^{-\frac{\pi \ii(\sigma_3\otimes I_3)}{3}}(K\otimes I_3)^{-1}E_0^{-1}.
$$
It can be selected up to right multiplication by block diagonal matrix, which corresponds to the fixing of extra functions $r_1$ and $r_2$. We are not concerned about them, so we do not specify this choice. 

The jumps on other arcs of the circle can be derived from $E_0$ using cyclic relations around the points of intersection on the circle.  
\item function $W(z)$ have the asymptotics \eqref{atinfty}, \eqref{atzero} at infinity and at zero respectively.
\end{itemize}
\end{rhp}%
The solution $W(z)$ of the Riemann-Hilbert problem \ref{rhplax} satisfies the Lax pair equations \eqref{lpq} and the symmetry condition \eqref{sym}.

\begin{figure}[h]
	\centering
	\includegraphics[scale=1]{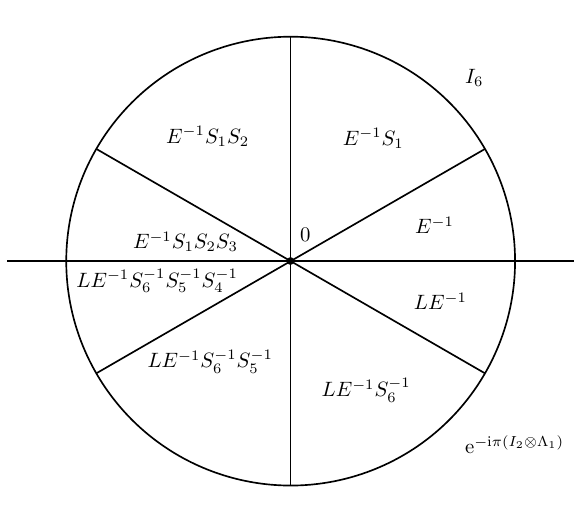}
	\caption{Matrix $M_1$}
    \label{fig:M1}
\end{figure}

The commutator $[q,p]$ given by \eqref{def:commutator} can be diagonalized. We have  $$
P_1^{-1}[q,p]P_1=\Lambda_1, \quad P_1=\left(\begin{array}{ccc}
\frac{2}{\sqrt{6}}&0&\frac{1}{\sqrt{3}}\\
-\frac{1}{\sqrt{6}}&-\frac{1}{\sqrt{2}}&\frac{1}{\sqrt{3}}\\
-\frac{1}{\sqrt{6}}&\frac{1}{\sqrt{2}}&\frac{1}{\sqrt{3}}\\
\end{array}\right).
$$
where $\Lambda_1$ is given by \eqref{eq:Lambda_1}. Conjugating $W(z)$ by $I_2\otimes P_1$ we can diagonalize the commutator in the asymptotic conditions of the Riemann-Hilbert problem. We also rearrange the jump near zero and infinity using matrix $M_1$ defined on Figure \ref{fig:M1} $$\Psi(z)=(I_2\otimes P_1)^{-1}W(z)(I_2\otimes P_1)M_1$$
\noindent where $S_j$ are given by \eqref{stokes} and 
$$
E_1=(I_2\otimes P_1)^{-1}E_0(I_2\otimes P_1), \quad A=P_1^{-1}\widetilde{A} P_1,\quad B= P_1^{-1}\widetilde{B} P_1,\quad C= P_1^{-1}\widetilde{C} P_1.
$$ As the result $\Psi(z)$ satisfies the Riemann-Hilbert problem \ref{mainrhp} with $E=(K\times I_3)^{-1}E_1^{-1}$.
We also can notice that
$$
m_1=(I_2\otimes P_1)^{-1}n_1(I_2\otimes P_1)
$$
and the relation \eqref{qquant} holds with
$$
q_0(t)=P_1^{-1}q(t)P_1.
$$
According to \cite{BCR} the cyclic relation \eqref{cyclic} implies 

\begin{equation} 
\begin{gathered}
\label{cyclic2}
(A+C+ABC)\ee^{-\ii\pi\Lambda_1}+\ee^{\ii\pi\Lambda_1}B+\ii I_3=0,\\
(AB+I_3)\ee^{-\ii\pi\Lambda_1}-\ee^{\ii\pi\Lambda_1}(BA+I_3)=0,\\
C\ee^{-\ii\pi\Lambda_1}A-A\ee^{\ii\pi\Lambda_1}C+\ee^{-\ii\pi\Lambda_1}-\ee^{\ii\pi\Lambda_1}=0,\\
(BC+I_3)\ee^{-\ii\pi\Lambda_1}-\ee^{\ii\pi\Lambda_1}(CB+I_3)=0,\\
B\ee^{-\ii\pi\Lambda_1}+\ee^{\ii\pi\Lambda_1}(A+C+CBA)+\ii I_3=0.
\end{gathered}
\end{equation}
We will use two different ansatzes for for matrices $A, B, C$. For asymptotics at $\infty$ we will assume
\begin{equation}\label{ansatzpinf}
 A=\left(\begin{array}{ccc}
a_{11}&a_{12}&a_{13}\\
a_{21}&a_{22}&a_{23}\\
0&a_{32}&a_{33}
\end{array}\right),\quad{B}=\left(\begin{array}{ccc}
0&b_{12}&b_{13}\\
0&0&b_{23}\\
0&0&0
\end{array}\right), \quad
C=\left(\begin{array}{ccc}
c_{11}&c_{12}&c_{13}\\
c_{21}&c_{22}&c_{23}\\
0&c_{32}&c_{33}
\end{array}\right).
\end{equation}
For asymptotics at $-\infty$ we will assume
\begin{equation}\label{ansatzminf}
 A=\left(\begin{array}{ccc}
a_{11}&0&0\\
a_{21}&a_{22}&0\\
a_{31}&a_{32}&a_{33}
\end{array}\right),\quad{B}=\left(\begin{array}{ccc}
b_{11}&b_{12}&0\\
b_{21}&b_{22}&b_{23}\\
b_{31}&b_{32}&b_{33}
\end{array}\right), \quad
C=\left(\begin{array}{ccc}
c_{11}&0&0\\
c_{21}&c_{22}&0\\
c_{31}&c_{32}&c_{33}
\end{array}\right).
\end{equation}

The motivation for this choice can be explained by simplification of asymptotic analysis and is mentioned at the end of Sections \ref{prelimp}, and \ref{prelimn}.
We notice that performing conjugation $A,B,C\to \Lambda_2^{-1}A\Lambda_2,\,\Lambda_2^{-1}B\Lambda_2,\,\Lambda_2^{-1}C\Lambda_2$ with
$$
\Lambda_2=\begin{pmatrix}
 b_{12}^{\frac{2}{3}}b_{23}^{\frac{1}{3}}&0&0\\
 0&b_{12}^{-\frac{1}{3}}b_{23}^{\frac{1}{3}}&0\\
 0&0&b_{12}^{-\frac{1}{3}}b_{23}^{-\frac{2}{3}}
\end{pmatrix}
$$
we can normalize $b_{12}$ and $b_{23}$ to be equal to $1$. This operation corresponds to the conjugation of the solution $\Psi(z)\to (I_2\otimes \Lambda_2)^{-1}\Psi(z)(I_2\otimes \Lambda_2)$. As a result, without loss of generality, we can put $b_{12}=b_{23}=1$. 

As a result the monodromy at $\infty$ can be parametrized in the case considered in this paper by $a_{13}$, $a_{23}$, $a_{33}$, $b_{13}$ as
\begin{equation}\aligned\label{eq:monodromy_plus_infty}
    &a_{21}={\sqrt{3}\ee^{-\frac{5\pi\ii}{6}}}{},\quad a_{32}={\sqrt{3}\ee^{\frac{5\pi\ii}{6}}}{},\quad a_{11}=\ee^{\frac{\ii\pi}{3}}a_{33}+\sqrt{3}\ee^{-\frac{5\pi\ii}{6}}b_{13},\\ & a_{22}=\ee^{-\frac{\ii\pi}{3}}a_{33}+\sqrt{3}\ee^{\frac{\ii\pi}{6}}b_{13},\quad a_{12}=-\ii\sqrt{3}b_{13}a_{33}+\ee^{-\frac{\ii\pi}{3}}a_{23}+\sqrt{3}\ee^{\frac{\ii\pi}{6}}b_{13}^2,\\
   &  c_{21}={\sqrt{3}\ee^{\frac{5\pi\ii}{6}}}{},\quad c_{32}={\sqrt{3}\ee^{-\frac{5\pi\ii}{6}}}{},\quad c_{11}=a_{33}+\ee^{-\frac{\ii\pi}{6}}+\sqrt{3}\ee^{\frac{5\pi\ii}{6}}b_{13},\quad c_{22}=\ii+\ee^{\frac{2\pi\ii}{3}}a_{33}+\sqrt{3}\ee^{-\frac{\ii\pi}{6}}b_{13},\\&c_{33}=\ee^{\frac{\ii\pi}{3}}a_{33}+\ee^{\frac{\ii\pi}{6}},\quad c_{12}=\ee^{-\frac{\ii\pi}{3}}+a_{23}+ \ee^{-\frac{\ii\pi}{6}}a_{33}+a_{33}^2+\sqrt{3}\ee^{\frac{2\pi\ii}{3}}b_{13}+\sqrt{3}\ee^{\frac{5\pi\ii}{6}}a_{33}b_{13}+\sqrt{3}\ee^{-\frac{\ii\pi}{6}}b_{13}^2,\\
   &c_{23}=\ee^{-\frac{2\pi\ii}{3}}+\ee^{-\frac{\ii\pi}{3}}a_{23}-\ii a_{33}+\ee^{-\frac{\ii\pi}{3}}a_{33}^2,\\& c_{13}=-a_{13}+\ee^{\frac{5\ii\pi}{6}}a_{23}+\ee^{\frac{2\pi\ii}{3}}a_{33}-2a_{23}a_{33}+\ee^{\frac{5\pi\ii}{6}}a_{33}^2-a_{33}^3+\ee^{-\frac{2\pi\ii}{3}}b_{13}+\sqrt{3}\ee^{-\frac{\ii\pi}{6}}a_{23}b_{13}-\ii a_{33}b_{13}\\&+\ee^{-\frac{\ii\pi}{3}}a_{33}^2b_{13}.\endaligned
\end{equation}
For the asymptotics at $-\infty$ the monodromy can be parametrized by $b_{11}$, $b_{22}$, $b_{21}$, $b_{31}$ and $a_{33}\in\{\ii,\ee^{-\frac{5\pi\ii}{6}}\}$.
\begin{equation}\aligned
\label{eq:monodromy_minus_infty}&a_{11}=a_{33}\ee^{\frac{\ii\pi}{3}},\quad a_{22}=a_{33}\ee^{-\frac{\ii\pi}{3}},\quad a_{21}=-\ii\sqrt{3}a_{33}b_{11}+\sqrt{3}\ee^{-\frac{5\pi\ii}{6}},\\& a_{32}=\sqrt{3}\ee^{\frac{5\pi\ii}{6}}+\sqrt{3}\ee^{\frac{5\pi\ii}{6}}a_{33}b_{11}+\sqrt{3}\ee^{\frac{5\pi\ii}{6}}a_{33}b_{22},\\
    & a_{31}=\ii\sqrt{3}b_{11}+\sqrt{3}\ee^{\frac{5\pi\ii}{6}}a_{33}b_{11}^2+\sqrt{3}\ee^{-\frac{5\pi\ii}{6}}a_{33}b_{21}+\sqrt{3}\ee^{\frac{\ii\pi}{6}}b_{22}+\ii\sqrt{3}a_{33}b_{11}b_{22},\\&b_{33}=b_{11}+b_{22},\quad b_{32}=b_{11}^2+b_{11}b_{22}+b_{22}^2+b_{21},\quad c_{11}=a_{33}+\ee^{-\frac{\ii\pi}{6}},\\& c_{22}=\ee^{\frac{2\pi\ii}{3}}a_{33} +\ii,\quad c_{33}=\ee^{\frac{\ii\pi}{3}}a_{33}+\ee^{\frac{\ii\pi}{6}},\quad 
    c_{21}=\sqrt{3}\ee^{\frac{2\pi\ii}{3}}b_{11}+\sqrt{3}\ee^{\frac{5\pi\ii}{6}}a_{33}b_{11}+\sqrt{3}\ee^{\frac{5\pi\ii}{6}},\\& c_{32}=\sqrt{3}\ee^{-\frac{5\pi\ii}{6}}+\sqrt{3}\ee^{-\frac{2\pi\ii}{3}}b_{11}-\ii\sqrt{3}a_{33}b_{11}+\sqrt{3}\ee^{-\frac{2\pi\ii}{3}}b_{22}-\ii\sqrt{3}a_{33}b_{22},\\&
    c_{31}=-\ii\sqrt{3}b_{11}+\sqrt{3}\ee^{-\frac{2\pi\ii}{3}}b_{11}^2-\ii\sqrt{3}a_{33}b_{11}^2-\sqrt{3}b_{21}+\sqrt{3}\ee^{-\frac{5\pi\ii}{6}}a_{33}b_{21}+\sqrt{3}\ee^{-\frac{\ii\pi}{6}}b_{22}+\sqrt{3}\ee^{-\frac{\ii\pi}{3}}b_{11}b_{22}\\&+\sqrt{3}\ee^{-\frac{\ii\pi}{6}}a_{33}b_{11}b_{22}.
\endaligned
\end{equation}
We can notice that the intersection of the parametrizations \eqref{ansatzpinf}, \eqref{eq:monodromy_plus_infty} and \eqref{ansatzminf}, \eqref{eq:monodromy_minus_infty} is given by
\begin{equation} \label{S1S2}
A=\left(\begin{array}{ccc}
\ee^{\frac{\ii\pi }{3}}a_{33}&0&0\\
{\sqrt{3}\ee^{-\frac{5\pi \ii}{6}}}&\ee^{-\frac{\ii \pi}{3}}a_{33}&0\\
0&{\sqrt{3}\ee^{\frac{5\pi \ii}{6}}}&a_{33}
\end{array}\right), \quad B=\left(\begin{array}{ccc}
0&1&0\\
0&0&1\\
0&0&0
\end{array}\right), \quad
C=\left(\begin{array}{ccc}
\ee^{\frac{2\pi\ii}{3}}a_{33}^{-1}&0&0\\[1mm]
{\sqrt{3}\ee^{\frac{5\pi \ii}{6}}}&\ee^{-\frac{2\pi\ii}{3}}a_{33}^{-1}&0\\[1mm]
0&{\sqrt{3}\ee^{-\frac{5\pi \ii}{6}}}{}&-a_{33}^{-1}
\end{array}\right)
\end{equation}
with $a_{33}^2+a_{33}\ee^{-\frac{\ii \pi }{6}}+\ee^{-\frac{\ii \pi }{3}}=0$ or equivalently $a_{33}\in\{\ii,\ee^{-\frac{5\pi\ii}{6}}\}$.

\section{Asymptotics as \texorpdfstring{$t\to\infty$}{t->infty}}{\label{RH1}}

\subsection{Preliminary transformations}\label{prelimp}

Consider scaling change of variables $$
\Phi(\lambda)=t^{-\frac{1}{2}I_2\otimes\Lambda_1}\Psi(\lambda\sqrt{t}).
$$
We have the behavior at infinity and zero changed to
\begin{equation}\label{atinfphi} 
\Phi(\lambda)=\left(I+O\left(\dfrac{1}{\lambda}\right)\right)\lambda^{I_2\otimes\Lambda_1}\ee^{\frac{\ii}{2}t^\frac{3}{2}\left(\frac{\lambda^3}{3}+\lambda\right)({\sigma}_3\otimes I_3)},\quad \lambda\to \infty,
\end{equation}
\begin{equation}
\aligned
\setlength{\tabcolsep}{1pt}\label{atzerophi}
\Phi(\lambda)=(K\otimes I_3)\begin{pmatrix}t^{-\frac{\Lambda_1}{2}}P_1^{-1}r_1P_1t^{\frac{\Lambda_1}{2}}&0\\0&t^{-\frac{\Lambda_1}{2}}P_1^{-1}r_2P_1t^{\frac{\Lambda_1}{2}}\end{pmatrix}\left(I_6+O\left(\lambda\right)\right)\lambda^{\frac{\sigma_3\otimes I_3}{6}}t^{-\frac{1}{2}I_2\otimes \Lambda_1+\frac{1}{12}\sigma_3\otimes I_3}E, \\\lambda \to 0,\,\,\,\, \,|\arg \lambda| < \frac{\pi}{6}.
\endaligned
\end{equation}

\begin{figure}[H]
\centering 
\includegraphics[scale=1]{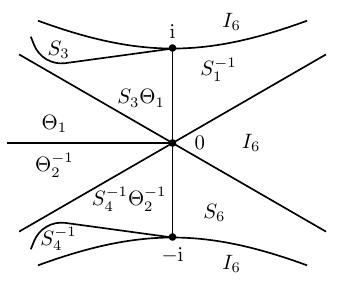}
	\caption{Constant matrix $M_2$}
	\label{fig:M2}	
\end{figure}

We see that the critical points for the method of nonlinear steepest descent are $\lambda=\pm \ii$. So we will move the jumps towards them. On the first step we want to factorize the diagonal jump $\ee^{-\frac{2\pi \ii}{3}}I_6=\Theta_1\Theta_2,$ with
$$\Theta_1=\ee^{-2\pi \ii \Lambda_3},\quad \Theta_2=\ee^{-2\pi \ii \Lambda_4},$$
$$
\Lambda_3=\left(\begin{array}{cccccc}
0&0&0&0&0&0\\
0& {-\frac{1}{3}}&0&0&0&0\\
0&0& -{\frac{2}{3}}&0&0&0\\
0&0&0&{\frac{1}{3}}&0&0\\
0&0&0&0&	{\frac{2}{3}}&0\\
0&0&0&0&0&0\\
\end{array} \right),\quad \Lambda_4=(\sigma_1\otimes I_3 )\Lambda_3 (\sigma_1\otimes I_3).
$$
The hyperbola $y^2-\dfrac{x^2}{3}=1$ is the antistokes curve  $\mathrm{Im}\left(\frac{\ii}{2}\left(\frac{\lambda^3}{3}+\lambda\right)\right)=0$ for the nonlinear steepest descent method.
We rearrange the jumps towards it by multiplying $\Phi(\lambda)$ by the constant matrix $M_2$ defined in Figure \ref{fig:M2}. 
\noindent As a result, we have the following Riemann-Hilbert problem for $Z(\lambda)={\Phi}(\lambda)M_2$.

\begin{figure}[H]
	\centering
	\includegraphics[scale=1]{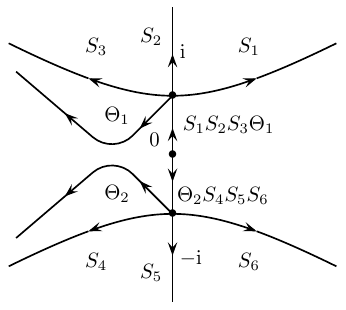}
	\caption{Contour for the Riemann-Hilbert problem \ref{pinfrhp1}.}
	\label{contpinfrhp1}
\end{figure}

\begin{rhp}\label{pinfrhp1}
Consider the contour $\Gamma$ shown on Figure \ref{contpinfrhp1}. The $6\times 6$ matrix valued function $Z(\lambda)$ satisfies the following conditions 
\begin{itemize}
    \item $Z(\lambda)$ is holomorphic on $\mathbb{C}\setminus \Gamma$.
    \item $Z(\lambda)$ has finite boundary values on the contour $\Gamma$ and satisfies the jump condition  indicated on Figure \ref{contpinfrhp1}.
\item $Z(\lambda)$ has the asymptotic \eqref{atinfphi}

at infinity and \eqref{atzerophi}
at zero respectively.
\end{itemize}
\end{rhp}

On the next step we perform the g-function transformation for this problem.
Consider $$S(\lambda)=\ee^{(I_2\otimes \Lambda_6)t^\frac{3}{2}}t^{(I_2\otimes  \Lambda_5)}Z(\lambda)\ee^{-\frac{\ii}{2}t^\frac{3}{2}\left(\frac{\lambda^3}{3}+\lambda\right)({\sigma}_3\otimes I_3)}\ee^{-(I_2\otimes \Lambda_6)t^\frac{3}{2}}t^{-I_2\otimes \Lambda_5}$$ where
$$
\Lambda_5=\left(\begin{array}{ccc}
\frac{2}{3}&0&0\\
0&\frac{1}{6}&0\\
0&0&-\frac{5}{6}\\
\end{array}\right),\quad 
 \Lambda_6=\left(\begin{array}{ccc}
-\frac{2}{3}&0&0\\
0&0&0\\
0&0&\frac{2}{3}\\
\end{array}\right).
$$
Terms involving $\Lambda_5$ will be used further in \eqref{matchi}, \eqref{matchmi}  for matching of parametrices.

We have the following sign charts describing behavior of  $\mathrm{Re}\left(\frac{\ii}{2}\left(\frac{\lambda^3}{3}+\lambda\right)\right)$.
\begin{figure}[H]
	\centering
\begin{subfigure}[b]{0.3\textwidth}\includegraphics[scale=1]{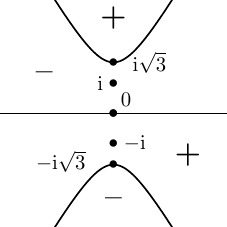}
	\caption{Sign chart for $\mathrm{Re}\left(\frac{\ii}{2}\left(\frac{\lambda^3}{3}+\lambda\right)\right)$}
\end{subfigure} 
\begin{subfigure}[b]{0.3\textwidth}
\includegraphics[scale=1]{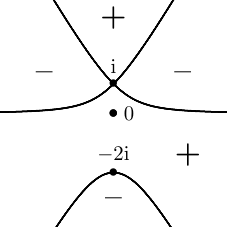}
	\caption{Sign chart for $\mathrm{Re}\left(\frac{\ii}{2}\left(\frac{\lambda^3}{3}+\lambda\right)\right)+\frac{1}{3}$}
\end{subfigure} 
\begin{subfigure}[b]{0.3\textwidth}
	\centering
	\includegraphics[scale=1]{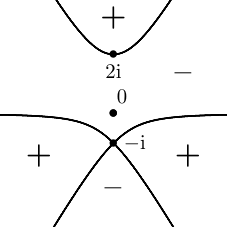}
	\caption{Sign chart for $\mathrm{Re}\left(\frac{\ii}{2}\left(\frac{\lambda^3}{3}+\lambda\right)\right)-\frac{1}{3}$}
 \end{subfigure} 
 \caption{}
\end{figure} 

For the matrix

$$\Upsilon=\left(\begin{array}{cccccc}
1&1&1&1&1&1\\
1&1&1&1&1&1\\
1&1&1&1&1&1\\
1&1&1&1&1&1\\
1&1&1&1&1&1\\
1&1&1&1&1&1
\end{array}\right)$$
we have
\begin{equation}
    \label{yps}\begin{gathered} 
\ee^{\frac{\ii}{2}\left(\frac{\lambda^3}{3}+\lambda\right){(\sigma}_3\otimes I_3)}\ee^{(I_2\otimes \Lambda_6)}\Upsilon \ee^{-\frac{\ii}{2}\left(\frac{\lambda^3}{3}+\lambda\right)({\sigma}_3\otimes I_3)}\ee^{-(I_2\otimes \Lambda_6)}
\\
=\left(\begin{array}{cccccc}
1&\ee^{-\frac{2}{3}}&\ee^{-\frac{4}{3}}&\ee^{{\ii}\left(\frac{\lambda^3}{3}+\lambda\right)}&\ee^{{\ii}\left(\frac{\lambda^3}{3}+\lambda\right)-\frac{2}{3}}&\ee^{{\ii}\left(\frac{\lambda^3}{3}+\lambda\right)-\frac{4}{3}}\\
\ee^{\frac{2}{3}}&1&\ee^{-\frac{2}{3}}&\ee^{{\ii}\left(\frac{\lambda^3}{3}+\lambda\right)+\frac{2}{3}}&\ee^{{\ii}\left(\frac{\lambda^3}{3}+\lambda\right)}&\ee^{{\ii}\left(\frac{\lambda^3}{3}+\lambda\right)-\frac{2}{3}}\\
\ee^{\frac{4}{3}}&\ee^{\frac{2}{3}}&1&\ee^{{\ii}\left(\frac{\lambda^3}{3}+\lambda\right)+\frac{4}{3}}&\ee^{{\ii}\left(\frac{\lambda^3}{3}+\lambda\right)+\frac{2}{3}}&\ee^{{\ii}\left(\frac{\lambda^3}{3}+\lambda\right)}\\
\ee^{-{\ii}\left(\frac{\lambda^3}{3}+\lambda\right)}&\ee^{-{\ii}\left(\frac{\lambda^3}{3}+\lambda\right)-\frac{2}{3}}&\ee^{-{\ii}\left(\frac{\lambda^3}{3}+\lambda\right)-\frac{4}{3}}&1&\ee^{-\frac{2}{3}}&\ee^{-\frac{4}{3}}\\
\ee^{-{\ii}\left(\frac{\lambda^3}{3}+\lambda\right)+\frac{2}{3}}&\ee^{-{\ii}\left(\frac{\lambda^3}{3}+\lambda\right)}&\ee^{-{\ii}\left(\frac{\lambda^3}{3}+\lambda\right)-\frac{2}{3}}&\ee^{\frac{2}{3}}&1&\ee^{-\frac{2}{3}}\\
\ee^{-{\ii}\left(\frac{\lambda^3}{3}+\lambda\right)+\frac{4}{3}}&\ee^{-{\ii}\left(\frac{\lambda^3}{3}+\lambda\right)+\frac{2}{3}}&\ee^{-{\ii}\left(\frac{\lambda^3}{3}+\lambda\right)}&\ee^{\frac{4}{3}}&\ee^{\frac{2}{3}}&1
\end{array}\right).
\end{gathered}
\end{equation}
We can see that after conjugation the jump matrices are exponentially close to the identity matrices except for the neighborhood of $\lambda=0,\pm\ii$ and parts with jumps $\Theta_1$, $\Theta_2$. This outcome has motivated us to choose ansatz \eqref{ansatzpinf}.  To cancel the remaining jumps we need to introduce parametrices solving the model Riemann-Hilbert problems.
\subsection{Global parametrix.}
\begin{figure}[H]
	\centering
	\includegraphics[scale=1]{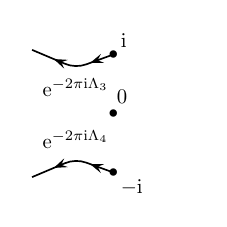}
	\caption{Contour for the Riemann-Hilbert problem \ref{pinfpinfrhp}}
	\label{pinfpinfcont}
\end{figure}

We start with the following Riemann-Hilbert problem
\begin{rhp}
\label{pinfpinfrhp}
Consider the contour $\Gamma$ shown on Figure \ref{pinfpinfcont}. The $6\times 6$ matrix valued function $P_{\infty}(\lambda)$ satisfies the following conditions 
\begin{itemize}
    \item $P_{\infty}(\lambda)$ is holomorphic on $\mathbb{C}\setminus \Gamma$.
    \item $P_{\infty}(\lambda)$ has finite boundary values on the contour $\Gamma$ and satisfies the jump condition  indicated on Figure \ref{pinfpinfcont}.
\item $P_{\infty}(\lambda)$ has the asymptotic $$P_{\infty}(\lambda)=(I_6+O(\lambda^{-1}))\lambda^{I_2\otimes\Lambda_1}, \quad \lambda\to \infty$$ at infinity and satisfies the estimate $$P_{\infty}(\lambda)=O((\lambda\mp \ii)^{-\frac{2}{3}}),\quad \lambda \to \pm \ii$$ at $\lambda=\pm \ii$.
\end{itemize}
\end{rhp}
\noindent The solution to it is given by
$
P_\infty=(\lambda-\ii)^{\Lambda_3}(\lambda+\ii)^{\Lambda_4}.
$

\subsection{Local parametrix near \texorpdfstring{$\lambda=\ii$}{lambda=i}}
\begin{figure}[H]
	\centering
	\includegraphics[scale=1]{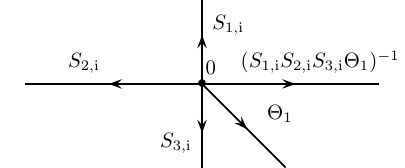}
	\caption{Contour for the Riemann-Hilbert problem \ref{plocirhp}}
	\label{plocicont}
\end{figure}

To describe parametrix near point $\lambda=\ii$ we introduce the local coordinate
$$
\dfrac{\zeta^2}{4}=\dfrac{\ii}{2}t^\frac{3}{2}\left(\frac{\lambda^3}{3}+\lambda\right)+\dfrac{t^\frac{3}{2}}{3},\quad \zeta= \ii\sqrt{2}(\lambda-\ii)t^\frac{3}{4}+O((\lambda-\ii)^2),\quad \lambda\to \ii.
$$
We introduce the parts of the jump matrices which are non-vanishing at the neighborhood of $\lambda=\ii$.

\begin{align}
  &  S_{1,\ii}=\left(\begin{array}{cccccc}
1&0&0&0&0&0\\
0&1&0&\sqrt{3}\ee^{-\frac{5\pi\ii}{6}}&0&0\\ 0&0&1&0&\sqrt{3}\ee^{\frac{5\pi\ii}{6}}&0\\
0&0&0&1&0&0\\
0&0& 0 &0& 1&0\\
0&0&0&0&0& 1\\
\end{array}\right),\quad  S_{2,\ii}=\left(\begin{array}{cccccc}
1&0&0&0&0&0\\
0&1&0&0&0&
0\\ 0&0&1&0&0&0\\
0&1&0&1&0&0\\
0&0& 1 &0& 1&0\\
0&0&0&0&0& 1\\
\end{array}\right),\\  &  S_{3,\ii}=\left(\begin{array}{cccccc}
1&0&0&0&0&0\\
0&1&0&\sqrt{3}\ee^{\frac{5\pi\ii}{6}}&0&0\\ 0&0&1&0&\sqrt{3}\ee^{-\frac{5\pi\ii}{6}}&0\\
0&0&0&1&0&0\\
0&0& 0 &0& 1&0\\
0&0&0&0&0& 1\\
\end{array}\right).
\end{align}
We will need solution to the following model Riemann-Hilbert problem
\begin{rhp}\label{plocirhp}
Consider the contour $\Gamma$ shown on Figure \ref{plocicont}. It consists of rays starting at zero in directions $\arg(z)=0,\frac{\pi}{2},\pi,\frac{3\pi}{2}, \frac{7\pi}{4}$. The $6\times 6$ matrix valued function $\Phi_\ii(z)$ satisfies the following conditions 
\begin{itemize}
    \item $\Phi_\ii(z)$ is holomorphic on $\mathbb{C}\setminus \Gamma$.
    \item $\Phi_\ii(z)$ has finite boundary values on the contour $\Gamma$ and satisfies the jump condition  indicated on Figure \ref{plocicont}. 
\item $\Phi_{\ii}(z)$ has the asymptotic $$\Phi_{\ii}(z)=\left(I_6+\frac{m_{1,\ii}}{z}+O(z^{-2})\right)z^{\Lambda_3}\ee^{\frac{z^2}{4}(\sigma_3\otimes I_3)}, \quad z\to \infty$$ at infinity and satisfies the estimate $\Phi_{\ii}(z)=O(1)$ at $z=0$. 

\end{itemize}
\end{rhp}
\noindent The solution to Riemann-Hilbert problem \ref{plocirhp} is constructed using parabolic cylinder functions in Appendix \ref{parcyli}.
It is possible to evaluate
$$
m_{1,\ii}=\left(\begin{array}{cccccc}
0&0&0&0&0&0\\
0&0&0&\dfrac{\sqrt{2\pi} \ee^{-\frac{\ii\pi }{6}}}{\Gamma\left(\frac{1}{3}\right)}&0&
0\\ 0&0&0&0&\dfrac{\sqrt{2\pi} \ee^{-\frac{5\pi \ii}{6}}}{\Gamma\left(\frac{2}{3}\right)}&0\\
0&\dfrac{\Gamma\left(\frac{4}{3}\right)\ee^{\frac{\ii\pi }{6}}}{\sqrt{2\pi} }&0&0&0&0\\
0&0& \dfrac{\Gamma\left(\frac{5}{3}\right)\ee^{\frac{5\pi \ii}{6}}}{\sqrt{2\pi} } &0& 0&0\\
0&0&0&0&0& 0\\
\end{array}\right).
$$
Using $\Phi_\ii$ we construct the parametrix near $\lambda=\ii$.
$$ P_{\ii}(\lambda)=A_\ii(\lambda)\Phi_\ii(\zeta(\lambda)) \ee^{-\frac{\zeta^2}{4}({\sigma}_3\otimes I_3)}t^{-(I_2\otimes \Lambda_5)},$$
Here $A_i(\lambda)$ is the holomorphic multiplier. It has the following asymptotic at $\lambda=\ii$
$$
A_{\ii}(\lambda)=P_\infty\zeta^{-\Lambda_3}t^{(I_2\otimes \Lambda_5)}= A_{\ii,0}+O(\lambda-\ii),\quad \lambda\to \ii,
$$
$$
A_{\ii,0}=\ee^{\frac{\ii\pi}{2}(\Lambda_4-\Lambda_3)}2^{\Lambda_4-\frac{\Lambda_3}{2}}t^{I_2\otimes \Lambda_5-\frac{3}{4}\Lambda_3}.
$$
The parametrix satisfy the matching condition
\begin{equation}\label{matchi} 
P_{\ii}P_\infty^{-1}=\left(I+O\left(\dfrac{1}{t^\frac{3}{4}}\right)\right).
\end{equation}
\subsection{Local parametrix near \texorpdfstring{$\lambda=-\ii$}{lambda=-i}}
\begin{figure}[H]
	\centering
	\includegraphics[scale=1]{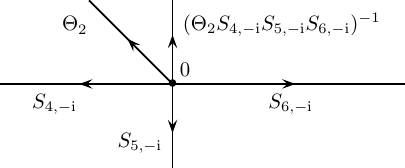}
	\caption{Contour for the Riemann-Hilbert problem \ref{plocmirhp}}
	\label{plocmicont}
\end{figure}

Now we proceed with parametrix near point $\lambda=-\ii$. We introduce local coordinate
$$
\dfrac{\zeta^2}{4}=\dfrac{\ii}{2}t^\frac{3}{2}\left(\frac{\lambda^3}{3}+\lambda\right)-\dfrac{t^\frac{3}{2}}{3},\quad \zeta= \sqrt{2}(\lambda+\ii)t^\frac{3}{4}+O((\lambda+\ii)^2),\quad \lambda\to -\ii.
$$
We introduce the parts of the jump matrices which are non-vanishing at the neighborhood of $\lambda=-\ii$.

\begin{align}
  &  S_{4,-\ii}=\left(\begin{array}{cccccc}
1&0&0&0&0&0\\
0&1&0&0&0&0\\ 0&0&1&0&0&0\\
0&0&0&1&0&0\\
\sqrt{3}\ee^{-\frac{5\pi\ii}{6}}&0& 0 &0& 1&0\\
0&\sqrt{3}\ee^{-\frac{\ii\pi}{6}}&0&0&0& 1\\
\end{array}\right),\quad  S_{5,-\ii}=\left(\begin{array}{cccccc}
1&0&0&0&1&0\\
0&1&0&0&0&
-1\\ 0&0&1&0&0&0\\
0&0&0&1&0&0\\
0&0& 0 &0& 1&0\\
0&0&0&0&0& 1\\
\end{array}\right),\\  &  S_{6,-\ii}=\left(\begin{array}{cccccc}
1&0&0&0&0&0\\
0&1&0&0&0&0\\ 0&0&1&0&0&0\\
0&0&0&1&0&0\\
\sqrt{3}\ee^{\frac{5\pi\ii}{6}}&0& 0 &0& 1&0\\
0&\sqrt{3}\ee^{\frac{\ii\pi}{6}}&0&0&0& 1\\
\end{array}\right).
\end{align}
We will need solution to the following model Riemann-Hilbert problem
\begin{rhp}\label{plocmirhp}
Consider the contour $\Gamma$ shown on Figure \ref{plocmicont}. It consists of rays starting at zero in directions $\arg(z)=0,\frac{\pi}{2},\frac{3\pi}{4},-\pi,-\frac{\pi}{2}$. The $6\times 6$ matrix valued function $\Phi_{-\ii}(z)$ satisfies the following conditions 
\begin{itemize}
    \item $\Phi_{-\ii}(z)$ is holomorphic on $\mathbb{C}\setminus \Gamma$.
    \item $\Phi_{-\ii}(z)$  has finite boundary values on the contour $\Gamma$ and satisfies the jump condition  indicated on Figure \ref{plocmicont}.
\item $\Phi_{-\ii}(z)$ has the asymptotic $$\Phi_{-\ii}(z)=\left(I_6+\frac{m_{1,-\ii}}{z}+O(z^{-2})\right)z^{\Lambda_4}\ee^{\frac{z^2}{4}(\sigma_3\otimes I_3)}, \quad z\to \infty$$ at infinity and satisfies the estimate $\Phi_{-\ii}(z)=O(1)$ at $z=0$.

\end{itemize}
\end{rhp}
\noindent The solution to Riemann-Hilbert problem \ref{plocmirhp} is constructed using parabolic cylinder functions in Appendix \ref{parcylmi}. It is possible to evaluate
$$
m_{1,-\ii}=\left(\begin{array}{cccccc}
0&0&0&0&\dfrac{\Gamma\left(\frac{4}{3}\right) \ee^{-\frac{\ii\pi }{3}}}{\sqrt{2\pi}}&0\\
0&0&0&0&0&\dfrac{\Gamma\left(\frac{5}{3}\right) \ee^{\frac{\ii\pi }{3}}}{\sqrt{2\pi}}\\ 0&0&0&0&0&0\\
0&0&0&0&0&0\\
\dfrac{\sqrt{2\pi}\ee^{-\frac{2\pi \ii}{3}}}{\Gamma\left(\frac{1}{3}\right) }&0&0 &0& 0&0\\
0& \dfrac{\sqrt{2\pi}\ee^{\frac{2\pi \ii}{3}}}{\Gamma\left(\frac{2}{3}\right) }&0&0&0& 0\\
\end{array}\right).
$$
Using $\Phi_{-\ii}$ we construct the parametrix near $\lambda=-\ii$$$ P_{-\ii}=A_{-\ii}(\lambda)\Phi_{-\ii}(\zeta(\lambda))\ee^{-\frac{\zeta^2}{4}({\sigma}_3\otimes I_3)}t^{-(I_2\otimes \Lambda_5)}.$$
Here $A_{-\ii}(\lambda)$ is the holomorphic multiplier. It has the following asymptotic at $\lambda=-\ii$
$$
A_{-\ii}(\lambda)=P_\infty\zeta^{- \Lambda_4}t^{(I_2\otimes \Lambda_5)}= A_{-\ii,0}+O(\lambda+\ii),\quad \lambda\to -\ii,
$$
$$
A_{-\ii,0}=\ee^{-\frac{\ii\pi}{2}\Lambda_3}2^{\Lambda_3-\frac{\Lambda_4}{2}}t^{I_2\otimes \Lambda_5-\frac{3\Lambda_4}{4}}.
$$
The parametrix satisfy the matching condition
\begin{equation}\label{matchmi}
P_{-
\ii}P_\infty^{-1}=\left(I+O\left(\dfrac{1}{t^\frac{3}{4}}\right)\right).
\end{equation}
\subsection{Local parametrix near \texorpdfstring{$\lambda=0$}{lambda=0}}
\begin{figure}[H]
	\centering
	\includegraphics[scale=1]{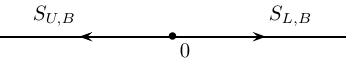}
	\caption{Contour for the Riemann-Hilbert problem \ref{ploczrhp}}
	\label{ploczcont}
\end{figure}

Finally to describe parametrix near $\lambda=0$ we introduce local coordinate
$$
\zeta=\dfrac{\ii t^\frac{3}{2}}{2}\left(\lambda+\frac{\lambda^3}{3}\right).
$$
We introduce the parts of the jump matrices which are non-vanishing at the neighborhood of $\lambda=0$.

\begin{align}
  &  S_{U,B}=\left(\begin{array}{cccccc}
1&0&0&\ee^{-\frac{5\pi\ii}{6}}&0&0\\
0&1&0&0&\ii&0\\ 0&0&1&0&0&\ee^{\frac{5\pi\ii}{6}}\\
0&0&0&1&0&0\\
0&0& 0 &0& 1&0\\
0&0&0&0&0& 1\\
\end{array}\right),\quad  S_{L,B}=\left(\begin{array}{cccccc}
1&0&0&0&0&0\\
0&1&0&0&0&0\\ 0&0&1&0&0&0\\
\ee^{-\frac{\ii\pi}{6}}&0&0&1&0&0\\
0&\ii& 0 &0& 1&0\\
0&0&\ee^{\frac{\ii\pi}{6}}&0&0& 1\\
\end{array}\right).
\end{align}
We will need solution to the following model Riemann-Hilbert problem
\begin{rhp}\label{ploczrhp}
Consider the contour $\Gamma$ shown on Figure \ref{ploczcont}. The $6\times 6$ matrix valued function $\Phi_{0}(z)$ satisfies the following conditions 
\begin{itemize}
    \item $\Phi_{0}(z)$ is holomorphic on $\mathbb{C}\setminus \Gamma$.
    \item $\Phi_{0}(z)$  has finite boundary values on the contour $\Gamma$ and satisfies the jump condition  indicated on Figure \ref{ploczcont}.
\item $\Phi_{0}(z)$ has the asymptotic $$\Phi_{0}(z)=\left(I_6+O(z^{-1})\right)\ee^{{z}{}(\sigma_3\otimes I_3)}, \quad z\to \infty$$ at infinity and $\Phi_{0}(z)=O(z^{-\frac{1}{6}})$ at zero.

\end{itemize}
\end{rhp}
\noindent The solution to Riemann-Hilbert problem \ref{ploczrhp} is constructed using Bessel functions in Appendix \ref{bessel}. Using $\Phi_{0}$ we construct the parametrix near $\lambda=0$
$$
P_{0}=P_\infty \Phi_{0}(\zeta(\lambda))\ee^{-\zeta({\sigma}_3\otimes I_3)}.
$$
The parametrix satisfy the matching condition
\begin{equation}\label{matchz}
    P_{0}P_\infty^{-1}=\left(I+O\left(\dfrac{1}{t^\frac{3}{2}}\right)\right).
\end{equation}

\subsection{Small norm theorem and leading terms of asymptotic}
\begin{figure}[H]
	\centering
\includegraphics[scale=1]{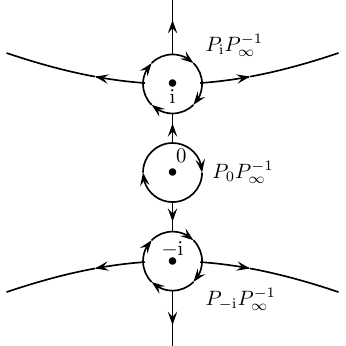}
	\caption{Contour for the Riemann-Hilbert problem \ref{pinfrrhp}}
	\label{pinfrcont}
\end{figure}

Now, having all parametrices constructed we define

$$
S_{as}=\left\{\begin{array}{cc}
P_{0},&\quad |\lambda|<\delta,\\
P_{\ii},&\quad |\lambda-\ii|<\delta,\\
P_{-\ii},&\quad |\lambda+\ii|<\delta,\\
P_{\infty},&\quad \mbox{otherwise}.
\end{array} \right.
$$
for small number $\delta$. Then we arrive to the Riemann Hilbert problem for 
$
R=SS_{as}^{-1}
$
with small jump.
\begin{rhp}\label{pinfrrhp}
Consider the contour $\Gamma$ shown on Figure \ref{pinfrcont}. The $6\times 6$ matrix valued function $R(\lambda)$ satisfies the following conditions 
\begin{itemize}
    \item $R(\lambda)$ is holomorphic on $\mathbb{C}\setminus \Gamma$.
    \item $R(\lambda)$  has finite boundary values on the contour $\Gamma$ and satisfies the jump condition  indicated on Figure \ref{pinfrcont}. On the non-labeled parts of contour the jump is obtained from the jump for $S(\lambda)$ conjugating with $S_{as}(\lambda)$. 
\item $R(\lambda)$ has the asymptotic at infinity $$R(\lambda)=I_6+ \frac{l_1}{\lambda}+O(\lambda^{-2}), \quad \lambda\to \infty.$$

\end{itemize}
\end{rhp}
Since $P_\infty(\lambda)$ is growing at infinity, we get the terms of sort $\lambda \ee^{t^{\frac{3}{2}}\left(-\frac{2}{3}+y-\frac{y^3}{3}\right)}$ in the jump for $R(\lambda)$ along the positive imaginary axis with $y=\mathrm{Im}\lambda$. We have for $y>3$ and for $t>2$ that
\begin{equation} \label{estimate}
\left|y\, \ee^{t^{\frac{3}{2}}\left(-\frac{2}{3}+y-\frac{y^3}{3}\right)}\right|\leq y \ee^{-t^{\frac{3}{2}}\frac{y^3}{6}}\leq 3\ee^{-\frac{9t^{\frac{3}{2}}}{2}}.
\end{equation}
Similar estimate needs to be done for the jumps along the other contours approaching infinity. As the result, using estimates \eqref{matchi}, \eqref{matchmi}, \eqref{matchz} we get the estimate for jump matrix
\begin{equation}\label{jest} 
\|J(\mu)-I_6\|_{L_p(\Gamma)}=O(t^{-\frac{3}{4}}),\quad p=1,2,\infty.
\end{equation}
Now following the standard procedure we write the singular integral equation for $R_-(\lambda)$.

$$
R_-(\lambda)=I+\intop_{\Gamma}\frac{R_-(\mu)(J(\mu)-I)}{\mu-\lambda+i0}\frac{d\mu}{2\pi \ii},
$$
where $J(\lambda)$ is the jump matrix. Using \eqref{jest} we get

\begin{equation}\label{rest}
R(\lambda)=I+O(t^{-\frac{3}{4}}).
\end{equation}
We also have the formula for $R(\lambda)$ away from contour $\Gamma$
$$
R(\lambda)=I+\intop_{\Gamma}\frac{R_-(\mu)(J(\mu)-I)}{\mu-\lambda}\frac{d\mu}{2\pi \ii}.
$$
Expanding it at infinity we get $$
l_1=-\intop_{\Gamma}{R_-(\mu)(J(\mu)-I)}\frac{d\mu}{2\pi \ii}.
$$
Using the estimates \eqref{rest}, \eqref{jest} we arrive at

$$
l_1=-\intop_{\Gamma}{(J(\mu)-I)}\frac{d\mu}{2\pi \ii}+O(t^{-\frac{3}{2}}).
$$

The main part of this integral comes from circles around $\lambda=\pm \ii $. We orient these circles clockwise, so that $J=P_{\pm\ii}P_\infty^{-1}$ on them. The plus side is outside each circle:
$R_+=SP_\infty^{-1}$ and $R_-=SP_{\pm\ii}^{-1}$.
Thus the minus sign in the Cauchy formula for $l_1$ cancels the minus sign
from clockwise integration, and each leading contribution is the positive
residue of $P_{\pm\ii}P_\infty^{-1}-I$.
$$
l_1=-\intop_{|\mu-\ii|=\delta}{(J(\mu)-I)}\frac{d\mu}{2\pi \ii}-\intop_{|\mu+\ii|=\delta}{(J(\mu)-I)}\frac{d\mu}{2\pi \ii}+O(t^{-\frac{3}{2}}).
$$
Using the expansion for $\Phi_\ii,\Phi_{-\ii}$ at infinity we get
$$
l_1=\dfrac{A_{\ii,0}m_{1,\ii}A_{\ii,0}^{-1}}{\ii\sqrt{2}t^{\frac{3}{4}}}+\dfrac{A_{-\ii,0}m_{1,-\ii}A_{-\ii,0}^{-1}}{\sqrt{2}t^{\frac{3}{4}}}+O(t^{-\frac{3}{2}}).
$$
Tracing back the transformations from Section \ref{prelimp} we have
$$
(m_1)_{\mbox{offdiag}}=\sqrt{t}(I_2\otimes P_1)\ee^{-(I_2\otimes \Lambda_6)t^{\frac{3}{2}}}t^{-I_2\otimes \Lambda_5+\frac{1}{2}I_2\otimes \Lambda_1}l_1\ee^{(I_2\otimes \Lambda_6)t^{\frac{3}{2}}}t^{I_2\otimes \Lambda_5-\frac{1}{2}I_2\otimes \Lambda_1}(I_2\otimes P_1)^{-1}.
$$
Using the formula $q(t)=-\ii\hat{m}_{1,12}$ we have
$$
\det(q(t)-xI_3)=-x^3+O(t^{-1})x^2+\left(-\frac{1}{2\sqrt{t}}+O(t^{-\frac{5}{4}})\right)x+O(t^{-\frac{3}{2}}).
$$
Put $x=t^{-1/4}\nu$ and multiply the characteristic polynomial by $t^{3/4}$. The resulting equation has the form
\[
P(\nu,\mathbf c(t))=-\nu^3+c_2(t)\nu^2+
\left(-\frac12+c_1(t)\right)\nu+c_0(t)=0,
\qquad \mathbf c(t)=O(t^{-3/4}).
\]
At $\mathbf c=0$ its roots are $\nu_1^0=0$ and
$\nu_{2,3}^0=\pm\ii/\sqrt2$. They are simple: the derivative
$\partial_\nu P(\nu,0)=-3\nu^2-1/2$ equals $-1/2$ at $0$ and $1$ at the other two roots.
Since $P$ is a polynomial in $\nu$ and $\mathbf c$, the implicit
function theorem gives a root $\nu_j(\mathbf c)$ near each $\nu_j^0$, with
$\nu_j(\mathbf c)=\nu_j^0+O(\|\mathbf c\|)$.
Applying this to the actual coefficients $\mathbf c(t)$ gives
$\nu_j(t)=\nu_j^0+O(t^{-3/4})$.
Undoing the rescaling, we obtain the following eigenvalue asymptotics up to permutation
$$
x_1=O(t^{-1}),\quad x_2=\frac{\ii\sqrt{2}}{{2t^{\frac{1}{4}}}}+O(t^{-1}),\quad x_3=-\frac{\ii\sqrt{2}}{{2t^{\frac{1}{4}}}}+O(t^{-1}).
$$
\subsection{Recurrence relation and subleading terms of asymptotics}
We can see that the full asymptotic have form
$$
x_1=\sum_{j=0}^\infty c_j t^{-1-\frac{3}{4}j},\quad x_2=\frac{\ii\sqrt{2}}{{2t^{\frac{1}{4}}}}+\sum_{j=0}^\infty d_j t^{-1-\frac{3}{4}j},\quad x_3=-\frac{\ii\sqrt{2}}{{2t^{\frac{1}{4}}}}+\sum_{j=0}^\infty e_j t^{-1-\frac{3}{4}j}.
$$
We can rewrite the Calogero-Painlev\'e system in the following form
\begin{equation} \label{rewrcp}
\begin{gathered} 
3(x_1-x_2)^3(x_1-x_3)^3(-6x_1''+12x_1^3+6x_1 t-1)-4((x_1-x_2)^3+(x_1-x_3)^3)=0,
\\
3(x_2-x_3)^3(x_2-x_1)^3(-6x_2''+12x_2^3+6x_2 t-1)-4((x_2-x_1)^3+(x_2-x_3)^3)=0,
\\
3(x_3-x_2)^3(x_3-x_1)^3(-6x_3''+12x_3^3+6x_3 t-1)-4((x_3-x_2)^3+(x_3-x_1)^3)=0.
\end{gathered} 
\end{equation}
We have 

$$
(x_1-x_2)^3=\frac{\ii}{2\sqrt{2}t^\frac{3}{4}}-\frac{3}{2}\sum_{k=2}^\infty (c_{k-2}-d_{k-2})t^{-\frac{3}{4}k}-\frac{3i}{\sqrt{2}}\sum_{k=3}^\infty \sum_{j=0}^{k-3}(c_{j}-d_{j})(c_{k-j-3}-d_{k-j-3})t^{-\frac{3}{4}k}$$
$$+\sum_{k=4}^\infty \sum_{j_1=0}^{k-4}\sum_{j_2=0}^{k-4-j_1}(c_{j_1}-d_{j_1})(c_{j_2}-d_{j_2})(c_{k-4-j_1-j_2}-d_{k-4-j_1-j_2})t^{-\frac{3}{4}k},
$$
$$
(x_1-x_3)^3=-\frac{\ii}{2\sqrt{2}t^\frac{3}{4}}-\frac{3}{2}\sum_{k=2}^\infty (c_{k-2}-e_{k-2})t^{-\frac{3}{4}k}+\frac{3i}{\sqrt{2}}\sum_{k=3}^\infty \sum_{j=0}^{k-3}(c_{j}-e_{j})(c_{k-j-3}-e_{k-j-3})t^{-\frac{3}{4}k}$$
$$+\sum_{k=4}^\infty \sum_{j_1=0}^{k-4}\sum_{j_2=0}^{k-4-j_1}(c_{j_1}-e_{j_1})(c_{j_2}-e_{j_2})(c_{k-4-j_1-j_2}-e_{k-4-j_1-j_2})t^{-\frac{3}{4}k},
$$
$$
(x_2-x_3)^3=-\frac{4\ii}{\sqrt{2}t^\frac{3}{4}}-6\sum_{k=2}^\infty (d_{k-2}-e_{k-2})t^{-\frac{3}{4}k}+\frac{6i}{\sqrt{2}}\sum_{k=3}^\infty \sum_{j=0}^{k-3}(d_{j}-e_{j})(d_{k-j-3}-e_{k-j-3})t^{-\frac{3}{4}k}$$
$$+\sum_{k=4}^\infty \sum_{j_1=0}^{k-4}\sum_{j_2=0}^{k-4-j_1}(d_{j_1}-e_{j_1})(d_{j_2}-e_{j_2})(d_{k-4-j_1-j_2}-e_{k-4-j_1-j_2})t^{-\frac{3}{4}k}
$$
$$
x_1''=\sum_{k=4}^{\infty}\left(2-\frac{3}{4}k\right)\left(1-\frac{3}{4}k\right)c_{k-4}t^{-\frac{3}{4}k},
$$
$$
x_2''= \frac{5\ii}{16\sqrt{2}}t^{-\frac{9}{4}}+\sum_{k=4}^{\infty}\left(2-\frac{3}{4}k\right)\left(1-\frac{3}{4}k\right)d_{k-4}t^{-\frac{3}{4}k},
$$
$$
x_3''= -\frac{5\ii}{16\sqrt{2}}t^{-\frac{9}{4}}+\sum_{k=4}^{\infty}\left(2-\frac{3}{4}k\right)\left(1-\frac{3}{4}k\right)e_{k-4}t^{-\frac{3}{4}k},
$$
$$
x_1^3=\sum_{k=4}^\infty \sum_{j_1=0}^{k-4}\sum_{j_2=0}^{k-4-j_1}c_{j_1}c_{j_2}c_{k-4-j_1-j_2}t^{-\frac{3}{4}k},
$$
$$
x_2^3=-\frac{\ii}{2\sqrt{2}t^\frac{3}{4}}-\frac{3}{2}\sum_{k=2}^\infty d_{k-2}t^{-\frac{3}{4}k}+\frac{3\ii}{\sqrt{2}}\sum_{k=3}^\infty \sum_{j=0}^{k-3}d_{j}d_{k-j-3}t^{-\frac{3}{4}k}+\sum_{k=4}^\infty \sum_{j_1=0}^{k-4}\sum_{j_2=0}^{k-4-j_1}d_{j_1}d_{j_2}d_{k-4-j_1-j_2}t^{-\frac{3}{4}k},
$$
$$
x_3^3=\frac{\ii}{2\sqrt{2}t^\frac{3}{4}}-\frac{3}{2}\sum_{k=2}^\infty e_{k-2}t^{-\frac{3}{4}k}-\frac{3\ii}{\sqrt{2}}\sum_{k=3}^\infty \sum_{j=0}^{k-3}e_{j}e_{k-j-3}t^{-\frac{3}{4}k}+\sum_{k=4}^\infty \sum_{j_1=0}^{k-4}\sum_{j_2=0}^{k-4-j_1}e_{j_1}e_{j_2}e_{k-4-j_1-j_2}t^{-\frac{3}{4}k}.
$$
We can see that in \eqref{rewrcp} in the coefficient near $t^{-\frac{3}{4}k}$ the terms with highest indices $c_{k-2}$, $d_{k-2}$, $e_{k-2}$ come from the terms
\begin{equation} 
\begin{gathered} 
18(x_1-x_2)^3(x_1-x_3)^3x_1 t-4((x_1-x_2)^3+(x_1-x_3)^3),
\\
18(x_2-x_1)^3(x_2-x_3)^3x_2 t-4((x_2-x_1)^3+(x_2-x_3)^3),
\\
18(x_3-x_2)^3(x_3-x_1)^3x_3 t-4((x_3-x_2)^3+(x_3-x_1)^3).
\end{gathered} 
\end{equation}
These coefficients have form

$$
\left(\begin{array}{ccc}
     \frac{57}{4}&-6&-6  \\
     48&-69&3\\
     48&3&-69
\end{array}\right)\left(\begin{array}{c}c_{k-2}\\d_{k-2}\\e_{k-2}\end{array}\right)+\mbox{terms with smaller indices}.
$$
The determinant of the matrix above is 26244, so we derived the recurrence relation for the coefficients $c_k,d_k,e_k$. It determines all the coefficients starting from $c_0, d_0, e_0$ uniquely. The first few terms are given in \eqref{x2as}.

\begin{remark} We can notice that other power asymptotics are possible for the solution of \eqref{rewrcp}, for example 
$$
x_1=\sum_{j=0}^\infty c_j t^{-1-\frac{3}{4}j},\quad x_2=\frac{\sqrt{2}}{{2t^{\frac{1}{4}}}}+\sum_{j=0}^\infty d_j t^{-1-\frac{3}{4}j},\quad x_3=-\frac{\sqrt{2}}{{2t^{\frac{1}{4}}}}+\sum_{j=0}^\infty e_j t^{-1-\frac{3}{4}j},
$$
but they don't produce desired estimate \eqref{con16}.
\end{remark}

\section{Asymptotics as \texorpdfstring{$t\to-\infty$}{t->-infty} for \texorpdfstring{$a_{33}=\ii$}{a33=i}}\label{RH2}
In this section we use the parametrization \eqref{ansatzminf}, \eqref{eq:monodromy_minus_infty} with $b_{12}=b_{23}=1$, $a_{33}=\ii$, and the remaining monodromy parameters fixed. This includes the data \eqref{S1S2}. The other root $a_{33}=\ee^{-\frac{5\pi\ii}{6}}$ leads to different asymptotics and is treated in Section~\ref{RH3}, where the constructions of this section are reused.

\subsection{Preliminary transformations.}\label{prelimn}
Consider scaling change of variables $$
\Phi(\lambda)=(-t)^{-\frac{1}{2}I_2\otimes\Lambda_1}\Psi(\lambda\sqrt{-t}).
$$
We have the behavior at infinity and zero changed to

\begin{equation}\label{atinfphi2} 
\Phi(\lambda)=\left(I+O\left(\dfrac{1}{\lambda}\right)\right)\lambda^{I_2\otimes\Lambda_1}\ee^{\frac{\ii}{2}(-t)^\frac{3}{2}\left(\frac{\lambda^3}{3}-\lambda\right)({\sigma}_3\otimes I_3)},\quad \lambda\to \infty,
\end{equation}
\begin{equation} \aligned\label{atzerophi2}\setlength\arraycolsep{2pt}
\Phi(\lambda)=(K\otimes I_3)\left(\begin{array}{cc}(-t)^{-\frac{\Lambda_1}{2}}P_1^{-1}r_1P_1(-t)^{\frac{\Lambda_1}{2}}&0\\0&(-t)^{-\frac{\Lambda_1}{2}}P_1^{-1}r_2P_1(-t)^{\frac{\Lambda_1}{2}}\end{array}\right)\\\times\left(I_6+O\left(\lambda\right)\right)\lambda^{\frac{\sigma_3\otimes I_3}{6}}(-t)^{-\frac{1}{2}I_2\otimes \Lambda_1+\frac{1}{12}\sigma_3\otimes I_3}E, \, \lambda\to 0.\endaligned
\end{equation}
Before doing the deformation of contours, we introduce the g-function
$$
g(\lambda)=\frac{1}{6}\left(2-\lambda^2\right)^\frac{3}{2}=\frac{1}{6}\ee^{\frac{3\pi}{2}\ii+\frac{3}{2}\ln(\lambda-\sqrt{2})+\frac{3}{2}\ln(\lambda+\sqrt{2})}.
$$

We introduce contours $\gamma_1, \gamma_3$ in such a way so along them $\mathrm{Re} (g(\lambda))=\mathrm{Re} (g(\sqrt{2}))=0$. They are parts of hyperbola
$$
\frac{2}{3}\,xy\sqrt {3}+{x}^{2}-{y}^{2}-2=0.
$$

We introduce contours $\gamma_{2,\pm}$ in such a way that along them $|\mathrm{Re} (g(\lambda))|=\dfrac{\sqrt{2}}{3}$. They are parts of algebraic curve
\begin{equation}
\begin{aligned}\label{eq:algebraic}
&-9\,{x}^{10}{y}^{2}+60\,{x}^{8}{y}^{4}-118\,{x}^{6}{y}^{6}+60\,{x}^{4}
{y}^{8}-9\,{x}^{2}{y}^{10}+72\,{x}^{8}{y}^{2}-312\,{x}^{6}{y}^{4}\\&+312
\,{x}^{4}{y}^{6}-72\,{x}^{2}{y}^{8}-216\,{x}^{6}{y}^{2}+528\,{x}^{4}{y
}^{4}
-216\,{x}^{2}{y}^{6}+8\,{x}^{6}+168\,{x}^{4}{y}^{2}\\&-168\,{x}^{2}{
y}^{4}-8\,{y}^{6}-48\,{x}^{4}+144\,{x}^{2}{y}^{2}-48\,{y}^{4}+96\,{x}^
{2}-96\,{y}^{2}
=0.
\end{aligned}
\end{equation}
Actually denote $g=u+\ii v$, $\lambda=x+\ii z$. The equation $36g(\lambda)^2=(2-\lambda^2)^3$ gives quadratic equations on $u$ and $v$. Squaring the condition $|\mathrm{Re} (g(\lambda))|=\dfrac{\sqrt{2}}{3}$ and eliminating radicals we obtain \eqref{eq:algebraic}. 
\begin{figure}[H]
	\centering
\includegraphics[scale=1]{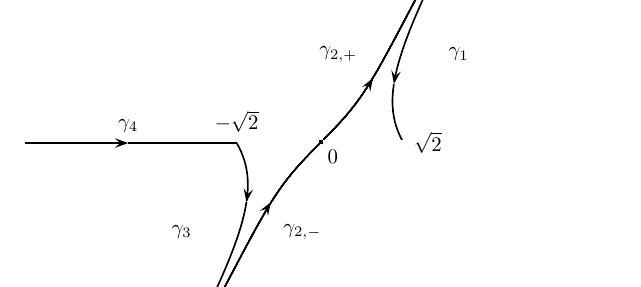}
	\caption{Branch cuts for $\arg(\lambda-\sqrt{2})$ and $\arg(\lambda+\sqrt{2})$}
	\label{cuts}
\end{figure}

We take $\arg(\lambda-\sqrt{2})=0$ for real $\lambda>\sqrt{2}$. We assume the branch cut for $\arg(\lambda-\sqrt{2})$ on the curves $\gamma_1$ ,$\gamma_{2,\pm}$, $\gamma_{3}$, $\gamma_4$. Similarly we assume $\arg(\lambda+\sqrt{2})=0$ for real $\lambda>-\sqrt{2}$ and we take the branch cut for $\arg(\lambda+\sqrt{2})$ along the curve $\gamma_4$. 

We can notice that along $\gamma_{2,\pm}$ we have $\mathrm{Re} (g_+(\lambda))=-\dfrac{\sqrt{2}}{3}$ and $\mathrm{Re} (g_-(\lambda))=\dfrac{\sqrt{2}}{3}$. Moreover on contours $\gamma_1, \gamma_{2,\pm}, \gamma_3$ we have
$
g_+(\lambda)=-g_-(\lambda ).
$

We also have $\arg(\lambda-\sqrt{2})_+=\arg(\lambda-\sqrt{2})_-+2\pi$ on $\gamma_1$ ,$\gamma_{2,\pm}$, $\gamma_{3}$, $\gamma_4$ and $\arg(\lambda+\sqrt{2})_+=\arg(\lambda+\sqrt{2})_-+2\pi$ on $\gamma_4$.

\begin{figure}[H]
	\centering
\includegraphics[scale=1]{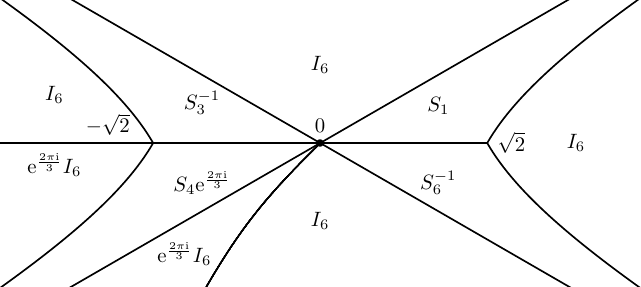}
	\caption{Constant matrix $M_3$}
    \label{fig:M3}
\end{figure}

The antistokes curves $\mathrm{Im}\left(g(\lambda)\right)=\mathrm{Im}\left(g(\pm\sqrt{2})\right)=0$ are parts of the hyperbolas $
\pm\frac{2}{3}\,xy\sqrt {3}+{x}^{2}-{y}^{2}-2=0.
$
We rearrange the jumps towards it by multiplying $\Phi(\lambda)$ by constant matrix $M_3$ defined on Figure \ref{fig:M3}.  
We also moved the branch cut for $\lambda^{I_2\otimes \Lambda_1}$ towards contour $\gamma_{2,-}$. As the result we have the following Riemann-Hilbert problem for $Z(\lambda)={\Phi}(\lambda)M_3$.

\begin{figure}[H]
	\centering
	\includegraphics[scale=1]{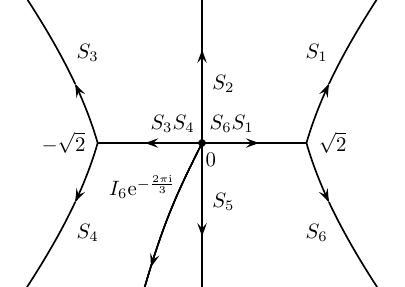}
	\caption{Contour for the Riemann-Hilbert problem \ref{pinfzrhp}.}
	\label{contpinfzrhp}
\end{figure}

\begin{rhp}\label{pinfzrhp}
Consider the contour $\Gamma$ shown on Figure \ref{contpinfzrhp}. The $6\times 6$ matrix valued function $Z(\lambda)$ satisfies the following conditions 
\begin{itemize}
    \item $Z(\lambda)$ is holomorphic on $\mathbb{C}\setminus \Gamma$.
    \item $Z(\lambda)$ has finite boundary values on the contour $\Gamma$ and satisfies the jump condition  indicated on Figure \ref{contpinfzrhp}.
\item $Z(\lambda)$ has the asymptotic \eqref{atinfphi2}

at infinity and \eqref{atzerophi2}
at zero respectively.
\end{itemize}
\end{rhp}

Next step is the ``opening of lenses". We introduce
\begin{align}
   & \Sigma_1=\left(\begin{array}{cccccc}
0&0&0&\ee^{-\frac{2\pi\ii}{3}}a_{33}&0&0\\
0&0&0&0&\ee^{\frac{2\pi\ii}{3}}a_{33}&0\\ 0&0&0&0&0&-a_{33}\\
\ee^{-\frac{\ii\pi}{3}}a_{33}^{-1}&0&0&0&0&0\\
0&\ee^{\frac{\ii\pi}{3}}a_{33}^{-1}& 0 &0& 0&0\\
0&0&a_{33}^{-1}&0&0& 0\\
\end{array}\right)=\ee^{-\ii\pi \sigma_3\otimes\Lambda_7}(-\ii\sigma_1\otimes I_3)\ee^{\ii\pi \sigma_3\otimes\Lambda_7},\\&   \Lambda_7=\left(\begin{array}{ccc}\dfrac{1}{12}&0&0\\
0&\dfrac{5}{12}&0\\
0&0&\dfrac{1}{4}\\
\end{array}\right)-\frac{\ln(a_{33})}{2\pi \ii}I_3,\quad    \Sigma_2=(\sigma_1\otimes I_3 )\Sigma_1 (\sigma_1\otimes I_3).
\end{align}

\begin{figure}[H]
	\centering
	\includegraphics[scale=1]{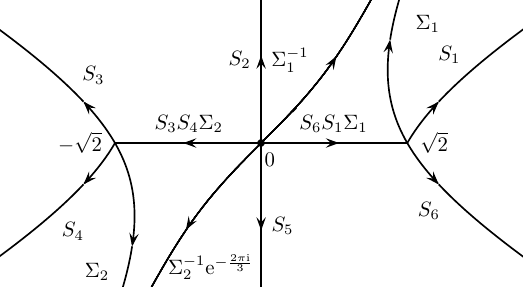}
	\caption{Contour for the Riemann-Hilbert problem \ref{pinftrhp2}}
\label{contpinftrhp2}
\end{figure}

We perform the deformation towards contours $\gamma_1,\gamma_{2,\pm},\gamma_3$ using matrix $M_4$.
\begin{figure}[H]
	\centering
	\includegraphics[scale=1]{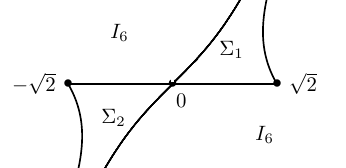}
	\caption{Constant matrix $M_4$}
\end{figure} 
We arrive to the Riemann-Hilbert problem for $T(\lambda)=Z(\lambda)M_4$.
\begin{rhp}
\label{pinftrhp2}
Consider the contour $\Gamma$ shown on Figure \ref{contpinftrhp2}. The $6\times 6$ matrix valued function $T(\lambda)$ satisfies the following conditions 
\begin{itemize}
    \item $T(\lambda)$ is holomorphic on $\mathbb{C}\setminus \Gamma$.
    \item $T(\lambda)$ has finite boundary values on the contour $\Gamma$ and satisfies the jump condition  indicated on Figure \ref{contpinftrhp2}.
\item $T(\lambda)$ has the asymptotic \eqref{atinfphi2}
at infinity and \eqref{atzerophi2}
at zero respectively.
\end{itemize}
\end{rhp}

Finally we perform the g-function transformation for this problem. $$S(\lambda)=\ee^{(-t)^\frac{3}{2}(I_2\otimes  \Lambda_8)}(-t)^{-I_2 \otimes  \Lambda_9}T(\lambda)\ee^{(-t)^\frac{3}{2}g(\lambda)({\sigma}_3\otimes I_3)}\ee^{-(-t)^\frac{3}{2}(I_2\otimes  \Lambda_8)}(-t)^{I_2 \otimes \Lambda_9 },$$
where
$$
 \Lambda_8=\dfrac{2\sqrt{2}}{3}\left(\begin{array}{ccc}1&0&0\\
0&0&0\\
0&0&-1\\
\end{array}\right),\quad \Lambda_9=\left(\begin{array}{ccc}
-\frac{11}{6}-\frac{3\ii}{2\pi}\ln(a_{33})&0&0\\
0&-\frac{1}{3}&0\\
0&0&\frac{13}{6}+\frac{3\ii}{2\pi}\ln(a_{33})
\end{array}\right).$$
Terms with $\Lambda_9$ will be used further in \eqref{matchz2} for matching of parametrices.

We have the following sign charts describing behavior of  $\mathrm{Re}\left(g(\lambda)\right)$.
\begin{figure}[H]
	\centering
    \begin{subfigure}{0.3\textwidth}
\includegraphics[scale=0.8]{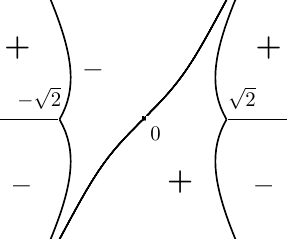}
	\caption{Sign chart for $\mathrm{Re}\left(g(\lambda)\right)$}
\end{subfigure}\hfill
\begin{subfigure}{0.3\textwidth}
	\centering
\includegraphics[scale=0.8]{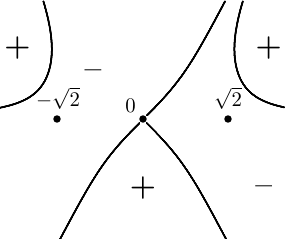}
	\caption{Sign chart for $\mathrm{Re}\left(g(\lambda)\right)-\frac{\sqrt{2}}{3}$}
\end{subfigure} \hfill
\begin{subfigure}{0.3\textwidth}
	\centering
\includegraphics[scale=0.8]{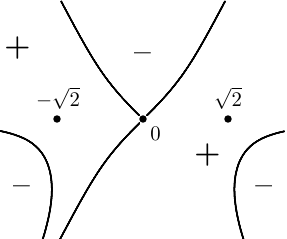}
	\caption{Sign chart for $\mathrm{Re}\left(g(\lambda)\right)+\frac{\sqrt{2}}{3}$}
    \end{subfigure}
    \caption{}
\end{figure} 

Similar to \eqref{yps} we have
$$
\ee^{-g({\sigma}_3\otimes I_3)}\ee^{I_2\otimes  \Lambda_8}\Upsilon \ee^{g({\sigma}_3\otimes I_3)}\ee^{-I_2\otimes  \Lambda_8}$$$$=\left(\begin{array}{cccccc}
1&\ee^{\frac{2\sqrt{2}}{3}}&\ee^{\frac{4\sqrt{2}}{3}}&\ee^{-2g}&\ee^{-2g+\frac{2\sqrt{2}}{3}}&\ee^{-2g+\frac{4\sqrt{2}}{3}}\\
\ee^{-\frac{2\sqrt{2}}{3}}&1&\ee^{\frac{2\sqrt{2}}{3}}&\ee^{-2g-\frac{2\sqrt{2}}{3}}&\ee^{-2g}&\ee^{-2g+\frac{2\sqrt{2}}{3}}\\
\ee^{-\frac{4\sqrt{2}}{3}}&\ee^{-\frac{2\sqrt{2}}{3}}&1&\ee^{-2g-\frac{4\sqrt{2}}{3}}&\ee^{-2g-\frac{2\sqrt{2}}{3}}&\ee^{-2g}\\
\ee^{2g}&\ee^{2g+\frac{2\sqrt{2}}{3}}&\ee^{2g+\frac{4\sqrt{2}}{3}}&1&\ee^{\frac{2\sqrt{2}}{3}}&\ee^{\frac{4\sqrt{2}}{3}}\\
\ee^{2g-\frac{2\sqrt{2}}{3}}&\ee^{2g}&\ee^{2g+\frac{2\sqrt{2}}{3}}&\ee^{-\frac{2\sqrt{2}}{3}}&1&\ee^{\frac{2\sqrt{2}}{3}}\\
\ee^{2g-\frac{4\sqrt{2}}{3}}&\ee^{2g-\frac{2\sqrt{2}}{3}}&\ee^{2g}&\ee^{-\frac{4\sqrt{2}}{3}}&\ee^{-\frac{2\sqrt{2}}{3}}&1
\end{array}\right).
$$

We can see that after conjugation the jump matrices are exponentially close to identity if you step away from points $\lambda=\pm \sqrt{2},0$ and contours $\gamma_1$, $\gamma_{2,\pm}$, $\gamma_3$. This outcome has motivated us to choose ansatz \eqref{ansatzminf}.  To cancel the remaining jumps we need to introduce parametrices solving the model Riemann-Hilbert problems.
\subsection{Global parametrix}
\begin{figure}[H]
	\centering
	\includegraphics[scale=1]{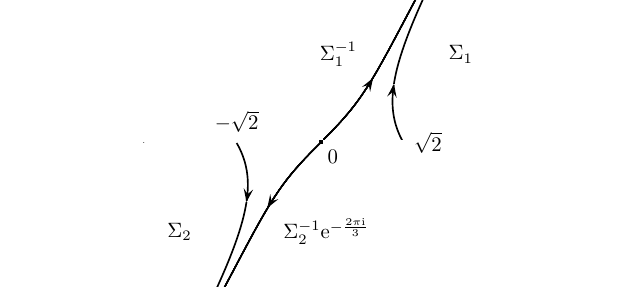}
	\caption{Contour for the Riemann-Hilbert problem \ref{pinfpinfrhp2}}
	\label{pinfpinfcont2}
\end{figure}

We start with global parametrix.
\begin{rhp}
\label{pinfpinfrhp2}
Consider the contour $\Gamma$ shown on Figure \ref{pinfpinfcont2}. The $6\times 6$ matrix valued function $P_{\infty}(\lambda)$ satisfies the following conditions 
\begin{itemize}
    \item $P_{\infty}(\lambda)$ is holomorphic on $\mathbb{C}\setminus \Gamma$.
    \item $P_{\infty}(\lambda)$ satisfies the jump condition  indicated on Figure \ref{pinfpinfcont2}.
\item $P_{\infty}(\lambda)$ has the asymptotic $$P_{\infty}(\lambda)=\left(I_6+\frac{m_{1,\infty}}{\lambda}+O\left(\frac{1}{\lambda^2}\right)\right)\lambda^{I_2\otimes\Lambda_1}, \quad \lambda\to \infty$$ at infinity and satisfies the estimates $$P_{\infty}(\lambda)=O(1)(\lambda-\sqrt{2})^{\frac{1}{4}(\sigma_3 \otimes I_3)}(K\otimes I_3)^{-1}\ee^{{\ii\pi}{}(\sigma_3\otimes \Lambda_{7})},\quad \lambda \to  \sqrt{2},$$
$$P_{\infty}(\lambda)=O(1)(\lambda+\sqrt{2})^{-\frac{1}{4}(\sigma_3 \otimes I_3)}(K\otimes I_3)^{-1}\ee^{-{\ii\pi}{}(\sigma_3\otimes \Lambda_{7})}(\sigma_3\otimes I_3),\quad \lambda \to-\sqrt{2},$$
$$P_{\infty}(\lambda)=(K\otimes I_3) \ee^{{\ii\pi}{}(\sigma_3\otimes \Lambda_{10})}2^{3{}(\sigma_3\otimes \Lambda_{10})}(I_6+O(\lambda))\lambda^{-2\sigma_3\otimes  \Lambda_{10}+I_2\otimes\Lambda_1},\quad \lambda \to 0,\quad \frac{\pi}{4}<\arg \lambda <\frac{5\pi}{4},$$
$$P_{\infty}(\lambda)=(K\otimes I_3)^{-1} \ee^{{\ii\pi}{}(\sigma_3\otimes \Lambda_{10})}2^{-3{}(\sigma_3\otimes \Lambda_{10})}(I_6+O(\lambda))\lambda^{2\sigma_3\otimes  \Lambda_{10}+I_2\otimes\Lambda_1},\quad \lambda \to 0,\quad -\frac{3\pi}{4}<\arg \lambda <\frac{\pi}{4}$$where
$$
\Lambda_{10}= \Lambda_7+\frac{I_3}{4},$$
\begin{align}
m_{1,\infty}=2\ii\sqrt{2}\sigma_3\otimes\Lambda_{10}+\frac{1}{\sqrt{2}}\sigma_2\otimes I_3.
\end{align}

\end{itemize}
\end{rhp}
The solution Riemann-Hilbert problem \ref{pinfpinfrhp2} is constructed explicitly in Appendix \ref{globali}.

\subsection{Local parametrix near \texorpdfstring{$\lambda=\sqrt{2}$}{lambda=sqrt(2)}}

To describe the parametrix near $\lambda=\sqrt{2}$ we introduce

$$
\frac{2}{3}\zeta^\frac{3}{2}=(-t)^\frac{3}{2}g(\lambda);\quad \zeta=(-t)\dfrac{1}{2^\frac{4}{3}}(2-\lambda^2)\simeq (-t){2^\frac{1}{6}}\ee^{\ii\pi}(\lambda-\sqrt{2}),\quad \lambda\to \sqrt{2}.
$$
We introduce the parts of the jump matrices which are non-vanishing at the neighborhood of $\lambda=\sqrt{2}$.
\begin{align}
  &  S_{1,\sqrt{2}}=\left(\begin{array}{cccccc}
1&0&0&\ee^{\frac{\ii\pi}{3}}a_{33}&0&0\\
0&1&0&0&\ee^{-\frac{\ii\pi}{3}}a_{33}&0\\ 0&0&1&0&0&a_{33}\\
0&0&0&1&0&0\\
0&0& 0 &0& 1&0\\
0&0&0&0&0& 1\\
\end{array}\right),\\&  S_{6,\sqrt{2}}=\left(\begin{array}{cccccc}
1&0&0&0&0&0\\
0&1&0&0&0&0\\ 0&0&1&0&0&0\\
\ee^{\frac{2\pi\ii}{3}}a_{33}^{-1}&0&0&1&0&0\\
0&\ee^{-\frac{2\pi\ii}{3}}a_{33}^{-1}& 0 &0& 1&0\\
0&0&-a_{33}^{-1}&0&0& 1\\
\end{array}\right).
\end{align}

\begin{figure}[H]
	\centering
	\includegraphics[scale=1]{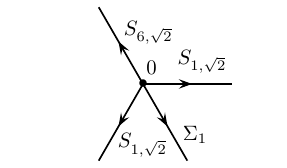}
	\caption{Contour for the Riemann-Hilbert problem \ref{plocsqu2rhp}}
	\label{plocsqu2cont}
\end{figure}

We will need solution to the following model Riemann-Hilbert problem
\begin{rhp}\label{plocsqu2rhp}
Consider the contour $\Gamma$ shown on Figure \ref{plocsqu2cont}. It consists of rays starting at zero in directions $\arg(z)=0,\frac{2\pi}{3},\frac{4\pi}{3},-\frac{\pi}{3}$. The $6\times 6$ matrix valued function $\Phi_{\sqrt{2}}(z)$ satisfies the following conditions 
\begin{itemize}
    \item $\Phi_{\sqrt{2}}(z)$ is holomorphic on $\mathbb{C}\setminus \Gamma$.
    \item $\Phi_{\sqrt{2}}(z)$ has finite boundary values on contour $\Gamma$ and satisfies the jump condition  indicated on Figure \ref{plocsqu2cont}.
\item $\Phi_{\sqrt{2}}(z)$ has the asymptotic $$\Phi_{\sqrt{2}}(z)=z^{-\frac{(\sigma_3\otimes I_3)}{4}}(K\otimes I_3)^{-1}\left(I_6+O(z^{-1})\right)\ee^{-\frac{2}{3}{z^\frac{3}{2}}(\sigma_3\otimes I_3)}\ee^{{\ii\pi}(\sigma_3\otimes \Lambda_7)}(\sigma_3\otimes I_3), \quad z\to \infty$$ at infinity and satisfies the estimate $\Phi_{\sqrt{2}}(z)=O(1)$ at $z=0$.

\end{itemize}
\end{rhp}

The solution to Riemann-Hilbert problem \ref{plocsqu2rhp} is constructed using Airy functions in Appendix \ref{solplocsqu2rhp}. Using it we construct the parametrix near $\lambda=\sqrt{2}$
$$
P_{\sqrt{2}}=P_\infty \ee^{-{\ii\pi}(\sigma_3\otimes \Lambda_7)}(\sigma_3\otimes I_3)(K\otimes I_3)\zeta^{\frac{({\sigma}_3\otimes I_3)}{4}}\Phi_{\sqrt{2}}(\zeta)\ee^{\frac{2}{3}\zeta^\frac{3}{2}({\sigma}_3\otimes I_3)}.
$$

The cut for $\zeta^\frac{1}{4}$ is taken along  $\gamma_1$.
The expression $P_\infty \ee^{-{\ii\pi}(\sigma_3\otimes \Lambda_7)}(\sigma_3\otimes I_3)(K\otimes I_3)\zeta^{\frac{({\sigma}_3\otimes I_3)}{4}}$ is holomorphic near $\lambda=\sqrt{2}$. The parametrix satisfies matching condition.
\begin{equation} \label{matchsq2}
P_{\sqrt{2}}=P_\infty \left(I+O\left(\dfrac{1}{(-t)^\frac{3}{2}}\right)\right),\quad t\to -\infty.\end{equation} 
\subsection{Local parametrix near \texorpdfstring{$\lambda=-\sqrt{2}$}{lambda=-sqrt(2)}}
To describe the parametrix near $\lambda=-\sqrt{2}$ we introduce

$$
\frac{2}{3}\zeta^\frac{3}{2}=-(-t)^\frac{3}{2}g(\lambda);\quad \zeta=(-t)\dfrac{1}{2^\frac{4}{3}}(2-\lambda^2)\simeq (-t){2^\frac{1}{6}}(\lambda+\sqrt{2}),\quad \lambda\to -\sqrt{2}.
$$
We introduce the parts of the jump matrices which are non-vanishing at the neighborhood of $\lambda=\sqrt{2}$.
\begin{align}
  &  S_{3,-\sqrt{2}}= (\sigma_1\otimes I_3 )S_{6,\sqrt{2}} (\sigma_1\otimes I_3),\quad  S_{4,-\sqrt{2}}=(\sigma_1\otimes I_3 )S_{1,\sqrt{2}} (\sigma_1\otimes I_3).
\end{align}

\begin{figure}[H]
	\centering
	\includegraphics[scale=1]{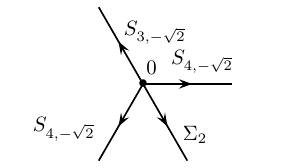}
	\caption{Contour for the Riemann-Hilbert problem \ref{plocmsqu2rhp}}
	\label{plocmsqu2cont}
\end{figure}

We will need solution to the following model Riemann-Hilbert problem
\begin{rhp}\label{plocmsqu2rhp}
Consider the contour $\Gamma$ shown on Figure \ref{plocmsqu2cont}. It consists of rays starting at zero in directions $\arg(z)=0,\frac{2\pi}{3},\frac{4\pi}{3},-\frac{\pi}{3}$. The $6\times 6$ matrix valued function $\Phi_{-\sqrt{2}}(z)$ satisfies the following conditions 
\begin{itemize}
    \item $\Phi_{-\sqrt{2}}(z)$ is holomorphic on $\mathbb{C}\setminus \Gamma$.
    \item $\Phi_{-\sqrt{2}}(z)$ has finite boundary values on contour $\Gamma$ and satisfies the jump condition  indicated on Figure \ref{plocmsqu2cont}.
\item $\Phi_{-\sqrt{2}}(z)$ has the asymptotic $$\Phi_{-\sqrt{2}}(z)=z^{-\frac{(\sigma_3\otimes I_3)}{4}}(K\otimes I_3)^{-1}\left(I_6+O(z^{-1})\right)\ee^{-\frac{2}{3}{z^\frac{3}{2}}(\sigma_3\otimes I_3)}\ee^{-{\ii\pi}(\sigma_3\otimes \Lambda_7)}(\ii\sigma_2\otimes I_3), \quad z\to \infty,$$ at infinity and satisfies the estimate $\Phi_{-\sqrt{2}}(z)=O(1)$ at $z=0$.

\end{itemize}
\end{rhp}

The solution to Riemann-Hilbert problem \ref{plocmsqu2rhp} is constructed using Airy functions in Appendix \ref{solplocmsqu2rhp}.

 We construct the parametrix near $\lambda=-\sqrt{2}$
$$
P_{-\sqrt{2}}=P_\infty (\sigma_1\otimes I_3)(\sigma_3\otimes I_3)\ee^{-\ii\pi(\sigma_3\otimes \Lambda_7)}(K\otimes I_3)\zeta^{\frac{({\sigma}_3\otimes I_3)}{4}}\Phi_{\sqrt{2}}(\zeta)\ee^{\frac{2}{3}\zeta^\frac{3}{2}({\sigma}_3\otimes I_3)}(\sigma_1\otimes I_3).
$$
Here $\Phi_{\sqrt2}$ is the solution of RHP \ref{plocsqu2rhp}: in the local variable $\zeta$ the jumps of $S$ near $\lambda=-\sqrt2$ are the $(\sigma_1\otimes I_3)$-conjugates of the jumps near $\lambda=\sqrt2$. The cut for $\zeta^\frac{1}{4}$ is taken along $\gamma_3$. The expression $P_\infty (\sigma_1\otimes I_3)(\sigma_3\otimes I_3)\ee^{-\ii\pi(\sigma_3\otimes \Lambda_7)}(K\otimes I_3)\zeta^{\frac{({\sigma}_3\otimes I_3)}{4}}$ is holomorphic near $\lambda=-\sqrt{2}$. The parametrix satisfies matching condition.
\begin{equation} \label{matchmsq2}
P_{-\sqrt{2}}=P_\infty \left(I+O\left(\dfrac{1}{(-t)^\frac{3}{2}}\right)\right),\quad t\to -\infty.\end{equation}

\subsection{Local parametrix near \texorpdfstring{$\lambda=0$}{lambda=0}}
To describe the parametrix near $\lambda=0$ we introduce
$$
\dfrac{\zeta^2}{4}=(-t)^{\frac{3}{2}}g(\lambda)\dfrac{\arg(\lambda-\sqrt{2})}{\pi}+(-t)^{\frac{3}{2}}\dfrac{\sqrt{2}}{3},\quad \zeta\simeq\lambda 2^\frac{1}{4} (-t)^\frac{3}{4},\quad \lambda\to 0.
$$
We introduce the parts of the jump matrices which are non-vanishing at the neighborhood of $\lambda=0$.
\begin{align}
  &  S_{2,0}=\left(\begin{array}{cccccc}
1&0&0&0&0&0\\
0&1&0&0&0&0\\ 0&0&1&0&0&0\\
0&1&0&1&0&0\\
0&0& 1 &0& 1&0\\
0&0&0&0&0& 1\\
\end{array}\right),\quad  S_{5,0}=\left(\begin{array}{cccccc}
1&0&0&0&1&0\\
0&1&0&0&0&-1\\ 0&0&1&0&0&0\\
0&0&0&1&0&0\\
0&0& 0 &0& 1&0\\
0&0&0&0&0& 1\\
\end{array}\right),\\&
S_{L,CH}=\left(\begin{array}{cccccc}
1&0&0&0&0&0\\
0&1&0&0&0&0\\ 0&0&1&0&0&0\\
0&0&0&1&0&0\\
\sqrt{3}\ee^{\frac{2\pi\ii}{3}}(a_{33}-\ee^{-\frac{5\pi\ii}{6}})&0& 0 &0& 1&0\\
0&\sqrt{3}\ee^{\frac{2\pi\ii}{3}}(a_{33}-\ii)&0&0&0& 1\\
\end{array}\right),\\&S_{U,CH}=\left(\begin{array}{cccccc}
1&0&0&0&0&0\\
0&1&0&\sqrt{3}\ee^{\frac{2\pi\ii}{3}}(a_{33}-\ee^{-\frac{5\pi\ii}{6}})&0&0\\ 0&0&1&0&\sqrt{3}\ee^{-\frac{\ii\pi}{3}}(a_{33}-\ii)&0\\
0&0&0&1&0&0\\
0&0& 0 &0& 1&0\\
0&0&0&0&0& 1\\
\end{array}\right).
\end{align}

\begin{figure}[H]
	\centering
	\includegraphics[scale=1]{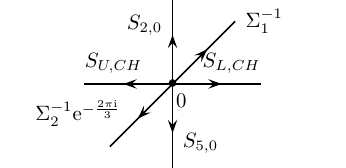}
	\caption{Contour for the Riemann-Hilbert problem \ref{plocz2rhp}}
	\label{plocz2cont}
\end{figure}

We will need solution to the following model Riemann-Hilbert problem
\begin{rhp}\label{plocz2rhp}Consider the contour $\Gamma$ shown on Figure \ref{plocz2cont}. It consists of rays starting in directions $\arg(z)=0,\frac{\pi}{4},\frac{\pi}{2},{\pi}{},\frac{5\pi}{4}, \frac{3\pi}{2}$. The $6\times 6$ matrix valued function $\Phi_{0}(z)$ satisfies the following conditions 
\begin{itemize}
    \item $\Phi_{0}(z)$ is holomorphic on $\mathbb{C}\setminus \Gamma$.
    \item $\Phi_{0}(z)$ has finite boundary values on contour $\Gamma$ and satisfies the jump condition  indicated on Figure \ref{plocz2cont}.
\item $\Phi_{0}(z)$ has the asymptotic $$\Phi_{0}(z)=\left(I_6+\frac{m_{1,0}}{z}+O\left(\frac{1}{z^2}\right)\right)\ee^{\frac{z^2}{4}(\sigma_3\otimes I_3)}z^{2(\sigma_3\otimes \Lambda_{10})+I_2\otimes \Lambda_1}M_5^{-1}, \quad z\to \infty$$ at infinity and satisfies the estimate $\Phi_{0}(z)=O(z^{-\frac{1}{6}})$ at $z=0$. The constant matrix $M_5$ is given by the Figure \ref{M7}.
\begin{figure}[H]  
	\centering
	\includegraphics[scale=1]{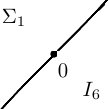}
	\caption{Constant matrix $M_5$}
	\label{M7}
\end{figure}

\end{itemize}
\end{rhp}

The solution to Riemann-Hilbert problem \ref{plocz2rhp} is constructed using confluent hypergeometric functions in Appendix \ref{solplocz2rhp}. In particular we have
$$
m_{1,0}=\left(\begin{array}{cccccc}
0&0&0&0&-\frac{2\ii\sqrt{3}}{9}&0\\
0&0&0&0&0&\frac{{2^{\frac{2}{3}}{} \ee^{-\frac{\ii \pi}{6}}\Gamma\left(\frac{5}{3}\right)}}{\pi}\\ 0&0&0&0&0&0\\
0&0& 0&0&0&0\\
-\ii\sqrt{3}&0&0 &0& 0&0\\
0&\frac{2^\frac{1}{3}\pi \ee^{-\frac{5\pi\ii}{6}}}{\Gamma\left(\frac{2}{3}\right)}&0&0&0& 0\\
\end{array}\right)
$$

We construct the parametrix near $\lambda=0$
$$
P_{0}=A_0(\lambda)\Phi_{0}(\zeta(\lambda))\ee^{\varepsilon\frac{\zeta^2}{4}({\sigma}_3\otimes I_3)}(-t)^{I_2\otimes \Lambda_{9}}
$$
where $\varepsilon=1$ to the left of $\gamma_{2,\pm}$ and $\varepsilon=-1$ to the right of $\gamma_{2,\pm}$.

Here $A_0(\lambda)$ is the holomorphic function
$$
A_0(\lambda)=P_\infty M_5 \zeta^{-2\sigma_3\otimes \Lambda_{10}}\zeta^{-I_2\otimes \Lambda_1}(-t)^{-I_2\otimes \Lambda_{9}}=A_{0,0}+O(\lambda).
$$
The matrix $A_{0,0}$ is given by  
$$
A_{0,0}= (K\otimes I_3)^{-1}2^{-\frac{1}{4}(14\sigma_3\otimes \Lambda_{10} +I_2\otimes \Lambda_1)}\ee^{\ii \pi \sigma_3\otimes \Lambda_{10}}(-t)^{-\frac{1}{4}(6\sigma_3\otimes \Lambda_{10}+3I_2\otimes \Lambda_1+4I_2\otimes \Lambda_9)}.
$$
We have the matching condition
\begin{equation} \label{matchz2}
P_{0}=P_\infty \left(I+O\left(\dfrac{1}{(-t)^\frac{3}{4}}\right)\right).\end{equation}
\subsection{Small norm theorem and leading terms of asymptotic}
\begin{figure}[H]
	\centering
\includegraphics[scale=1]{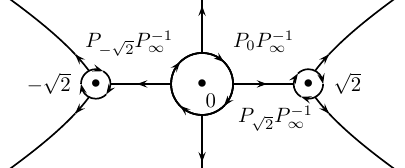}
	\caption{Contour for the Riemann-Hilbert problem \ref{pinfrrhp2}}
	\label{pinfrcont2}
\end{figure}

Now, having all parametrices constructed we define

$$
S_{as}=\left\{\begin{array}{cc}
P_{0},&\quad |\lambda|<\delta,\\
P_{\sqrt{2}},&\quad |\lambda-\sqrt{2}|<\delta,\\
P_{-\sqrt{2}},&\quad |\lambda+\sqrt{2}|<\delta,\\
P_{\infty},&\quad \mbox{otherwise}.
\end{array} \right.
$$
for small number $\delta$. Then we arrive to the Riemann Hilbert problem for 
$
R=SS_{as}^{-1}
$
with small jump.
\begin{rhp}\label{pinfrrhp2}
Consider the contour $\Gamma$ shown on Figure \ref{pinfrcont2}. The $6\times 6$ matrix valued function $R(\lambda)$ satisfies the following conditions 
\begin{itemize}
    \item $R(\lambda)$ is holomorphic on $\mathbb{C}\setminus \Gamma$.
    \item $R(\lambda)$  has finite boundary values on the contour $\Gamma$ and satisfies the jump condition  indicated on Figure \ref{pinfrcont2}. On the non-labeled parts of contour the jump is obtained from the jump for $S(\lambda)$ conjugating with $S_{as}(\lambda)$. 
\item $R(\lambda)$ has the asymptotic at infinity $$R(\lambda)=I_6+ \frac{l_1}{\lambda}+O(\lambda^{-2}), \quad \lambda\to \infty.$$

\end{itemize}
\end{rhp}

We have again $P_\infty(\lambda)$ growing at infinity and we can do the estimates similar to \eqref{estimate}. After that, using \eqref{matchsq2}, \eqref{matchmsq2},\eqref{matchz2} we get 
\begin{equation}\label{jest2} 
|J(\mu)-I_6|=O((-t)^{-\frac{3}{4}}).
\end{equation}

Now following the standard procedure we write the singular integral equation for $R_-(\lambda)$.

$$
R_-(\lambda)=I+\intop_{\Gamma}\frac{R_-(\mu)(J(\mu)-I)}{\mu-\lambda+i0}\frac{d\mu}{2\pi \ii},
$$
where $J(\lambda)$ is the jump matrix. Using \eqref{jest2} we get
\begin{equation}\label{rest2}
R(\lambda)=I+O((-t)^{-\frac{3}{4}}).
\end{equation}
We also have the formula for $R(\lambda)$ away from contour $\Gamma$.
$$
R(\lambda)=I+\intop_{\Gamma}\frac{R_-(\mu)(J(\mu)-I)}{\mu-\lambda}\frac{d\mu}{2\pi \ii}.
$$
Expanding it at infinity we get $$
l_1=-\intop_{\Gamma}{R_-(\mu)(J(\mu)-I)}\frac{d\mu}{2\pi \ii}.
$$
Using the estimates \eqref{rest2}, \eqref{jest2} we arrive at
$$
l_1=-\intop_{\Gamma}{(J(\mu)-I)}\frac{d\mu}{2\pi \ii}+O((-t)^{-\frac{3}{2}}).
$$

The main part of this integral comes from circle around $\lambda=0 $, oriented clockwise, so that $J=P_{0}P_\infty^{-1}$ on it. As in Section~\ref{RH1}, the leading contribution to
$l_1$ is the positive residue of $P_0P_\infty^{-1}-I$.
$$
l_1=-\intop_{|\mu|=\delta}{(J(\mu)-I)}\frac{d\mu}{2\pi \ii}+O((-t)^{-\frac{3}{2}}).
$$
Using the expansion for $\Phi_0$ at infinity we get
$$
l_1=\dfrac{A_{0,0}m_{1,0}A_{0,0}^{-1}}{{2^\frac{1}{4}}(-t)^{\frac{3}{4}}}+O((-t)^{-\frac{3}{2}}).
$$
Tracing back the transformations from Section \ref{prelimp} we have 
$$
(m_1)_{\mbox{offdiag}}=\sqrt{-t}(I_2\otimes P_1)\ee^{-(I_2\otimes \Lambda_8)(-t)^{\frac{3}{2}}}(-t)^{I_2\otimes \Lambda_9+\frac{1}{2}I_2\otimes \Lambda_1}\left(\frac{\sigma_2 \otimes I_3}{\sqrt{2}}+l_1\right)$$
$$\times\ee^{(I_2\otimes \Lambda_8)(-t)^{\frac{3}{2}}}(-t)^{-I_2\otimes \Lambda_9-\frac{1}{2}I_2\otimes \Lambda_1}(I_2\otimes P_1)^{-1}.
$$
The matrices conjugating the $(1,2)$ block are block diagonal with equal blocks, and the $(1,2)$ block of
$\frac{\sigma_2\otimes I_3}{\sqrt{2}}$ is $-\frac{\ii}{\sqrt2}I_3$. Hence, using the formula $q(t)=-\ii\hat{m}_{1,12}$,
we get that $q(t)$ is similar to
$$
-\sqrt{\frac{-t}{2}}\,I_3+\frac{N}{(-t)^{\frac{1}{4}}}+O\left(\frac{1}{t}\right),\quad
N=-\frac{\ii}{2^{\frac{1}{4}}}\left(A_{0,0}m_{1,0}A_{0,0}^{-1}\right)_{12},
$$
where $(\cdot)_{12}$ denotes the $(1,2)$ block. Substituting the explicit
expressions for $A_{0,0}$ and $m_{1,0}$, we compute
$$
\det(N-\nu I_3)=-\nu^3-\frac{\nu}{2\sqrt{2}}.
$$
The roots $0,\pm\ii2^{-3/4}$ are simple. The residue sign fixes the sign of $N$;
replacing $N$ by $-N$ only interchanges the two nonzero eigenvalues.
The particle labels below are chosen so that $x_2$ has the positive
imaginary leading correction. After adding
$\sqrt{-t/2}I_3$ and multiplying by $(-t)^{1/4}$, the matrix is
$N+\mathcal E(t)$, where $\mathcal E(t)=O((-t)^{-3/4})$.
Consider
\[
P(\nu,\mathcal E)=\det(N+\mathcal E-\nu I_3).
\]
This is a polynomial in $\nu$ and the entries of $\mathcal E$.
At $\mathcal E=0$, its roots are $\nu_1^0=0$ and
$\nu_{2,3}^0=\pm\ii2^{-3/4}$, and
$\partial_\nu P(\nu,0)=-3\nu^2-1/(2\sqrt2)$ is nonzero at each root.
The implicit function theorem therefore gives roots
$\nu_j(\mathcal E)=\nu_j^0+O(\|\mathcal E\|)$ near these three distinct values.
Substituting $\mathcal E=\mathcal E(t)$ yields an error $O((-t)^{-3/4})$
in the scaled eigenvalues.
Undoing the shift and rescaling, we obtain, up to permutation,
$$
x_1=-\sqrt{\frac{-t}{2}}+O(t^{-1}),\quad x_2=-\sqrt{\frac{-t}{2}}+\frac{\ii}{{2^\frac{3}{4}(-t)^{\frac{1}{4}}}}+O(t^{-1}),\quad x_3=-\sqrt{\frac{-t}{2}}-\frac{\ii}{{2^\frac{3}{4}(-t)^{\frac{1}{4}}}}+O(t^{-1}).
$$
\subsection{Recurrence relation and subleading terms of asymptotics}
The asymptotic of solutions have form
$$
x_1= -\sqrt{\frac{-t}{2}}+\sum_{j=0}^\infty c_j (-t)^{-1-\frac{3}{4}j},\, x_2=-\sqrt{\frac{-t}{2}} +\frac{\ii}{{2^{\frac{3}{4}}(-t)^{\frac{1}{4}}}}+\sum_{j=0}^\infty d_j (-t)^{-1-\frac{3}{4}j},$$
$$x_3=-\sqrt{\frac{-t}{2}} -\frac{\ii}{{2^{\frac{3}{4}}(-t)^{\frac{1}{4}}}}+\sum_{j=0}^\infty e_j (-t)^{-1-\frac{3}{4}j}.
$$
We rewrite the Calogero-Painlev\'e system in the following form
\begin{equation} \label{rewrcpa}
\begin{gathered} 
3(x_1-x_2)^3(x_1-x_3)^3(-6x_1''+12x_1^3+6x_1 t-1)-4((x_1-x_2)^3+(x_1-x_3)^3)=0,
\\
3(x_2-x_3)^3(x_2-x_1)^3(-6x_2''+12x_2^3+6x_2 t-1)-4((x_2-x_1)^3+(x_2-x_3)^3)=0,
\\
3(x_3-x_2)^3(x_3-x_1)^3(-6x_3''+12x_3^3+6x_3 t-1)-4((x_3-x_2)^3+(x_3-x_1)^3)=0.
\end{gathered} 
\end{equation}
We have 

$$
(x_1-x_2)^3=\frac{\ii}{2^{\frac{9}{4}}(-t)^\frac{3}{4}}-\frac{3}{2\sqrt{2}}\sum_{k=2}^\infty (c_{k-2}-d_{k-2})(-t)^{-\frac{3}{4}k}-\frac{3\ii}{2^{\frac{3}{4}}}\sum_{k=3}^\infty \sum_{j=0}^{k-3}(c_{j}-d_{j})(c_{k-j-3}-d_{k-j-3})(-t)^{-\frac{3}{4}k}$$
$$+\sum_{k=4}^\infty \sum_{j_1=0}^{k-4}\sum_{j_2=0}^{k-4-j_1}(c_{j_1}-d_{j_1})(c_{j_2}-d_{j_2})(c_{k-4-j_1-j_2}-d_{k-4-j_1-j_2})(-t)^{-\frac{3}{4}k},
$$
$$
(x_1-x_3)^3=-\frac{\ii}{2^{\frac{9}{4}}(-t)^\frac{3}{4}}-\frac{3}{2\sqrt{2}}\sum_{k=2}^\infty (c_{k-2}-e_{k-2})(-t)^{-\frac{3}{4}k}+\frac{3\ii}{{2^\frac{3}{4}}}\sum_{k=3}^\infty \sum_{j=0}^{k-3}(c_{j}-e_{j})(c_{k-j-3}-e_{k-j-3})(-t)^{-\frac{3}{4}k}$$
$$+\sum_{k=4}^\infty \sum_{j_1=0}^{k-4}\sum_{j_2=0}^{k-4-j_1}(c_{j_1}-e_{j_1})(c_{j_2}-e_{j_2})(c_{k-4-j_1-j_2}-e_{k-4-j_1-j_2})(-t)^{-\frac{3}{4}k},
$$
$$
(x_2-x_3)^3=-\frac{\ii2^\frac{3}{4}}{(-t)^\frac{3}{4}}-3\sqrt{2}\sum_{k=2}^\infty (d_{k-2}-e_{k-2})(-t)^{-\frac{3}{4}k}+{3\ii2^\frac{1}{4}}{}\sum_{k=3}^\infty \sum_{j=0}^{k-3}(d_{j}-e_{j})(d_{k-j-3}-e_{k-j-3})(-t)^{-\frac{3}{4}k}$$
$$+\sum_{k=4}^\infty \sum_{j_1=0}^{k-4}\sum_{j_2=0}^{k-4-j_1}(d_{j_1}-e_{j_1})(d_{j_2}-e_{j_2})(d_{k-4-j_1-j_2}-e_{k-4-j_1-j_2})(-t)^{-\frac{3}{4}k}
$$
$$
x_1''=\frac{1}{4\sqrt{2}{(-t)^\frac{3}{2}}}+\sum_{k=4}^{\infty}\left(2-\frac{3}{4}k\right)\left(1-\frac{3}{4}k\right)c_{k-4}(-t)^{-\frac{3}{4}k}
$$
$$
x_2''=\frac{1}{4\sqrt{2}{(-t)^\frac{3}{2}}}+ \frac{5\ii}{2^{\frac{3}{4}}16(-t)^{\frac{9}{4}}}+\sum_{k=4}^{\infty}\left(2-\frac{3}{4}k\right)\left(1-\frac{3}{4}k\right)d_{k-4}(-t)^{-\frac{3}{4}k}
$$
$$
x_3''= \frac{1}{4\sqrt{2}{(-t)^\frac{3}{2}}}-\frac{5\ii}{2^{\frac{3}{4}}16(-t)^{\frac{9}{4}}}+\sum_{k=4}^{\infty}\left(2-\frac{3}{4}k\right)\left(1-\frac{3}{4}k\right)e_{k-4}(-t)^{-\frac{3}{4}k}
$$
$$
x_1^3=-\frac{(-t)^\frac{3}{2}}{2\sqrt{2}}+\frac{3}{2}\sum_{k=0}^\infty c_{k}(-t)^{-\frac{3}{4}k}-\frac{3}{\sqrt{2}}\sum_{k=2}^\infty \sum_{j=0}^{k-2}c_{j}c_{k-j-2}(-t)^{-\frac{3}{4}k}$$
$$+\sum_{k=4}^\infty \sum_{j_1=0}^{k-4}\sum_{j_2=0}^{k-4-j_1}c_{j_1}c_{j_2}c_{k-4-j_1-j_2}(-t)^{-\frac{3}{4}k}
$$
\begin{multline*}
x_2^3=-\frac{(-t)^\frac{3}{2}}{2\sqrt{2}}+\frac{3\ii(-t)^{\frac{3}{4}}}{2^{\frac{7}{4}}}+\frac{3}{2}\sum_{k=0}^{\infty}d_{k}(-t)^{-\frac{3}{4}k}+\frac{3}{4}\\
-\frac{3\ii}{2^{\frac{1}{4}}}\sum_{k=1}^{\infty}d_{k-1}(-t)^{-\frac{3}{4}k}-\frac{3}{\sqrt{2}}\sum_{k=2}^\infty \sum_{j=0}^{k-2}d_{j}d_{k-j-2}(-t)^{-\frac{3}{4}k}-\frac{i}{2^{\frac{9}{4}}(-t)^\frac{3}{4}}\\
-\frac{3}{2\sqrt{2}}\sum_{k=2}^\infty d_{k-2}(-t)^{-\frac{3}{4}k}+\frac{3\ii}{{2^\frac{3}{4}}}\sum_{k=3}^\infty \sum_{j=0}^{k-3}d_{j}d_{k-j-3}(-t)^{-\frac{3}{4}k}+\sum_{k=4}^\infty \sum_{j_1=0}^{k-4}\sum_{j_2=0}^{k-4-j_1}d_{j_1}d_{j_2}d_{k-4-j_1-j_2}(-t)^{-\frac{3}{4}k}
\end{multline*}
\begin{multline*}
x_3^3=-\frac{(-t)^\frac{3}{2}}{2\sqrt{2}}-\frac{3\ii(-t)^{\frac{3}{4}}}{2^{\frac{7}{4}}}+\frac{3}{2}\sum_{k=0}^{\infty}e_{k}(-t)^{-\frac{3}{4}k}+\frac{3}{4}\\
+\frac{3\ii}{2^{\frac{1}{4}}}\sum_{k=1}^{\infty}e_{k-1}(-t)^{-\frac{3}{4}k}-\frac{3}{\sqrt{2}}\sum_{k=2}^\infty \sum_{j=0}^{k-2}e_{j}e_{k-j-2}(-t)^{-\frac{3}{4}k}+\frac{i}{2^{\frac{9}{4}}(-t)^\frac{3}{4}}\\
-\frac{3}{2\sqrt{2}}\sum_{k=2}^\infty e_{k-2}(-t)^{-\frac{3}{4}k}-\frac{3\ii}{{2^\frac{3}{4}}}\sum_{k=3}^\infty \sum_{j=0}^{k-3}e_{j}e_{k-j-3}(-t)^{-\frac{3}{4}k}+\sum_{k=4}^\infty \sum_{j_1=0}^{k-4}\sum_{j_2=0}^{k-4-j_1}e_{j_1}e_{j_2}e_{k-4-j_1-j_2}(-t)^{-\frac{3}{4}k}
\end{multline*}

We can see that in \eqref{rewrcpa} in the coefficient near $(-t)^{-\frac{3}{4}k}$ the terms with highest indices $c_{k-2}$, $d_{k-2}$, $e_{k-2}$ come from the terms
\begin{equation} 
\begin{gathered} 
18(x_1-x_2)^3(x_1-x_3)^3(2x_1^3+x_1 t)-4((x_1-x_2)^3+(x_1-x_3)^3),
\\
18(x_2-x_3)^3(x_2-x_1)^3(2x_2^3+x_2 t)-4((x_2-x_1)^3+(x_2-x_3)^3),
\\
18(x_3-x_2)^3(x_3-x_1)^3(2x_3^3+x_3 t)-4((x_3-x_2)^3+(x_3-x_1)^3).
\end{gathered} 
\end{equation}
It has form

$$
\left(\begin{array}{ccc}
     \frac{57\sqrt{2}}{8}&-3\sqrt{2}&-3\sqrt{2}  \\
     24\sqrt{2}&-\frac{69\sqrt{2}}{2}&\frac{3\sqrt{2}}{{2}}\\
       24\sqrt{2}&\frac{3\sqrt{2}}{{2}}&-\frac{69\sqrt{2}}{2}
\end{array}\right)\left(\begin{array}{c}c_{k-2}\\d_{k-2}\\e_{k-2}\end{array}\right)+\mbox{lower terms}.
$$
The determinant of the matrix above is $6561\sqrt{2}$, so we derived the recurrence relation for the coefficients $c_k,e_k,d_k$. We don't write the explicit formula for it. It determines all the coefficients uniquely.
\begin{remark} We can notice that other power asymptotics are possible for the solution of \eqref{rewrcpa}, for example 
$$
x_1= \sqrt{\frac{-t}{2}}+\sum_{j=0}^\infty c_j (-t)^{-1-\frac{3}{4}j},\quad x_2=\sqrt{\frac{-t}{2}} +\frac{\ii}{{2^{\frac{3}{4}}(-t)^{\frac{1}{4}}}}+\sum_{j=0}^\infty d_j (-t)^{-1-\frac{3}{4}j},$$
$$x_3=\sqrt{\frac{-t}{2}} -\frac{\ii}{{2^{\frac{3}{4}}(-t)^{\frac{1}{4}}}}+\sum_{j=0}^\infty e_j (-t)^{-1-\frac{3}{4}j},
$$
or
$$
x_1= -\sqrt{\frac{-t}{2}}+\sum_{j=0}^\infty c_j (-t)^{-1-\frac{3}{4}j},\quad x_2=-\sqrt{\frac{-t}{2}} +\frac{1}{{2^{\frac{3}{4}}(-t)^{\frac{1}{4}}}}+\sum_{j=0}^\infty d_j (-t)^{-1-\frac{3}{4}j},$$
$$x_3=-\sqrt{\frac{-t}{2}} -\frac{1}{{2^{\frac{3}{4}}(-t)^{\frac{1}{4}}}}+\sum_{j=0}^\infty e_j (-t)^{-1-\frac{3}{4}j},
$$
but they don't produce desired estimate \eqref{expconstweak}. The first of them is realized by the choice $a_{33}=\ee^{-\frac{5\pi\ii}{6}}$ in \eqref{S1S2}, see Section \ref{RH3}.
\end{remark}

% ===========================================================================
% New Section 6 for arXiv:2010.06733: asymptotics as t -> -infinity for the
% second root a_33 = e^{-5 pi i/6}.  \input before \appendix.
% The solutions of RHP \ref{pinfpinfrhp3} and RHP \ref{plocz3rhp} are given in
% the appendices (subsections \ref{globalii} and \ref{solplocz3rhp}).
% ===========================================================================
\section{Asymptotics as \texorpdfstring{$t\to-\infty$}{t->-infty} for \texorpdfstring{$a_{33}=\ee^{-\frac{5\pi\ii}{6}}$}{a33=exp(-5 pi i/6)}}\label{RH3}
In this section we consider the parametrization \eqref{ansatzminf},
\eqref{eq:monodromy_minus_infty} with $b_{12}=b_{23}=1$ and
$a_{33}=\ee^{-\frac{5\pi\ii}{6}}$. The remaining monodromy parameters are fixed.
This includes the data \eqref{S1S2}. We follow the transformations and the
order of the parametrix construction in Section \ref{RH2}.

The local model at $\lambda=0$ requires a different normalization for this
root. The uncorrected $\mathbf W^{(2)}$ construction has the required jumps,
but local exponents $\pm7/6$ instead of $\pm1/6$ at the origin; see Appendix
\ref{solplocz3rhp}. Correcting these exponents also changes the normalization
at infinity. We therefore modify the global parametrix and the diagonal
matrix used in the $g$-function transformation. These modifications lead
to the asymptotics \eqref{x3as}.

For the present value of $a_{33}$ we have
$$
\Lambda_7=\left(\begin{array}{ccc}\frac{1}{2}&0&0\\0&\frac{5}{6}&0\\0&0&\frac{2}{3}\end{array}\right),\quad
\Lambda_{10}=\left(\begin{array}{ccc}\frac{3}{4}&0&0\\0&\frac{13}{12}&0\\0&0&\frac{11}{12}\end{array}\right).
$$
We also introduce
\begin{equation}\label{L12L13}
\Lambda_{12}=\left(\begin{array}{ccc}-\frac{19}{12}&0&0\\0&-\frac{1}{3}&0\\0&0&\frac{23}{12}\end{array}\right),\quad
\Lambda_{13}=\Lambda_{10}-\frac{I_3}{2}=\left(\begin{array}{ccc}\frac{1}{4}&0&0\\0&\frac{7}{12}&0\\0&0&\frac{5}{12}\end{array}\right).
\end{equation}
We use $\ln a_{33}=-5\pi\ii/6$. Relative to the value of $\Lambda_9$
in Subsection \ref{prelimn},
$$
\Lambda_{12}=\Lambda_9+\begin{pmatrix}\frac32&0&0\\0&0&0\\0&0&-\frac32\end{pmatrix}.
$$
This is a change of normalization, not a change of branch of the logarithm:
branches of $\ln a_{33}$ differ by multiples of $2\pi\ii$.

\subsection{Preliminary transformations}
The transformations $\Psi\mapsto\Phi\mapsto Z\mapsto T$, the $g$-function and the contours
$\gamma_1,\gamma_{2,\pm},\gamma_3,\gamma_4$ are the same as in Subsection \ref{prelimn}. In the $g$-function
transformation we replace $\Lambda_9$ by $\Lambda_{12}$,
$$
S(\lambda)=\ee^{(-t)^\frac{3}{2}(I_2\otimes  \Lambda_8)}(-t)^{-I_2 \otimes  \Lambda_{12}}T(\lambda)\ee^{(-t)^\frac{3}{2}g(\lambda)({\sigma}_3\otimes I_3)}\ee^{-(-t)^\frac{3}{2}(I_2\otimes  \Lambda_8)}(-t)^{I_2 \otimes \Lambda_{12} }.
$$
The factors $(-t)^{\pm I_2\otimes\Lambda_{12}}$ are constant diagonal matrices, so, as in Section \ref{RH2}, the jump
matrices are exponentially close to the identity if we step away from the points $\lambda=\pm \sqrt{2},0$ and the
contours $\gamma_1$, $\gamma_{2,\pm}$, $\gamma_3$. The terms with $\Lambda_{12}$ will be used in \eqref{matchz3}
for matching of parametrices.

\subsection{Global parametrix}
\begin{rhp}
\label{pinfpinfrhp3}
Consider the contour $\Gamma$ shown on Figure \ref{pinfpinfcont2}. The $6\times 6$ matrix valued function $P_{\infty}(\lambda)$ satisfies the following conditions
\begin{itemize}
    \item $P_{\infty}(\lambda)$ is holomorphic on $\mathbb{C}\setminus \Gamma$.
    \item $P_{\infty}(\lambda)$ satisfies the jump condition  indicated on Figure \ref{pinfpinfcont2}.
\item $P_{\infty}(\lambda)$ has the asymptotic $$P_{\infty}(\lambda)=\left(I_6+\frac{m_{1,\infty}}{\lambda}+O\left(\frac{1}{\lambda^2}\right)\right)\lambda^{I_2\otimes\Lambda_1}, \quad \lambda\to \infty$$ at infinity and satisfies the estimates $$P_{\infty}(\lambda)=O(1)(\lambda-\sqrt{2})^{\frac{1}{4}(\sigma_3 \otimes I_3)}(K\otimes I_3)^{-1}\ee^{{\ii\pi}{}(\sigma_3\otimes \Lambda_{7})},\quad \lambda \to  \sqrt{2},$$
$$P_{\infty}(\lambda)=O(1)(\lambda+\sqrt{2})^{-\frac{1}{4}(\sigma_3 \otimes I_3)}(K\otimes I_3)^{-1}\ee^{-{\ii\pi}{}(\sigma_3\otimes \Lambda_{7})}(\sigma_3\otimes I_3),\quad \lambda \to-\sqrt{2},$$
$$P_{\infty}(\lambda)=O(1)\lambda^{-2\sigma_3\otimes  \Lambda_{13}+I_2\otimes\Lambda_1},\quad \lambda \to 0,\quad \frac{\pi}{4}<\arg \lambda <\frac{5\pi}{4},$$
$$P_{\infty}(\lambda)=O(1)\lambda^{2\sigma_3\otimes  \Lambda_{13}+I_2\otimes\Lambda_1},\quad \lambda \to 0,\quad -\frac{3\pi}{4}<\arg \lambda <\frac{\pi}{4},$$
where $O(1)$ denotes matrix functions holomorphic and invertible at the corresponding point, and
\begin{align}\label{m1inf3}
m_{1,\infty}=2\ii\sqrt{2}\sigma_3\otimes\Lambda_{10}+\frac{1}{\sqrt{2}}\sigma_2\otimes I_3-\ii\sqrt{2}\left(\begin{array}{cc}I_3&-I_3\\I_3&-I_3\end{array}\right).
\end{align}
\end{itemize}
\end{rhp}
Riemann-Hilbert problem \ref{pinfpinfrhp3} differs from Riemann-Hilbert problem \ref{pinfpinfrhp2} only by the
behavior at $\lambda=0$, where $\Lambda_{10}$ is replaced by $\Lambda_{13}$. Notice that $\Lambda_{13}-\Lambda_{10}$
is not an integer matrix, so this change cannot be obtained by another choice of the branch of $\ln a_{33}$.
The solution of Riemann-Hilbert problem \ref{pinfpinfrhp3} is constructed explicitly in Appendix \ref{globalii}.

\subsection{Local parametrix near \texorpdfstring{$\lambda=\sqrt{2}$}{lambda=sqrt(2)}}
The rational factor relating the present global parametrix to that of
Appendix \ref{globali} is holomorphic and invertible near $\lambda=\sqrt2$.
We use the local coordinate and the solution $\Phi_{\sqrt2}$ of
Riemann-Hilbert problem \ref{plocsqu2rhp} from Section \ref{RH2}, and set
$$
P_{\sqrt2}=P_\infty\ee^{-\ii\pi(\sigma_3\otimes\Lambda_7)}
(\sigma_3\otimes I_3)(K\otimes I_3)\zeta^{(\sigma_3\otimes I_3)/4}
\Phi_{\sqrt2}(\zeta)\ee^{\frac23\zeta^{3/2}(\sigma_3\otimes I_3)}.
$$
The prefactor is holomorphic for the same reason as in Section \ref{RH2}:
it is the previous prefactor multiplied on the left by a holomorphic,
invertible matrix. The jumps are unchanged, and the Airy expansion gives
\eqref{matchsq2}, uniformly on the boundary of a fixed disc.

\subsection{Local parametrix near \texorpdfstring{$\lambda=-\sqrt{2}$}{lambda=-sqrt(2)}}
At $\lambda=-\sqrt2$ we use the reflected Airy construction of Section
\ref{RH2}, with the same convention for the branch of $\zeta^{1/4}$:
\begin{align*}
P_{-\sqrt2}={}&P_\infty(\sigma_1\otimes I_3)(\sigma_3\otimes I_3)
\ee^{-\ii\pi(\sigma_3\otimes\Lambda_7)}(K\otimes I_3)
\zeta^{(\sigma_3\otimes I_3)/4}\\
&\qquad\times\Phi_{\sqrt2}(\zeta)
\ee^{\frac23\zeta^{3/2}(\sigma_3\otimes I_3)}(\sigma_1\otimes I_3).
\end{align*}
The jumps are the same as in Section \ref{RH2}. The prefactor is the
corresponding prefactor from that section multiplied on the left by
$I_6+C_0/\lambda$, which is holomorphic and invertible near $-\sqrt2$.
The Airy expansion gives the matching condition \eqref{matchmsq2}.

\subsection{Local parametrix near \texorpdfstring{$\lambda=0$}{lambda=0}}
We use the same local variable $\zeta(\lambda)$ and the same jump matrices $S_{2,0}$, $S_{5,0}$, $S_{L,CH}$, $S_{U,CH}$ as
in Section \ref{RH2}. The model problem differs from Riemann-Hilbert problem \ref{plocz2rhp} only by the exponent at
infinity.
\begin{rhp}\label{plocz3rhp}Consider the contour $\Gamma$ shown on Figure \ref{plocz2cont}. The $6\times 6$ matrix valued function $\Phi_{0}(z)$ satisfies the following conditions
\begin{itemize}
    \item $\Phi_{0}(z)$ is holomorphic on $\mathbb{C}\setminus \Gamma$.
    \item $\Phi_{0}(z)$ has finite boundary values on contour $\Gamma$ and satisfies the jump condition  indicated on Figure \ref{plocz2cont}.
\item $\Phi_{0}(z)$ has the asymptotic $$\Phi_{0}(z)=\left(I_6+\frac{m_{1,0}}{z}+O\left(\frac{1}{z^2}\right)\right)\ee^{\frac{z^2}{4}(\sigma_3\otimes I_3)}z^{2(\sigma_3\otimes \Lambda_{13})+I_2\otimes \Lambda_1}M_5^{-1}, \quad z\to \infty$$ at infinity and satisfies the estimate $\Phi_{0}(z)=O(z^{-\frac{1}{6}})$ at $z=0$. The constant matrix $M_5$ is given by the Figure \ref{M7}.
\end{itemize}
\end{rhp}
The solution to Riemann-Hilbert problem \ref{plocz3rhp} is constructed using confluent hypergeometric functions in
Appendix \ref{solplocz3rhp}. In particular we have
\begin{equation}\label{m10third}
m_{1,0}=\left(\begin{array}{cccccc}
0&0&0&0&\frac{2^{\frac{1}{3}}\Gamma\left(\frac{4}{3}\right)\ee^{-\frac{5\pi\ii}{6}}}{\pi}&0\\
0&0&0&0&0&-\frac{8\ii\sqrt{3}}{27}\\ 0&0&0&0&0&0\\
0&0& 0&0&0&0\\
\frac{2^{\frac{2}{3}}\pi\ee^{-\frac{\ii\pi}{6}}}{\Gamma\left(\frac{1}{3}\right)}&0&0 &0& 0&0\\
0&-\frac{3\sqrt{3}\ii}{2}&0&0&0& 0\\
\end{array}\right).
\end{equation}

We construct the parametrix near $\lambda=0$
$$
P_{0}=A_0(\lambda)\Phi_{0}(\zeta(\lambda))\ee^{\varepsilon\frac{\zeta^2}{4}({\sigma}_3\otimes I_3)}(-t)^{I_2\otimes \Lambda_{12}},
$$
where $\Phi_0$ is the solution of Riemann-Hilbert problem \ref{plocz3rhp}, $\varepsilon=1$ to the left of $\gamma_{2,\pm}$ and $\varepsilon=-1$ to the right of $\gamma_{2,\pm}$.
Here $A_0(\lambda)$ is the holomorphic function
$$
A_0(\lambda)=P_\infty M_5 \zeta^{-2\sigma_3\otimes \Lambda_{13}}\zeta^{-I_2\otimes \Lambda_1}(-t)^{-I_2\otimes \Lambda_{12}}=A_{0,0}+O(\lambda).
$$
Since $\zeta=2^{\frac{1}{4}}(-t)^{\frac{3}{4}}\lambda(1+O(\lambda^2))$, we have
\begin{equation}\label{A00third}
A_{0,0}=H_0\,(-t)^{-\frac{1}{4}(6\sigma_3\otimes \Lambda_{13}+3I_2\otimes \Lambda_1+4I_2\otimes \Lambda_{12})},
\end{equation}
where
\[
\begin{aligned}
H_0={}&\begin{pmatrix}-2\ii I_3&-\frac{\ii}{4}I_3\\
-2\ii I_3&\frac{\ii}{4}I_3\end{pmatrix}
2^{-\frac14(12\sigma_3\otimes\Lambda_{10}
+2\sigma_3\otimes\Lambda_{13}+I_2\otimes\Lambda_1)}
\ee^{\ii\pi\sigma_3\otimes\Lambda_{10}}.
\end{aligned}
\]
Using the expansion of $\Phi_0$ at infinity and \eqref{A00third}, we obtain
the matching condition
\begin{equation} \label{matchz3}
P_{0}=P_\infty \left(I+O\left(\dfrac{1}{(-t)^\frac{3}{4}}\right)\right),\quad t\to -\infty.
\end{equation}

\subsection{Small norm theorem and leading terms of asymptotic}
We define $S_{as}$ and $R=SS_{as}^{-1}$ as in Section \ref{RH2}, using the parametrices of this section. The
function $R(\lambda)$ satisfies the error problem of the same form as
Riemann-Hilbert problem \ref{pinfrrhp2}, with the parametrices just constructed. Using \eqref{matchsq2},
\eqref{matchmsq2}, \eqref{matchz3} we get the estimate \eqref{jest2} for the jump, and as in Section \ref{RH2}
$$
l_1=\dfrac{A_{0,0}m_{1,0}A_{0,0}^{-1}}{{2^\frac{1}{4}}(-t)^{\frac{3}{4}}}+O((-t)^{-\frac{3}{2}}),
$$
where $m_{1,0}$ is given by \eqref{m10third}. By \eqref{A00third}, $A_{0,0}m_{1,0}A_{0,0}^{-1}=H_0m_{1,0}H_0^{-1}$
does not depend on $t$.

Tracing back the transformations we have
$$
(m_1)_{\mbox{offdiag}}=\sqrt{-t}(I_2\otimes P_1)\ee^{-(I_2\otimes \Lambda_8)(-t)^{\frac{3}{2}}}(-t)^{I_2\otimes \Lambda_{12}+\frac{1}{2}I_2\otimes \Lambda_1}\left(\left(m_{1,\infty}\right)_{\mbox{offdiag}}+l_1\right)$$
$$\times\ee^{(I_2\otimes \Lambda_8)(-t)^{\frac{3}{2}}}(-t)^{-I_2\otimes \Lambda_{12}-\frac{1}{2}I_2\otimes \Lambda_1}(I_2\otimes P_1)^{-1},
$$
where the $(1,2)$ block of $m_{1,\infty}$ is equal to $\left(-\frac{\ii}{\sqrt{2}}+\ii\sqrt{2}\right)I_3=\frac{\ii}{\sqrt{2}}I_3$,
see \eqref{m1inf3}. The matrices conjugating the $(1,2)$ block are block diagonal with equal blocks. Hence, using the formula $q(t)=-\ii\hat{m}_{1,12}$,
we get that $q(t)$ is similar to
$$
\sqrt{\frac{-t}{2}}\,I_3+\frac{N}{(-t)^{\frac{1}{4}}}+O\left(\frac{1}{t}\right),\quad
N=-\frac{\ii}{2^{\frac{1}{4}}}\left(H_0m_{1,0}H_0^{-1}\right)_{12},
$$
where $(\cdot)_{12}$ denotes the $(1,2)$ block. Substituting the explicit
expressions for $H_0$ and $m_{1,0}$, we compute
$$
\det(N-\nu I_3)=-\nu^3-\frac{\nu}{2\sqrt{2}}.
$$
The roots $0,\pm\ii2^{-3/4}$ are simple. The residue sign fixes the sign of $N$;
replacing $N$ by $-N$ only interchanges the two nonzero eigenvalues.
The particle labels below are chosen so that $x_2$ has the positive
imaginary leading correction. After subtracting
$\sqrt{-t/2}I_3$ and multiplying by $(-t)^{1/4}$, the matrix is
$N+\mathcal E(t)$, where $\mathcal E(t)=O((-t)^{-3/4})$.
Consider
\[
P(\nu,\mathcal E)=\det(N+\mathcal E-\nu I_3).
\]
This is a polynomial in $\nu$ and the entries of $\mathcal E$.
At $\mathcal E=0$, its roots are $\nu_1^0=0$ and
$\nu_{2,3}^0=\pm\ii2^{-3/4}$, and
$\partial_\nu P(\nu,0)=-3\nu^2-1/(2\sqrt2)$ is nonzero at each root.
The implicit function theorem therefore gives roots
$\nu_j(\mathcal E)=\nu_j^0+O(\|\mathcal E\|)$ near these three distinct values.
Substituting $\mathcal E=\mathcal E(t)$ yields an error $O((-t)^{-3/4})$
in the scaled eigenvalues.
Undoing the shift and rescaling, we obtain, up to permutation,
$$
x_1=\sqrt{\frac{-t}{2}}+O(t^{-1}),\quad x_2=\sqrt{\frac{-t}{2}}+\frac{\ii}{{2^\frac{3}{4}(-t)^{\frac{1}{4}}}}+O(t^{-1}),\quad x_3=\sqrt{\frac{-t}{2}}-\frac{\ii}{{2^\frac{3}{4}(-t)^{\frac{1}{4}}}}+O(t^{-1}).
$$

\subsection{Recurrence relation and subleading terms of asymptotics}
The model parametrices have expansions to every fixed order. The
block decomposition used above controls the full remainder at the origin,
and the error problem can be solved by a Neumann series for sufficiently
large $-t$. Expanding that series and then the simple eigenvalues gives
asymptotic expansions in powers of $(-t)^{-3/4}$. As in Section
\ref{RH2}, the differential equations determine their coefficients.
The expansions below are asymptotic series; convergence is not asserted.
They can be differentiated to the orders used in \eqref{rewrcpa} by
performing the same construction for the differentiated error equation.
We write
$$
x_1\sim \sqrt{\frac{-t}{2}}+\sum_{j=0}^\infty c_j (-t)^{-1-\frac{3}{4}j},\, x_2\sim\sqrt{\frac{-t}{2}} +\frac{\ii}{{2^{\frac{3}{4}}(-t)^{\frac{1}{4}}}}+\sum_{j=0}^\infty d_j (-t)^{-1-\frac{3}{4}j},$$
$$x_3\sim\sqrt{\frac{-t}{2}} -\frac{\ii}{{2^{\frac{3}{4}}(-t)^{\frac{1}{4}}}}+\sum_{j=0}^\infty e_j (-t)^{-1-\frac{3}{4}j}.
$$
We substitute these series into \eqref{rewrcpa} and expand in the same way as in Section \ref{RH2}; the only
difference is the sign of the leading term $\sqrt{\frac{-t}{2}}$. The particle differences are unchanged by changing the common leading
term from $-\sqrt{-t/2}$ to $+\sqrt{-t/2}$. Moreover, the linearization
of $2x^3+tx$ at either equilibrium is $6(-t/2)+t=-2t$.
Therefore the terms with highest
indices $c_{k-2}$, $d_{k-2}$, $e_{k-2}$ in the coefficient near $(-t)^{-\frac{3}{4}k}$ again come from the terms
\begin{equation*}
\begin{gathered}
18(x_1-x_2)^3(x_1-x_3)^3(2x_1^3+x_1 t)-4((x_1-x_2)^3+(x_1-x_3)^3),
\\
18(x_2-x_3)^3(x_2-x_1)^3(2x_2^3+x_2 t)-4((x_2-x_1)^3+(x_2-x_3)^3),
\\
18(x_3-x_2)^3(x_3-x_1)^3(2x_3^3+x_3 t)-4((x_3-x_2)^3+(x_3-x_1)^3),
\end{gathered}
\end{equation*}
and they have the form
$$
\left(\begin{array}{ccc}
     \frac{57\sqrt{2}}{8}&-3\sqrt{2}&-3\sqrt{2}  \\
     24\sqrt{2}&-\frac{69\sqrt{2}}{2}&\frac{3\sqrt{2}}{{2}}\\
       24\sqrt{2}&\frac{3\sqrt{2}}{{2}}&-\frac{69\sqrt{2}}{2}
\end{array}\right)\left(\begin{array}{c}c_{k-2}\\d_{k-2}\\e_{k-2}\end{array}\right)+\mbox{lower terms},
$$
with the same matrix as in Section \ref{RH2}. Its determinant is $6561\sqrt{2}$, so the recurrence relation determines all the
coefficients uniquely. In particular,
$$
c_0=\frac{19}{36},\quad d_0=e_0=\frac{11}{18},\quad c_1=0,\quad d_1=-e_1=-\frac{179\cdot2^{\frac{3}{4}}\ii}{864}.
$$
Since $c_1=0$, the next possible correction to $x_1$ is of order
$(-t)^{-5/2}$; for $x_2$ and $x_3$ the first omitted term is of order
$(-t)^{-7/4}$. Also $(-t)^{-1}=-1/t$. This gives exactly the coefficients
and remainder orders in \eqref{x3as} of Theorem \ref{ccCon2}.

\begin{remark}
The asymptotics \eqref{x3as} coincide with the first of the alternative power asymptotics mentioned in the remark at the
end of Section \ref{RH2}. They are different from \eqref{x1as}. Hence, among the data \eqref{S1S2}, only $a_{33}=\ii$ can
generate the Tracy-Widom distribution $F_6(t)$. As $t\to+\infty$, the parametrices of Section \ref{RH1} involve
only the entries of the jump matrices which do not depend on $a_{33}$, so both values of $a_{33}$ lead to the
asymptotics \eqref{x2as}.
\end{remark}

 \appendix
\section{Parabolic cylinder functions parametrices}
We introduce matrix functions 
\label{parab}

$$
\mathbf{D}_{\alpha,\beta}^{(1)}(z)=\begin{pmatrix}
     1&0\\-\frac{z}{2\alpha}&\frac{1}{\alpha} 
    \end{pmatrix}$$
$$\times\left\{\begin{array}{l}\begin{pmatrix}
     D_{\alpha\beta}(\ii z)&D_{-\alpha\beta-1}(z)\\\left(D_{\alpha\beta}(\ii z)\right)'&\left(D_{-\alpha\beta-1}(z)\right)'
    \end{pmatrix}\begin{pmatrix}
     \ee^{-\frac{\ii\pi}{2} \alpha\beta }&0\\0&-\alpha \end{pmatrix},\quad -\frac{\pi}{4}<\arg (z)<0,\\[0.35cm]
     \begin{pmatrix}
     D_{\alpha\beta}(-\ii z)&D_{-\alpha\beta-1}(z)\\\left(D_{\alpha\beta}(-\ii z)\right)'&\left(D_{-\alpha\beta-1}(z)\right)'
    \end{pmatrix}\begin{pmatrix}
     \ee^{\frac{\ii\pi}{2} \alpha\beta }&0\\0&-\alpha \end{pmatrix},\quad 0<\arg (z)<\frac{\pi}{2},\\[0.35cm]
     \begin{pmatrix}
     D_{\alpha\beta}(-\ii z)&D_{-\alpha\beta-1}(-z)\\\left(D_{\alpha\beta}(-\ii z)\right)'&\left(D_{-\alpha\beta-1}(-z)\right)'
    \end{pmatrix}\begin{pmatrix}
     \ee^{\frac{\ii\pi}{2} \alpha\beta }&0\\0&\alpha \ee^{-{\ii\pi}{} \alpha\beta } \end{pmatrix},\quad \frac{\pi}{2}<\arg (z)<\pi,\\[0.35cm]
    \begin{pmatrix}
     D_{\alpha\beta}(\ii z)&D_{-\alpha\beta-1}(-z)\\\left(D_{\alpha\beta}(\ii z)\right)'&\left(D_{-\alpha\beta-1}(-z)\right)'
    \end{pmatrix} \begin{pmatrix}
     \ee^{\frac{3\pi \ii}{2} \alpha\beta }&0\\0&\alpha \ee^{-{\ii\pi}{} \alpha\beta } \end{pmatrix},\quad {\pi}{}<\arg (z)<\frac{3\pi}{2},\\[0.35cm]
     \begin{pmatrix}
     D_{\alpha\beta}(\ii z)&D_{-\alpha\beta-1}(z)\\\left(D_{\alpha\beta}(\ii z)\right)'&\left(D_{-\alpha\beta-1}(z)\right)'
    \end{pmatrix}\begin{pmatrix}
     \ee^{\frac{3\pi \ii}{2} \alpha\beta }&0\\0&-\alpha \ee^{-{2\pi \ii}{} \alpha\beta } \end{pmatrix}.\quad \frac{3\pi}{2}<\arg (z)<\frac{7\pi}{4},
     \end{array}\right.
$$
$$
\mathbf{D}_{\alpha,\beta}^{(2)}(z)=\begin{pmatrix}
     1&0\\-\frac{z}{2\alpha}&\frac{1}{\alpha} 
    \end{pmatrix}$$
$$\times\left\{\begin{array}{l}\begin{pmatrix}
     D_{\alpha\beta}(-\ii z)&D_{-\alpha\beta-1}(-z)\\\left(D_{\alpha\beta}(-\ii z)\right)'&\left(D_{-\alpha\beta-1}(-z)\right)'
    \end{pmatrix}\begin{pmatrix}
     \ee^{-\frac{3\pi \ii}{2} \alpha\beta }&0\\0&\alpha \ee^{{\ii\pi}{} \alpha\beta } \end{pmatrix},\quad -\frac{5\pi}{4}{}{}<\arg (z)<-\pi,\\[0.35cm]
   \begin{pmatrix}
     D_{\alpha\beta}(\ii z)&D_{-\alpha\beta-1}(-z)\\\left(D_{\alpha\beta}(\ii z)\right)'&\left(D_{-\alpha\beta-1}(-z)\right)'
    \end{pmatrix}\begin{pmatrix}
     \ee^{-\frac{\ii\pi}{2} \alpha\beta }&0\\0&\alpha \ee^{{\ii\pi}{} \alpha\beta } \end{pmatrix},\quad -\pi<\arg (z)<-\frac{\pi}{2},\\[0.35cm]\begin{pmatrix}
     D_{\alpha\beta}(\ii z)&D_{-\alpha\beta-1}(z)\\\left(D_{\alpha\beta}(\ii z)\right)'&\left(D_{-\alpha\beta-1}(z)\right)'
    \end{pmatrix}\begin{pmatrix}
     \ee^{-\frac{\ii\pi}{2} \alpha\beta }&0\\0&-\alpha \end{pmatrix},\quad -\frac{\pi}{2}<\arg (z)<0,\\[0.35cm]
     \begin{pmatrix}
     D_{\alpha\beta}(-\ii z)&D_{-\alpha\beta-1}(z)\\\left(D_{\alpha\beta}(-\ii z)\right)'&\left(D_{-\alpha\beta-1}(z)\right)'
    \end{pmatrix}\begin{pmatrix}
     \ee^{\frac{\ii\pi}{2} \alpha\beta }&0\\0&-\alpha \end{pmatrix},\quad 0<\arg (z)<\frac{\pi}{2},\\[0.35cm]
     \begin{pmatrix}
     D_{\alpha\beta}(-\ii z)&D_{-\alpha\beta-1}(-z)\\\left(D_{\alpha\beta}(-\ii z)\right)'&\left(D_{-\alpha\beta-1}(-z)\right)'
    \end{pmatrix}\begin{pmatrix}
     \ee^{\frac{\ii\pi}{2} \alpha\beta }&0\\0&\alpha \ee^{-{\ii\pi}{} \alpha\beta } \end{pmatrix},\quad \frac{\pi}{2}<\arg (z)<\frac{3\pi}{4}, \end{array}\right.
$$
where $D_\nu(z)$ is the parabolic cylinder function. It is entire function and it solves differential equation
\begin{align}
   \dfrac{d^2}{dz^2}D_{\nu}(z)+\left(\nu+\frac{1}{2}-\frac{z^2}{4}\right)D_\nu(z)=0.
\end{align}
It implies the differential equation or matrix functions
$$
\dfrac{d}{dz}\mathbf{D}^{(j)}_{\alpha,\beta}(z)=\left(\frac{z}{2}\sigma_3+\begin{pmatrix}
0&\alpha\\
\beta&0\end{pmatrix}
\right)\mathbf{D}^{(j)}_{\alpha,\beta}(z).
$$

The asymptotic
\begin{align}
    D_{\nu}(z)=\left(1+\dfrac{\nu(1-\nu)}{2z^2}+O\left(\dfrac{1}{z^4}\right)\right) z^{\nu}\ee^{-\frac{z^{2}}{4}},\quad z\to \infty,\quad |\arg(z)|<\frac{3\pi}{4}.
\end{align}
implies the following asymptotics for the matrix functions
\begin{align}
    \mathbf{D}^{(j)}_{\alpha,\beta}(z)=\left(\mathbb{1}+\dfrac{1}{z}\begin{pmatrix}
        0&-\alpha\\\beta&0
    \end{pmatrix}+O\left(\dfrac{1}{z^2}\right)\right) z^{\alpha\beta\sigma_3} \ee^{\frac{z^2}{4}\sigma_3}.
\end{align}

\begin{flushleft}
\begin{figure}[h]
\begin{subfigure}[b]{0.5\textwidth}
\includegraphics[scale=1]{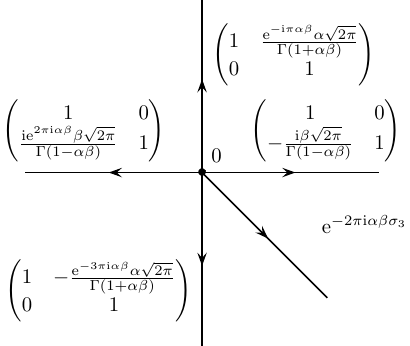}
	\caption{Jump for  $\mathbf{D}^{(1)}_{\alpha,\beta}(z)$}
\end{subfigure}
\begin{subfigure}[b]{0.5\textwidth}
\includegraphics[scale=1]{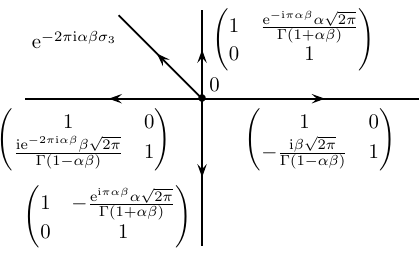}
	\caption{ Jump for $\mathbf{D}^{(2)}_{\alpha,\beta}(z)$}
\end{subfigure} 
\caption{}\label{fig:parabolic_jump}
\end{figure}
\end{flushleft}

Using the connection formulas
\begin{align}
    D_{\nu}(z)=\ee^{-\ii\pi\nu}D_{\nu}(-z)-\ii\ee^{-\frac{\ii\pi\nu}{2}}\dfrac{\sqrt{2\pi}}{\Gamma(-\nu)}D_{-\nu-1}(\ii z),\\
     D_{\nu}(z)=\ee^{\ii\pi\nu}D_{\nu}(-z)+\ii\ee^{\frac{\ii\pi\nu}{2}}\dfrac{\sqrt{2\pi}}{\Gamma(-\nu)}D_{-\nu-1}(-\ii z)
\end{align}
we get the jump for  $\mathbf{D}^{(j)}_{\alpha,\beta}(z)$ indicated on Figure \ref{fig:parabolic_jump}.
\subsection{Riemann-Hilbert problem \ref{plocirhp}}\label{parcyli}

The solution is given by
$$
\Phi_i(z)=P_5\begin{pmatrix}
     \mathbf{D}_{\alpha_1,\beta_1}^{(1)}(z)&0&0\\0&\ee^{\frac{z^2}{4}\sigma_3}&0\\
     0&0&\mathbf{D}_{\alpha_2,\beta_2}^{(1)}(z)
    \end{pmatrix}P_5^{-1},
$$
where

$$
\alpha_1=\dfrac{\sqrt{2\pi} \ee^{\frac{\ii\pi}{6}}}{\Gamma\left(\frac{2}{3}\right)},\quad \alpha_2=\dfrac{\sqrt{2\pi} \ee^{\frac{5\pi \ii}{6}}}{\Gamma\left(\frac{1}{3}\right)},\quad 
\beta_1=-\frac{2}{3\alpha_1},\quad  \beta_2=-\frac{1}{3\alpha_2},\quad 
P_5=\left(\begin{array}{cccccc}
0&0&1&0&0&0\\
0&0&0&0&1&0\\
1&0&0&0&0&0\\
0&0&0&0&0&1\\
0&1&0&0&0&0\\
0&0&0&1&0&0\\
\end{array}\right).
$$
\subsection{Riemann-Hilbert problem \ref{plocmirhp}}\label{parcylmi}
The solution is given by
$$
\Phi_{-\ii}(z)=P_6\begin{pmatrix}
     \ee^{\frac{z^2}{4}\sigma_3}&0&0\\0&\mathbf{D}_{\alpha_1,\beta_1}^{(2)}(z)&0\\
     0&0&\mathbf{D}_{\alpha_2,\beta_2}^{(2)}(z)
    \end{pmatrix}P_6^{-1},
$$
where

$$
\alpha_1=\frac{1}{3\beta_1},\quad  \alpha_2=\frac{2}{3\beta_2},\quad \beta_1=\dfrac{\sqrt{2\pi} \ee^{\frac{4\pi \ii}{3}}}{\Gamma\left(\frac{1}{3}\right)},\quad \beta_2=\dfrac{\sqrt{2\pi} \ee^{\frac{2\pi \ii}{3}}}{\Gamma\left(\frac{2}{3}\right)} 
,\quad 
P_6=\left(\begin{array}{cccccc}
0&0&1&0&0&0\\
0&0&0&0&1&0\\
1&0&0&0&0&0\\
0&1&0&0&0&0\\
0&0&0&1&0&0\\
0&0&0&0&0&1\\
\end{array}\right).
$$

\section{Modified Bessel functions parametrix}
\label{bessel_section}
Introduce matrix function
$$
\mathbf{K}_{\frac{2}{3}}(z)=\frac{\ee^{\frac{3\pi \ii}{4}}}{\sqrt{\pi}}\ee^{-\frac{\ii\pi}{4}\sigma_3}K^{-1}\sqrt{z}\left(\begin{array}{cc}
1&0\\
\dfrac{2}{3z}&1
\end{array}\right)\left\{\begin{array}{l}\begin{pmatrix}K_{\frac{2}{3}}\left(\ee^{{\ii\pi}{}}z\right)&K_{\frac{2}{3}}\left(z\right)\\\left(K_{\frac{2}{3}}\left(\ee^{{\ii\pi}{}}z\right)\right)'&K_{\frac{2}{3}}\left(z\right)'\end{pmatrix},\quad -{\pi}<\arg (z)<0,\\[0.35cm]
     \begin{pmatrix}K_{\frac{2}{3}}\left(\ee^{{-\ii\pi}{}}z\right)&K_{\frac{2}{3}}\left(z\right)\\\left(K_{\frac{2}{3}}\left(\ee^{-{\ii\pi}{}}z\right)\right)'&K_{\frac{2}{3}}\left(z\right)'\end{pmatrix}(-\sigma_3),\quad 0<\arg (z)<{\pi}.\\[0.35cm]
    \end{array}\right.
$$
Here $K_\nu(z)$ are modified Bessel functions and matrix $K$ is given in \eqref{K}.

The differential equation 
\begin{align}
    \dfrac{d^2}{dz^2}K_{\frac{2}{3}}(z)+\frac{1}{z}\dfrac{d}{dz}K_{\frac{2}{3}}(z)-\left(1+\frac{4}{9z^2}\right)K_{\frac{2}{3}}(z)=0
\end{align}
implies that the function $\mathbf{K}_{\frac{2}{3}}(z)$ satisfies $$
\dfrac{d}{dz}\mathbf{K}_{\frac{2}{3}}(z)=\left(\sigma_3+\frac{\sigma_2}{6z}\right)\mathbf{K}_{\frac{2}{3}}(z).
$$

The asymptotic 
\begin{align}
    K_{\frac{2}{3}}(z)=\left(1+O\left(\frac{1}{z}\right)\right)\sqrt{\dfrac{\pi}{2z}}\ee^{-z},\quad z\to \infty,\quad  |\arg(z)|<\frac{3\pi}{2}
\end{align}
implies 
\begin{align}
    \mathbf{K}_{\frac{2}{3}}(z)=\left(\mathbb{1}+O\left(\frac{1}{z}\right)\right)\ee^{z\sigma_3},\quad z\to \infty.
\end{align}

\begin{figure}[h]
    \centering
    \includegraphics[width=0.3\linewidth]{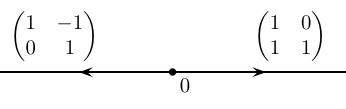}
    \caption{Jump for $\mathbf{K}_{\frac{2}{3}}(z)$}
    \label{fig:bessel_jump}
\end{figure}

The analytic continuation formula
\begin{align}
    K_{\frac{2}{3}}(\ee^{\ii\pi}z)+K_{\frac{2}{3}}(-\ee^{\ii\pi}z)+K_{\frac{2}{3}}(z)=0
\end{align}
implies the following jumps for function    $\mathbf{K}_{\frac{2}{3}}(z)$ indicated on Figure \ref{fig:bessel_jump}. 
\subsection{Riemann-Hilbert problem \ref{ploczrhp}}\label{bessel} 
The solution is given by
$$
\Phi_0(z)=\ee^{-\ii\pi(\sigma_3\otimes \Lambda_{11})}\left(\mathbf{K}_{\frac{2}{3}}(z) \otimes I_3\right)
\ee^{\ii\pi(\sigma_3\otimes \Lambda_{11})}
$$ 
where
$$
 \Lambda_{11}=\begin{pmatrix}
 -\frac{1}{12}&0&0\\
 0&\frac{1}{4}&0\\
 0&0&\frac{1}{12}
\end{pmatrix}.
$$

\section{Global parametrix for asymptotic \texorpdfstring{$t\to-\infty$}{t->-infty}}
\label{global}
Consider the function
$$
\delta(\lambda)=\left(\dfrac{\sqrt{\lambda^2-2}+\ii\sqrt{2}}{\sqrt{\lambda^2-2}-\ii\sqrt{2}}\right)^{\sigma_3\otimes \Lambda_{10}}.$$

 We choose cuts for logarithms in   $\delta(\lambda)$ as described below the Figure \ref{cuts}. It implies the following jump properties

$$
\delta_{+}(\lambda)\delta_-(\lambda)=\ee^{2\pi \ii \sigma_3\otimes \Lambda_{10}}, \quad \lambda \in \gamma_1, \, \gamma_{2,+},
$$
$$
\delta_{+}(\lambda)\delta_-(\lambda)=\ee^{-2\pi \ii \sigma_3\otimes \Lambda_{10}}, \quad \lambda \in \gamma_{2,-},\,\gamma_3.
$$
It also has the following behavior near zero
$$\delta(\lambda)= \ee^{{\ii\pi}{}(\sigma_3\otimes \Lambda_{10})}2^{3{}(\sigma_3\otimes \Lambda_{10})}(I_6+O(\lambda))\lambda^{-2\sigma_3\otimes  \Lambda_{10}},\quad \lambda \to 0,\quad \frac{\pi}{4}<\arg \lambda <\frac{5\pi}{4},$$
$$\delta(\lambda)= \ee^{{\ii\pi}{}(\sigma_3\otimes \Lambda_{10})}2^{-3{}(\sigma_3\otimes \Lambda_{10})}(I_6+O(\lambda))\lambda^{2\sigma_3\otimes  \Lambda_{10}},\quad \lambda \to 0,\quad -\frac{3\pi}{4}<\arg \lambda <\frac{\pi}{4}.$$

Consider another function

$$
 \upsilon=\left(\dfrac{\lambda-{\sqrt{2}}}{\lambda+{\sqrt{2}}}\right)^\frac{1}{4}.$$

 We choose cuts for logarithms in  $\upsilon(\lambda)$ as described below the Figure \ref{cuts}.
 We have

$$
\upsilon_{+}(\lambda)=\upsilon_{-}(\lambda)\ii,  \quad \lambda \in \gamma_1, \, \gamma_{2,\pm}, \,\gamma_3.
$$

We form the matrix 
 $$
 P_7=(K\upsilon^{{\sigma}_3}K^{-1})\otimes I_3,\quad
 $$
 where $K$ given by \eqref{K}. It provides us with the following jumps
 $$
P_{7,+}(\lambda)=P_{7,-}(\lambda)\ii({\sigma}_1\otimes I_3),  \quad \lambda \in \gamma_1, \, \gamma_{2,\pm},\,\gamma_3.
$$
\subsection{Riemann-Hilbert problem \ref{pinfpinfrhp2}}\label{globali}
 The solution is given by

$$
P_\infty(\lambda)= \ee^{\frac{\ii\pi}{4}({\sigma}_3\otimes I_3)}P_7\delta(\lambda)\lambda^{I_2\otimes \Lambda_1}\ee^{-\frac{\ii\pi}{4}({\sigma}_3\otimes I_3)},
$$
where the cut for $\lambda^{{I_2\otimes \Lambda_1}}$ is along $\gamma_{2,-}$.
\subsection{Riemann-Hilbert problem \ref{pinfpinfrhp3}}\label{globalii}
Let $P_\infty^{\mathrm{old}}$ denote the expression in Appendix \ref{globali},
with $\Lambda_{10}$ evaluated at $a_{33}=\ee^{-5\pi\ii/6}$.
We seek the new parametrix in the form
$$
P_\infty(\lambda)=\left(I_6+\frac{C_0}{\lambda}\right)
P_\infty^{\mathrm{old}}(\lambda),
$$
where the constant matrix $C_0$ is to be determined from the condition at
zero. The multiplier is single valued and tends to $I_6$ at infinity.
We first work in the sector
$-3\pi/4<\arg\lambda<\pi/4$, where $M_5=I_6$. Using the same matrix $K$ and function $\upsilon$ as in Appendix
\ref{globali}, write
$$
P_\infty^{\mathrm{old}}(\lambda)
=(Q(\lambda)\otimes I_3)\delta(\lambda)\lambda^{I_2\otimes\Lambda_1},
\qquad
Q(\lambda)=\ee^{\frac{\ii\pi}{4}\sigma_3}
K\upsilon(\lambda)^{\sigma_3}K^{-1}
\ee^{-\frac{\ii\pi}{4}\sigma_3}.
$$
In this sector $\upsilon(0)=\ee^{-\ii\pi/4}$. Expanding at zero gives
$$
Q(\lambda)=Q_0+\lambda Q_1+\lambda^2Q_2+O(\lambda^3),\qquad
Q_0=K^{-1},\quad Q_1=\frac{\ii}{2\sqrt2}K,\quad
Q_2=\frac1{16}K^{-1}.
$$
The required change from $\Lambda_{10}$ to
$\Lambda_{13}=\Lambda_{10}-I_3/2$ lowers the first exponent in each
$2\times2$ block by one and raises the second by one. Thus the second
column of the regular factor must vanish to order one after multiplication.
Put
$$
v=Q_0e_2=K^{-1}e_2,
\qquad w=Q_1e_2=\frac{\ii}{2\sqrt2}Ke_2.
$$
Expanding the corresponding three columns gives
$$
\left(I_6+\frac{C_0}{\lambda}\right)
\bigl((Q(\lambda)e_2)\otimes I_3\bigr)
=\frac{C_0(v\otimes I_3)}{\lambda}
+v\otimes I_3+C_0(w\otimes I_3)+O(\lambda).
$$
The pole and constant term vanish precisely when
$$
C_0(v\otimes I_3)=0,\qquad
C_0(w\otimes I_3)=-v\otimes I_3.
$$
Since $v,w$ are linearly independent, the columns of $v\otimes I_3$ and
$w\otimes I_3$ form a basis of $\mathbb C^6$. These equations therefore
determine $C_0$ uniquely. Solving them gives
$$
c=2\ii\sqrt2\,K\begin{pmatrix}0&1\\0&0\end{pmatrix}K^{-1},
\qquad
C_0=c\otimes I_3=-\ii\sqrt2
\begin{pmatrix}I_3&-I_3\\I_3&-I_3\end{pmatrix}.
$$
In particular, $C_0^2=0$, so the multiplier has determinant one and inverse
$I_6-C_0/\lambda$. It is holomorphic and invertible away from zero,
preserves all jumps and the endpoint behavior at $\pm\sqrt2$, and its
expansion at infinity gives \eqref{m1inf3}.

Consequently the second column of $(I_2+c/\lambda)Q$ vanishes to order
one at zero, whereas the first has a simple pole. More precisely,
$$
\left(I_2+\frac c\lambda\right)Q(\lambda)
\begin{pmatrix}\lambda&0\\0&\lambda^{-1}\end{pmatrix}
=H+O(\lambda),\qquad
H=K\begin{pmatrix}-2\ii\sqrt2&0\\0&\dfrac{\ii}{2\sqrt2}\end{pmatrix},\qquad \det H=1.
$$
The remaining factors in $P_\infty^{\mathrm{old}}$ are diagonal. After removing their
powers of $\lambda$, they are holomorphic and invertible at zero. Thus
multiplication by $I_6+C_0/\lambda$ lowers the exponent of the first
column by one and raises that of the second by one. This replaces
$2\sigma_3\otimes\Lambda_{10}+I_2\otimes\Lambda_1$ by
$2\sigma_3\otimes\Lambda_{13}+I_2\otimes\Lambda_1$ in this sector.
The holomorphic prefactor is invertible, since its determinant is unchanged
and nonzero. The other sector follows from the prescribed constant jumps.
This proves all conditions of Riemann-Hilbert problem \ref{pinfpinfrhp3}.

Finally, the local variable satisfies
$\zeta=2^{1/4}(-t)^{3/4}\lambda(1+O(\lambda^2))$.
Removing its powers and the diagonal $t$-normalization gives
\[
H_0=(H\otimes I_3)\,
2^{-\frac14(12\sigma_3\otimes\Lambda_{10}
+2\sigma_3\otimes\Lambda_{13}+I_2\otimes\Lambda_1)}
\ee^{\ii\pi\sigma_3\otimes\Lambda_{10}},
\]
as used in \eqref{A00third}.

\section{Airy functions parametrix}
\label{airy}
Introduce matrix function

$$
\mathbf{{Ai}}(z)=\sqrt{2\pi}\left\{\begin{array}{l}\left(\begin{array}{cc}\mathrm{Ai}\left(z\right)&\mathrm{Ai}\left(\ee^{\frac{2\pi \ii}{3}}z\right)\\\mathrm{Ai}\left(z\right)'&\left(\mathrm{Ai}\left(\ee^{\frac{2\pi \ii}{3}}z\right)\right)'\end{array}\right) \begin{pmatrix}
1&0\\0&\ee^{\frac{\ii\pi}{6}}
\end{pmatrix},\quad -\frac{\pi}{3}<\arg (z)<0,\\[0.35cm]
     \left(\begin{array}{cc}\mathrm{Ai}\left(z\right)&\mathrm{Ai}\left(\ee^{-\frac{2\pi \ii}{3}}z\right)\\\mathrm{Ai}\left(z\right)'&\left(\mathrm{Ai}\left(\ee^{-\frac{2\pi \ii}{3}}z\right)\right)'\end{array}\right) \begin{pmatrix}
1&0\\0&\ee^{-\frac{\ii\pi}{6}}
\end{pmatrix},\quad 0<\arg (z)<\frac{2\pi}{3},\\[0.35cm]
    \left(\begin{array}{cc}\mathrm{Ai}\left(\ee^{\frac{2\pi \ii}{3}}z\right)&\mathrm{Ai}\left(\ee^{-\frac{2\pi \ii}{3}}z\right)\\\mathrm{Ai}\left(\ee^{\frac{2\pi \ii}{3}}z\right)'&\left(\mathrm{Ai}\left(\ee^{-\frac{2\pi \ii}{3}}z\right)\right)'\end{array}\right) \begin{pmatrix}
\ee^{-\frac{\ii\pi}{3}}&0\\0&\ee^{-\frac{\ii\pi}{6}}
\end{pmatrix},\quad \frac{2\pi}{3}<\arg (z)<\frac{4\pi}{3},\\[0.35cm]
 \left(\begin{array}{cc}\mathrm{Ai}\left(\ee^{\frac{2\pi \ii}{3}}z\right)&\mathrm{Ai}\left(z\right)\\\mathrm{Ai}\left(\ee^{\frac{2\pi \ii}{3}}z\right)'&\left(\mathrm{Ai}\left(z\right)\right)'\end{array}\right) \begin{pmatrix}
\ee^{-\frac{\ii\pi}{3}}&0\\0&\ee^{-\frac{\ii\pi}{2}}
\end{pmatrix},\quad \frac{4\pi}{3}<\arg (z)<\frac{5\pi}{3}.\\[0.35cm]
    \end{array}\right.
$$
Here $\mathrm{Ai}(z)$ is Airy function, which is entire function of $z$. Function $\mathbf{Ai}(z)$ satisfies the differential equation
$$
\dfrac{d}{dz}\mathbf{Ai}(z)=\begin{pmatrix}
0&1\\z&0
\end{pmatrix}\mathbf{Ai}(z).
$$

The asymptotics of Airy function
\begin{align}
    \mathrm{Ai}(z)=\left(1+O\left(\frac{1}{z^{\frac{3}{2}}}\right)\right)\dfrac{\ee^{-\frac{2}{3}z^{\frac{3}{2}}}}{2\sqrt{\pi}z^{\frac{1}{4}}},\quad z\to\infty,\quad  |\arg{z}|<\pi
\end{align}
implies the asymptotics of function $\mathbf{Ai}(z)$
\begin{align}
    \mathbf{Ai}(z)=z^{-\frac{1}{4}\sigma_3}K^{-1}\left(\mathbb{1}+O\left(\frac{1}{z^{\frac{3}{2}}}\right)\right)\ee^{-\frac{2}{3}z^\frac{3}{2}\sigma_3}.
\end{align}

\begin{figure}[h]
    \centering
    \includegraphics[width=0.3\linewidth]{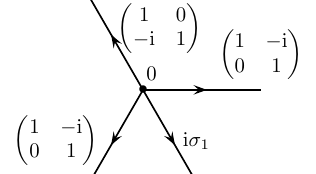}
    \caption{Jump for $\mathbf{Ai}(z)$}
    \label{fig:airy_jump}
\end{figure}

The analytic continuation formula 
\begin{align}
    \mathrm{Ai}(z)+\ee^{-\frac{2\pi\ii}{3}}\mathrm{Ai}(\ee^{-\frac{2\pi\ii}{3}}z)+\ee^{\frac{2\pi\ii}{3}}\mathrm{Ai}(\ee^{\frac{2\pi\ii}{3}}z)=0
\end{align}
implies the following jumps for function $\mathbf{Ai}(z)$ indicated on Figure \ref{fig:airy_jump}.

\subsection{ Riemann-Hilbert problem \ref{plocsqu2rhp}}\label{solplocsqu2rhp}
 The solution is given by
$$
\Phi_{\sqrt{2}}(z)= (\mathbf{Ai}(z)\otimes I_3)\ee^{{\ii\pi}(\sigma_3\otimes \Lambda_7)}(\sigma_3\otimes I_3).
$$

\subsection{ Riemann-Hilbert problem \ref{plocmsqu2rhp}}\label{solplocmsqu2rhp}
 The solution is given by
$$
\Phi_{-\sqrt{2}}(z)= (\mathbf{Ai}(z)\otimes I_3)\ee^{-{\ii\pi}(\sigma_3\otimes \Lambda_7)}(\ii\sigma_2\otimes I_3).
$$
\section{Confluent hypergeometric functions parametrix}
\label{confluent}

We introduce matrix functions
\begin{align}\setlength\arraycolsep{2pt}
&\mathbf{W}^{(1)}_{\alpha,\beta}(z)=\frac{1}{\sqrt{z}}\begin{pmatrix}
     1&0\\-\frac{z}{2\alpha}-\frac{1}{3\alpha z}&\frac{1}{\alpha} 
    \end{pmatrix}\\&\times\left\{\begin{array}{ll}
  &  \begin{pmatrix}
     W_{\nu^{(1)},\frac{1}{3}}\left(\frac{\ee^{\ii\pi}z^2}{2}\right)& W_{-\nu^{(1)},\frac{1}{3}}\left(\frac{\ee^{2\pi \ii}z^2}{2}\right)\\\left(W_{\nu^{(1)},\frac{1}{3}}\left(\frac{\ee^{\ii\pi}z^2}{2}\right)\right)'&\left(W_{-\nu^{(1)},\frac{1}{3}}\left(\frac{\ee^{2\pi \ii}z^2}{2}\right)\right)'
    \end{pmatrix}2^{\nu^{(1)}\sigma_3}\begin{pmatrix}
      \ee^{-{\ii\pi}\nu^{(1)} }&0\\0&-\alpha \ee^{{2\pi \ii}\nu^{(1)} }\end{pmatrix},\\&\hfill -\frac{3\pi}{4}<\arg (z)<-\frac{\pi}{2},\\[0.35cm]
    &\begin{pmatrix}
     W_{\nu^{(1)},\frac{1}{3}}\left(\frac{\ee^{\ii\pi}z^2}{2}\right)& W_{-\nu^{(1)},\frac{1}{3}}\left(\frac{z^2}{2}\right)\\\left(W_{\nu^{(1)},\frac{1}{3}}\left(\frac{\ee^{\ii\pi}z^2}{2}\right)\right)'&\left(W_{-\nu^{(1)},\frac{1}{3}}\left(\frac{z^2}{2}\right)\right)'
    \end{pmatrix}2^{\nu^{(1)}\sigma_3}\begin{pmatrix}
      \ee^{-{\ii\pi}\nu^{(1)} }&0\\0&-\alpha \end{pmatrix},\\&\hfill -\frac{\pi}{2}<\arg (z)<0,\\[0.35cm]
    & \begin{pmatrix}
     W_{\nu^{(1)},\frac{1}{3}}\left(\frac{\ee^{-\ii\pi}z^2}{2}\right)& W_{-\nu^{(1)},\frac{1}{3}}\left(\frac{z^2}{2}\right)\\\left(W_{\nu^{(1)},\frac{1}{3}}\left(\frac{\ee^{-\ii\pi}z^2}{2}\right)\right)'&\left(W_{-\nu^{(1)},\frac{1}{3}}\left(\frac{z^2}{2}\right)\right)'
    \end{pmatrix}2^{\nu^{(1)}\sigma_3}\begin{pmatrix}
      \ee^{{\ii\pi}\nu^{(1)} }&0\\0&-\alpha \end{pmatrix},\\&\hfill 0<\arg (z)<\frac{\pi}{2},\\[0.35cm]
   &\begin{pmatrix}
     W_{\nu^{(1)},\frac{1}{3}}\left(\frac{\ee^{-\ii\pi}z^2}{2}\right)& W_{-\nu^{(1)},\frac{1}{3}}\left(\frac{\ee^{-2\pi \ii}z^2}{2}\right)\\\left(W_{\nu^{(1)},\frac{1}{3}}\left(\frac{\ee^{-\ii\pi}z^2}{2}\right)\right)'&\left(W_{-\nu^{(1)},\frac{1}{3}}\left(\frac{\ee^{-2\pi \ii}z^2}{2}\right)\right)'
    \end{pmatrix} 2^{\nu^{(1)}\sigma_3}\begin{pmatrix}
      \ee^{{\ii\pi}\nu^{(1)} }&0\\0&-\alpha \ee^{-{2\pi \ii}\nu^{(1)} }\end{pmatrix},\\&\hfill \frac{\pi}{2}<\arg (z)<\pi,\\[0.35cm]
    &\begin{pmatrix}
     W_{\nu^{(1)},\frac{1}{3}}\left(\frac{\ee^{-3\pi \ii}z^2}{2}\right)& W_{-\nu^{(1)},\frac{1}{3}}\left(\frac{\ee^{-2\pi \ii}z^2}{2}\right)\\\left(W_{\nu^{(1)},\frac{1}{3}}\left(\frac{\ee^{-3\pi \ii}z^2}{2}\right)\right)'&\left(W_{-\nu^{(1)},\frac{1}{3}}\left(\frac{\ee^{-2\pi \ii}z^2}{2}\right)\right)'
    \end{pmatrix}2^{\nu^{(1)}\sigma_3}\begin{pmatrix}
      -\ee^{{3\pi}i\nu^{(1)} }&0\\0&\alpha \ee^{-{2\pi \ii}\nu^{(1)} }\end{pmatrix},\\&\hfill\quad \pi<\arg (z)<\frac{5\pi}{4},
     \end{array}\right.
\end{align}
\begin{align}\setlength\arraycolsep{2pt}
&\mathbf{W}^{(2)}_{\alpha,\beta}(z)=\frac{1}{\sqrt{z}}\begin{pmatrix}
     1&0\\-\frac{z}{2\alpha}-\frac{5}{3\alpha z}&\frac{1}{\alpha} 
    \end{pmatrix}\\&\times\left\{\begin{array}{ll}
    &\begin{pmatrix}
     W_{\nu^{(2)},\frac{1}{3}}\left(\frac{\ee^{\ii\pi}z^2}{2}\right)& W_{-\nu^{(2)},\frac{1}{3}}\left(\frac{\ee^{2\pi \ii}z^2}{2}\right)\\\left(W_{\nu^{(2)},\frac{1}{3}}\left(\frac{\ee^{\ii\pi}z^2}{2}\right)\right)'&\left(W_{-\nu^{(2)},\frac{1}{3}}\left(\frac{\ee^{2\pi \ii}z^2}{2}\right)\right)'
    \end{pmatrix}2^{\nu^{(2)}\sigma_3}\begin{pmatrix}
      \ee^{-{\ii\pi}\nu^{(2)} }&0\\0&-\alpha \ee^{{2\pi \ii}\nu^{(2)} }\end{pmatrix},\\&\hfill -\frac{3\pi}{4}<\arg (z)<-\frac{\pi}{2},\\[0.35cm]
   & \begin{pmatrix}
     W_{\nu^{(2)},\frac{1}{3}}\left(\frac{\ee^{\ii\pi}z^2}{2}\right)& W_{-\nu^{(2)},\frac{1}{3}}\left(\frac{z^2}{2}\right)\\\left(W_{\nu^{(2)},\frac{1}{3}}\left(\frac{\ee^{\ii\pi}z^2}{2}\right)\right)'&\left(W_{-\nu^{(2)},\frac{1}{3}}\left(\frac{z^2}{2}\right)\right)'
    \end{pmatrix}2^{\nu^{(2)}\sigma_3}\begin{pmatrix}
      \ee^{-{\ii\pi}\nu^{(2)} }&0\\0&-\alpha \end{pmatrix},\\&\hfill -\frac{\pi}{2}<\arg (z)<0,\\[0.35cm]
     &\begin{pmatrix}
     W_{\nu^{(2)},\frac{1}{3}}\left(\frac{\ee^{-\ii\pi}z^2}{2}\right)& W_{-\nu^{(2)},\frac{1}{3}}\left(\frac{z^2}{2}\right)\\\left(W_{\nu^{(2)},\frac{1}{3}}\left(\frac{\ee^{-\ii\pi}z^2}{2}\right)\right)'&\left(W_{-\nu^{(2)},\frac{1}{3}}\left(\frac{z^2}{2}\right)\right)'
    \end{pmatrix}2^{\nu^{(2)}\sigma_3}\begin{pmatrix}
      \ee^{{\ii\pi}\nu^{(2)} }&0\\0&-\alpha \end{pmatrix},\\&\hfill 0<\arg (z)<\frac{\pi}{2},\\[0.35cm]
   &\begin{pmatrix}
     W_{\nu^{(2)},\frac{1}{3}}\left(\frac{\ee^{-\ii\pi}z^2}{2}\right)& W_{-\nu^{(2)},\frac{1}{3}}\left(\frac{\ee^{-2\pi \ii}z^2}{2}\right)\\\left(W_{\nu^{(2)},\frac{1}{3}}\left(\frac{\ee^{-\ii\pi}z^2}{2}\right)\right)'&\left(W_{-\nu^{(2)},\frac{1}{3}}\left(\frac{\ee^{-2\pi \ii}z^2}{2}\right)\right)'
    \end{pmatrix} 2^{\nu^{(2)}\sigma_3}\begin{pmatrix}
      \ee^{{\ii\pi}\nu^{(2)} }&0\\0&-\alpha \ee^{-{2\pi \ii}\nu^{(2)} }\end{pmatrix},\\&\hfill \frac{\pi}{2}<\arg (z)<\pi,\\[0.35cm]
   & \begin{pmatrix}
     W_{\nu^{(2)},\frac{1}{3}}\left(\frac{\ee^{-3\pi \ii}z^2}{2}\right)& W_{-\nu^{(2)},\frac{1}{3}}\left(\frac{\ee^{-2\pi \ii}z^2}{2}\right)\\\left(W_{\nu^{(2)},\frac{1}{3}}\left(\frac{\ee^{-3\pi \ii}z^2}{2}\right)\right)'&\left(W_{-\nu^{(2)},\frac{1}{3}}\left(\frac{\ee^{-2\pi \ii}z^2}{2}\right)\right)'
    \end{pmatrix}2^{\nu^{(2)}\sigma_3}\begin{pmatrix}
      -\ee^{{3\pi}i\nu^{(2)} }&0\\0&\alpha \ee^{-{2\pi \ii}\nu^{(2)} }\end{pmatrix}.\\&\hfill \pi<\arg (z)<\frac{5\pi}{4},
     \end{array}\right.
\end{align}
where $W_{\kappa,\mu}(z)$ is the Whittaker's confluent hypergeometric function, $\sqrt{z}$ is understood in principal value sense, and \begin{align}\nu^{(1)}=\frac{\alpha\beta}{2}+\frac{1}{6},\quad \nu^{(2)}=\frac{\alpha\beta}{2}+\frac{5}{6}.\end{align}

The function $w_{\nu}(z)=\frac{1}{\sqrt{z}}  W_{\nu,\frac{1}{3}}\left(-\frac{z^2}{2}\right)$ satisfies the differential equation
\begin{align}
      \dfrac{d^2}{dz^2}w_{\nu}(z)-\left(2\nu+\frac{7}{36z^2}+\frac{z^2}{4}\right)w_\nu(z)=0.
\end{align}
That implies the differential equations for $\mathbf{W}^{(j)}_{\alpha,\beta}(z)$
\begin{align}
&\dfrac{d}{dz}\mathbf{W}^{(1)}_{\alpha,\beta}(z)=\left(\left(\frac{z}{2}-\frac{1}{6z}\right)\sigma_3+\begin{pmatrix}
0&\alpha\\
\beta&0\end{pmatrix}
\right)\mathbf{W}^{(1)}_{\alpha,\beta}(z),\\&
\dfrac{d}{dz}\mathbf{W}^{(2)}_{\alpha,\beta}(z)=\left(\left(\frac{z}{2}+\frac{7}{6z}\right)\sigma_3+\begin{pmatrix}
0&\alpha\\
\beta&0\end{pmatrix}
\right)\mathbf{W}^{(2)}_{\alpha,\beta}(z)
\end{align}

The asymptotics
\begin{align}
    W_{\nu,\frac{1}{3}}(z)=\left(1-\frac{(6\nu-1)(6\nu-5)}{36z}+O\left(\frac{1}{z^2}\right)\right)z^{\nu}\ee^{-\frac{z}{2}},\quad z\to\infty,\quad |\arg(z)|<\frac{3\pi}{2}.
\end{align}
implies that function $\mathbf{W}_{\alpha,\beta}(z)$ satisfies the asymptotics
\begin{align}
    \mathbf{W}^{(j)}_{\alpha,\beta}(z)=\left(\mathbb{1}+\dfrac{1}{z}\begin{pmatrix}
        0&-\alpha\\\beta&0
    \end{pmatrix}+O\left(\dfrac{1}{z^2}\right)\right)z^{\left(2\nu^{(j)}-\frac{1}{2}\right)\sigma_3}\ee^{\frac{z^2}{4}\sigma_3}.   
\end{align}

\begin{figure}[h]
    \centering
    \includegraphics[width=0.7\linewidth]{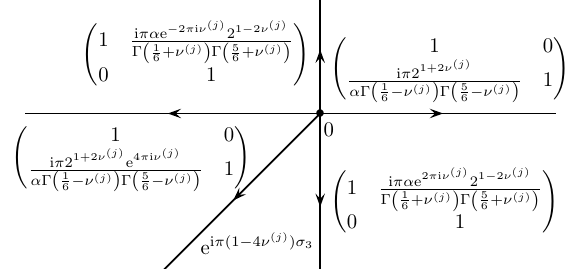}
    \caption{Jump for $\mathbf{W}_{\alpha,\beta}(z)$  }
    \label{fig:whittaker_jump}
\end{figure}

The identity
\begin{align}
    \dfrac{2\pi\ii}{\Gamma\left(\frac{5}{6}-\nu\right)\Gamma\left(\frac{1}{6}-\nu\right)}W_{-\nu,\frac{1}{3}}(z)=\ee^{-\ii\pi\nu}W_{\nu,\frac{1}{3}}(\ee^{\ii\pi}z)
-\ee^{\ii\pi\nu}W_{\nu,\frac{1}{3}}(\ee^{-\ii\pi}z)
\end{align}
implies the following jump for function  $\mathbf{W}_{\alpha,\beta}^{(j)}(z)$  indicated on Figure \ref{fig:whittaker_jump}.

\subsection{Riemann-Hilbert problem \ref{plocz2rhp}}\label{solplocz2rhp}
For $a_{33}=\ii$ the solution is given by
$$
\Phi_0(z)=P_7\begin{pmatrix}
   \mathbf{W}^{(1)}_{\alpha_1,\beta_1}(z)  &0&0\\0&\ee^{\frac{z^2}{4}\sigma_3}z^{-\frac{\sigma_3}{6}}&0\\
     0&0&\mathbf{W}^{(1)}_{\alpha_2,\beta_2}(z)
    \end{pmatrix}P_7^{-1}M_5^{-1},
$$
where

$$
\alpha_1=\dfrac{2\ii\sqrt{3}}{9},\quad\alpha_2=\dfrac{1}{\pi}2^\frac{2}{3}\Gamma\left(\frac{5}{3}\right)\ee^{\frac{5\pi \ii}{6}},\quad 
\beta_1=\frac{2}{3\alpha_1},\quad  \beta_2=\frac{4}{3\alpha_2},\quad 
P_7=\left(\begin{array}{cccccc}
1&0&0&0&0&0\\
0&0&0&0&1&0\\
0&0&1&0&0&0\\
0&0&0&1&0&0\\
0&1&0&0&0&0\\
0&0&0&0&0&1\\
\end{array}\right).
$$
\subsection{Riemann-Hilbert problem \ref{plocz3rhp}}\label{solplocz3rhp}
We first determine the rational multiplier needed to change the local
exponents of $\mathbf W^{(2)}_{\alpha,\beta}$ from $\pm7/6$ to $\pm1/6$.
Write
\[
A(z)=\left(\frac z2+\frac7{6z}\right)\sigma_3+
\begin{pmatrix}0&\alpha\\\beta&0\end{pmatrix},\qquad
\widetilde A(z)=\left(\frac z2+\frac1{6z}\right)\sigma_3+
\begin{pmatrix}0&p\\q&0\end{pmatrix},
\]
where $p,q$ are unknown constants. We seek
\[
R(z)=B_1z+B_0+\frac{B_{-1}}z,
\qquad R'(z)+R(z)A(z)=\widetilde A(z)R(z),
\]
with three arbitrary constant $2\times2$ matrices $B_1,B_0,B_{-1}$.
The coefficient of $z^2$ gives $B_1\sigma_3=\sigma_3B_1$, while the
coefficient of $z^{-2}$ gives
\[
-B_{-1}+\frac76B_{-1}\sigma_3-\frac16\sigma_3B_{-1}=0.
\]
The diagonal part of the coefficient of $z^{-1}$ then forces the diagonal
entries of $B_0$ to vanish. Hence
\[
B_1=\begin{pmatrix}h&0\\0&\ell\end{pmatrix},\qquad
B_0=\begin{pmatrix}0&u\\v&0\end{pmatrix},\qquad
B_{-1}=\begin{pmatrix}k&0\\0&0\end{pmatrix}.
\]
The remaining coefficient equations are
\[
\begin{aligned}
u&=\frac{3k\alpha}{4},&v&=\frac{3kq}{4},\\
u&=h\alpha-p\ell,&v&=qh-\ell\beta,\\
2h+u\beta-pv&=0,&v\alpha-qu&=0.
\end{aligned}
\]
At infinity the original solution has the form
\[
\mathbf W^{(2)}_{\alpha,\beta}(z)
=(I_2+O(z^{-1}))z^{r\sigma_3}\ee^{z^2\sigma_3/4},
\qquad r=\alpha\beta+\frac76.
\]
We require $R\mathbf W^{(2)}_{\alpha,\beta}$ to have leading coefficient
$I_2$ and power $z^{(r-1)\sigma_3}$. The term $hz$ would increase the
first exponent, so $h=0$; the leading coefficient in the second column
requires $\ell=1$. The equations above now give
\[
v=-\beta,\qquad u=\frac{3k\alpha}{4},\qquad
p=-\frac{3k\alpha}{4},\qquad q=-\frac{4\beta}{3k}.
\]
Both differential systems have trace zero. Their normalized fundamental
solutions have determinant one, so $\det R=1$. Since
\[
\det R=k-uv=k\left(1+\frac{3\alpha\beta}{4}\right),
\]
we obtain, with $d=4+3\alpha\beta$,
\[
k=\frac4d,\qquad p=-\frac{3\alpha}{d},\qquad
q=-\frac{\beta d}{3},\qquad
R_{\alpha,\beta}(z)=
\begin{pmatrix}\dfrac4{dz}&\dfrac{3\alpha}{d}\\[2mm]-\beta&z\end{pmatrix}.
\]
Thus the constant term of $\widetilde A$ is determined together with $R$.
For the two parameter pairs below, $d=6$ and $d=8$, respectively.

For $a_{33}=\ee^{-\frac{5\pi\ii}{6}}$ the solution is given by
$$
\Phi_0(z)=P_7\begin{pmatrix}
R_{\alpha_3,\beta_3}(z)\mathbf{W}^{(2)}_{\alpha_3,\beta_3}(z) &0&0\\0&   \ee^{\frac{z^2}{4}\sigma_3}z^{\frac{1}{6}\sigma_3} &0\\
     0&0&R_{\alpha_4,\beta_4}(z)\mathbf{W}^{(2)}_{\alpha_4,\beta_4}(z)
    \end{pmatrix}P_7^{-1}M_5^{-1},
$$
where
$$
\alpha_3=\dfrac{1}{\pi}2^\frac{4}{3}\Gamma\left(\frac{4}{3}\right)\ee^{-\frac{5\pi \ii}{6}},\quad \alpha_4=-\dfrac{64\ii\sqrt{3}}{81},\quad 
\beta_3=\frac{2}{3\alpha_3},\quad  \beta_4=\frac{4}{3\alpha_4},
$$
and
$$
R_{\alpha_3,\beta_3}(z)=\left(\begin{array}{cc}\dfrac{2}{3z}&\dfrac{\alpha_3}{2}\\[2mm]-\beta_3&z\end{array}\right),\quad
R_{\alpha_4,\beta_4}(z)=\left(\begin{array}{cc}\dfrac{1}{2z}&\dfrac{3\alpha_4}{8}\\[2mm]-\beta_4&z\end{array}\right).
$$
Without the rational factors $R_{\alpha,\beta}(z)$ and with $z^{\frac{7}{6}\sigma_3}$ in the middle block, this function has the jumps
of Riemann-Hilbert problem \ref{plocz2rhp}, but it has the local exponents $\pm\frac{7}{6}$ at $z=0$, because of the
term $\frac{7}{6z}$ in the differential equation for $\mathbf{W}^{(2)}_{\alpha,\beta}(z)$. The rational factors are single valued,
$\det R_{\alpha,\beta}(z)=1$, and with $d=4+3\alpha\beta$
$$
\left(\dfrac{d}{dz}R_{\alpha,\beta}(z)+R_{\alpha,\beta}(z)\left(\left(\frac{z}{2}+\frac{7}{6z}\right)\sigma_3+\begin{pmatrix}
0&\alpha\\
\beta&0\end{pmatrix}\right)\right)R_{\alpha,\beta}^{-1}(z)=\left(\frac{z}{2}+\frac{1}{6z}\right)\sigma_3+\begin{pmatrix}
0&-\frac{3\alpha}{d}\\
-\frac{\beta d}{3}&0\end{pmatrix}.
$$
Hence $\Phi_0(z)$ has the jumps of Riemann-Hilbert problem \ref{plocz3rhp} and the local exponents $\pm\frac{1}{6}$ at $z=0$. At infinity
$$
R_{\alpha,\beta}(z)\mathbf{W}^{(2)}_{\alpha,\beta}(z)=\left(\mathbb{1}+\dfrac{1}{z}\begin{pmatrix}
        0&\frac{3\alpha}{d}\\-\frac{\beta d}{3}&0
    \end{pmatrix}+O\left(\dfrac{1}{z^2}\right)\right)z^{\left(\alpha\beta+\frac{1}{6}\right)\sigma_3}\ee^{\frac{z^2}{4}\sigma_3},
$$
so the exponent at infinity is shifted by $-\sigma_3$ in each block, which replaces $\Lambda_{10}$ by $\Lambda_{13}$, and \eqref{m10third}
follows.

\section{Right tail behavior of the Tracy-Widom distribution function in  case \texorpdfstring{$\beta=2$}{beta=2}  via Calogero-Painlev\'e system}
\label{app:beta_two}

In case $\beta=2$, , the Calogero-Painlev\'e
system becomes just  one equation for $x_1(t) \equiv x(t)$  - a second Painlev\'e equation, with the parameter $\theta = 1/2$, i.e.,
\begin{equation}\label{p21/2}
x'' = 2x^3 +tx + \frac{1}{2}.
\end{equation}
Simultaneously,  formula \eqref{betacc1} transforms to the equation \footnote{for general $\beta = 2n$, we have $\theta = \frac{2-n}{2n}$,  $g =\frac{i}{n}$
in the Calogero-Painlev\'e system and the formulae for $F_{2n}$ reads:
$$
\log  F_{2n}(-2^{-1/3}t) = \int _{-\infty}^{t} \left(H -\frac{ns^2}{8} + \frac{1}{2}\sum_{k=1}^{n}x_k(s) \right)ds,
$$
where the Hamiltonian 
$$
H = \sum_{k=1}^{n}\left(\frac{(x_k'(s))^2}{2} - \frac{x_k^4(s)}{2} -\frac{sx_k^2(s)}{2}  -\theta x_k\right) + g^2\sum_{j<k}\frac{1}{(x_k-x_j)^2}
$$},
\begin{equation}\label{F2}\aligned
\log F_2(-2^{-1/3}t) = \int _{-\infty}^{t} \left(\frac{(x'(s))^2}{2} - \frac{x^4(s)}{2} -\frac{sx^2(s)}{2} -\frac{s^2}{8}\right)ds
\\= \frac{1}{2}\int_{-\infty}^{t}(t-s)\left(x'(s) -x^2(s) -\frac{s}{2}\right)ds.\endaligned
\end{equation}
The relevant solution $x(t)$, i.e. the $\beta =2$  analog of our special $\beta=6$  solution is characterized by the asymptotics
($\beta =2$ analogs of \eqref{x1as} and \eqref{x2as}),
\begin{equation}\label{1.20_2}
x(t) = -\sqrt{\frac{|t|}{2}} + \frac{1}{4t} + O(|t|^{-5/2}), \quad t \to -\infty,
\end{equation}
\begin{equation}\label{1.25_2}
x(t) = -\frac{1}{2t} - \frac{3}{4t^4} + O(t^{-7}), \quad t \to \infty.
\end{equation}
Important: in the case of $\beta =2$ we do know that asymptotics \eqref{1.20_2} and \eqref{1.25_2} determine the solution
$x(t)$ uniquely; indeed, this is the non-homogeneous analog (presence of the constant $\theta = 1/2$) of the classical Hastings-McLeod 
solution.  Moreover, the solution is known to be smooth for all real $t$. Also, as in the case of $\beta=6$, a direct substitution of
\eqref{1.25_2} into \eqref{F2} yields the correct (without the constant term)  behavior of $F_2$ for $t \to  -\infty$,
$$
\log F_2(t)  = -\frac{|t|^3}{12} - \frac{1}{8}\log |t| + \chi+ O(|t|^{-1}).
$$
At the same time, what we  get from the substitution of \eqref{1.20_2} into \eqref{F2}  is,
$$
F_2(t) = 1 + O\left(\frac{1}{t^p}\right), \quad t \to \infty,
$$
with $p$ a multiple of $\frac32$ determined by the number of terms kept, with the possible improvement to $p=\infty$. However, from the alternative formulae for $F_2$, 
\begin{equation}\label{F2alt}
\log F_2(t) = -\int _{t}^{\infty} (s-t)u^2(s) ds 
\end{equation}
where $u(t)$ is the classical Hastings-McLeod solution of the homogeneous Painlev\'e II equation,
$$
u'' = tu + 2u^3, \quad u \sim \frac{1}{2\sqrt{\pi}t^{1/4}}e^{-\frac{2}{3}t^{3/2}}, \quad t \to \infty,
$$
we know that
\begin{equation}\label{F21}
F_2(t)  \sim   1 - \frac{1}{16\pi t^{3/2}}e^{-\frac{4}{3}t^{3/2}}, \quad t \to \infty.
\end{equation}
Hence, if we start with the representation of $F_2$ -function by (\ref{F2} )  we would  have the 
same right tail problem as in the case of $\beta =6$. In the case of $\beta =2$ though  there is 
a way out without using the alternative formula \eqref{F2alt}.  In what follows we shall describe 
how indeed the problem can be solved in the case $\beta=2$.

We shall need a few general facts from \cite{FIKN} concerning the solutions of the second Painlev\'e equation \eqref{p21/2} with the asymptotic
behavior \eqref{1.20_2}  on the minus infinity.  
\begin{proposition}[\cite{FIKN}, Theorems 11.2 and 11.3]\label{proposition1}
There is a one - parameter family of the solutions of \eqref{p21/2}, real for real t,  with the asymptotic behavior,
\begin{equation}\label{1.20_2.1}
x(t) \sim -\sqrt{-\frac{t}{2}}\sum_{n=0}^{\infty}c_n(-t)^{-\frac{3n}{2}}, \quad |t| \rightarrow \infty, \quad \arg t = \pi + \arg(-t) \in \left(\frac{2\pi}{3}, \frac{4\pi}{3}\right),
\end{equation}
where the coefficients $c_n$ are given by the recurrent relations, 
$$
c_0 =1, \quad c_1 = \frac{1}{2\sqrt{2}},
$$
\begin{equation}\label{cn}
c_{n+2} = \frac{9n^2-1}{8}c_n -\sum_{m=1}^{n+1}c_{m}c_{n+2-m}
-\frac{1}{2}\sum_{l=1}^{n+1}\sum_{m=1}^{n+2-l}c_{l}c_{m}c_{n+2-l-m},\quad n\geq 0.
\end{equation}
These solutions are smooth for all sufficiently large negative $t$; the distinguished solution $x_0(t)$ is smooth for all $t\leq0$.
\end{proposition}
\begin{proposition}[\cite{FIKN}(11.4.16)]\label{proposition2}
There is a  unique member, $x_{0}(t)$,  of the above family that has asymptotics \eqref{1.20_2.1} valid in  the bigger sector,
\begin{equation}\label{1.20_2.2}
x_{0}(t) \sim -\sqrt{-\frac{t}{2}}\sum_{n=0}^{\infty}c_n(-t)^{-\frac{3n}{2}}, \quad |t| \rightarrow \infty, \quad \arg t = \pi + \arg(-t) \in \Bigl(0, 2\pi \Bigr).
\end{equation}
Moreover, this solution is given by the explicit formulae,
\begin{equation}\label{formula}
x_0(t) = 2^{-1/3}\frac{\mathrm{Ai}'(z_0)}{\mathrm{Ai}(z_0)}, \quad z_0 = -2^{-1/3} t,
\end{equation}
where  $\mathrm{Ai}(z)$ is the Airy function and $\mathrm{Ai}'(z) = \frac{d\mathrm{Ai}(z)}{dz}$.
\end{proposition}
\begin{proposition}[\cite{FIKN}, Theorems 11.2 and 11.3]\label{proposition3}
Every solution from  family \eqref{1.20_2.1} satisfies the asymptotic relation,
\begin{equation}\label{1.20_2.4}
x(t) - x_{0}(t) = \frac{C}{t}e^{-\frac{2\sqrt{2}}{3}(-t)^{3/2}}\Bigl( 1 + O(t^{-1/4})\Bigr), 
\quad |t| \rightarrow \infty, \quad \arg t = \pi + \arg(-t) \in \left(\frac{2\pi}{3}, \frac{4\pi}{3}\right),
\end{equation}
where the  real number $C$ can be taken as a parameter of the family \eqref{1.20_2.1}. 
\end{proposition}
\begin{proposition}[\cite{FIKN}, Theorems 11.5 and 11.6]\label{proposition4}
If in \eqref{1.20_2.4} 
\begin{equation}\label{C}
C= \frac{1}{8\pi},
\end{equation}
then the solution $x(t)$  also satisfies the  asymptotics,
\begin{equation}\label{x+infty}
x(t) \sim -\frac{1}{2}\sum_{n=0}^{\infty}a_{n}t^{-3n - 1}, \quad |t| \rightarrow \infty, \quad \arg t\in \left(-\frac{\pi}{3}, \frac{\pi}{3}\right),
\end{equation}
where  the coefficients $a_n$ are given by the recurrent relations, 
$$
a_0 =1,
$$
\begin{equation}\label{an}
a_{n+1} = (3n+1)(3n+2)a_n -\frac{1}{2}\sum_{k,l,m =0; k+l+m = n}^{n} a_{l}a_{m}a_{k}, \quad n \geq 0.
\end{equation}
In other words,  the special solution  \eqref{1.20_2} - \eqref{1.25_2} is the solution from  family  \eqref{1.20_2.1} whose $C$-parameter is 
given by \eqref{C}. 
\end{proposition} 
\begin{remark}
Strictly speaking Theorems 11.2, 11.3, 11.5, 11.6 are proven in \cite{FIKN} only for $\alpha-\frac{1}{2}\notin \mathbb{Z}$. The relevant extension can be produced after reasonable effort. 
\end{remark}
Let us proceed now  with showing that formula \eqref{F2} yields the estimate \eqref{F21}. We first observe that formulae
\eqref{formula} implies that the solution $x_0(t)$, in addition to the Painlev\'e equation \eqref{p21/2}, also satisfies the Riccati equation,
\begin{equation}\label{Riccati}
x_0'(t) -x_0^2(t) -\frac{t}{2} = 0.
\end{equation}
Note that with the replacement of $x_0(t)$ by $x(t)$, the  left hand side of this equation is exactly the expression appearing in the integrand in formula \eqref{F2}. Define,
$$
r(t) := x(t) - x_0(t).
$$
We will obtain then,
\begin{equation}\label{F22}
x'-x^2 -\frac{t}{2} = x_0'(t) -x_0^2(t) -\frac{t}{2} +r' -2x_0r - r^2 = r' -2x_0r - r^2,
\end{equation}
where in the last equality we made use of \eqref{Riccati}. 
Because of \eqref{1.20_2.4} and \eqref{C}  we have that
$$
r(t) =  \frac{1}{8\pi t}\ee^{-\frac{2\sqrt{2}}{3}(-t)^{3/2}}\Bigl( 1 + O(t^{-1/4})\Bigr), \quad |t| \rightarrow \infty, \quad \arg t = \pi + \arg(-t) \in \left(\frac{2\pi}{3}, \frac{4\pi}{3}\right).
$$
Plugging this asymptotics into \eqref{F22} we shall arrive at the estimate,
\begin{equation}\label{F23}
 x'(t)-x(t)^2-\frac t2
 =-\frac{1}{2\sqrt2\pi}(-t)^{-1/2}
 \ee^{-\frac{2\sqrt2}{3}(-t)^{3/2}}
 \bigl(1+O((-t)^{-1/4})\bigr),\qquad t\to-\infty.
\end{equation}
$$
|t| \rightarrow \infty, \quad \arg t = \pi + \arg(-t) \in \left(\frac{2\pi}{3}, \frac{4\pi}{3}\right).
$$
In its turn, asymptotics \eqref{F23} produces, via the integral formula \eqref{F2},  the needed estimate \eqref{F21} for the right  tail of the  Tracy-Widom
distribution $F_2(t)$.

Let's come back to $\beta =6$.  Keeping in mind the just described $\beta=2$ case, a natural idea of how to transform estimate \eqref{F6}
into estimate \eqref{expconst6} would be to find a $\beta=6$  analog of the  explicit solution $x_0(t)$. In more details, we are thinking
of the following program. 

First we notice that similar to \eqref{F2}, equation \eqref{betacc1} can be rewritten as,
\begin{equation}\label{F66}
\log F_6(-2^{-1/3}t) = \frac{1}{2}\int_{-\infty}^{t}(t-s)\mathrm{Tr} \left(q'(s) -q^2(s) -\frac{s}{2}\right)ds
\end{equation}
where $q(t)$ is the solution of the matrix Painlev\'e II equation \eqref{PIIv} whose eigenvalues, $x_1(t)$, $x_2(t)$, $x_3(t)$ form the
solution \eqref{cc6}, \eqref{x1as}. Similar to the case $\beta =2$, our goal is to show that 
\begin{equation}\label{goal}
 \operatorname{Tr}\left(q'(t)-q(t)^2-\frac t2 I_3\right)
 =-\frac{1}{96\sqrt2\pi}(-t)^{-7/2}
 \ee^{-2\sqrt2(-t)^{3/2}}\bigl(1+o(1)\bigr),\qquad t\to-\infty.
\end{equation}
The sign and coefficient are fixed by the normalization of the desired
right tail. Indeed, put $\tau=-t$ and $L(t)=\log F_6(-2^{-1/3}t)$.
The double-integral formula gives
$2L''(t)=\operatorname{Tr}(q'-q^2-tI_3/2)$. The required leading term is
$L(t)\sim-2^{3/2}(13824\pi)^{-1}\tau^{-9/2}
\ee^{-2\sqrt2\tau^{3/2}}$; its twice-differentiated leading coefficient
is $-1/(96\sqrt2\pi)$ in $2L''$. This calculation fixes the target
estimate and does not prove the exponentially small asymptotics.

Indeed,  similar \eqref{F23} in the case of $\beta =2$, this estimate  produces, via the integral formula \eqref{F66},  the needed estimate \eqref{expconst6} for the right  tail of the  Tracy-Widom distribution $F_6(t)$.

Using again the experience we have gained in the case $\beta=2$, a  natural way to proceed is to try to find a special, presumably explicit 
solution $(x_{01}(t), x _{02}(t), x_{03}(t))$ of the Calogero-Painlev\'e system \eqref{cc6} such that it belongs to the four parameter family  \eqref{ansatzminf}, \eqref{eq:monodromy_minus_infty}. of the
solutions and for the corresponding  solution $q_0(t)$ of the matrix Painlev\'e II  equation \eqref{PIIv} a matrix analog of the  Riccati equation
\eqref{Riccati},
\begin{equation}\label{Riccat6}
\mathrm{Tr} \left(q'_0(t) -q^2_0(t) -\frac{t}{2}\right) =0 
\end{equation}
holds. If, in addition, we will be able to show that for  a proper constant  matrix $C$,  the matrix Painlev\'e II  function  $q(t)$ corresponding
to the solution \eqref{S1S2} of the  Calogero-Painlev\'e system \eqref{cc6} satisfies the asymptotic relation  (cf. \eqref{1.20_2.4}),
\begin{equation}\label{goal2}
q(t) - q_0(t) = C\frac{1}{(-t)^{9/2}}\ee^{-2\sqrt{2}(-t)^{3/2}}\Bigl(1 + o(1)\Bigr),
\end{equation}
$$
|t| \rightarrow \infty, \quad \arg t = \pi + \arg(-t) \in \left(\frac{2\pi}{3}, \frac{4\pi}{3}\right),
$$
then, again similar to  $\beta=2$ case, we will reach our goal - estimate \eqref{goal}. We hope to be able to realize this program
in the forthcoming  publication.

\printbibliography
\end{document}